\documentclass[prx,twocolumn,superscriptaddress,longbibliography]{revtex4-1}
\usepackage[utf8]{inputenc}
\usepackage{color}
\usepackage{amsmath}
\usepackage{amsfonts}
\usepackage{hyperref}
\usepackage{comment}
\usepackage{amsthm}
\usepackage{bm}
\usepackage{float}
\usepackage{accents}
\usepackage[percent]{overpic}
\usepackage{natbib}
\usepackage[final]{changes}

\hypersetup{
  colorlinks = true,
  allcolors= blue,
}
\usepackage{tikz,pgfplots}
\usepackage{blkarray}
\usetikzlibrary{patterns}
\usetikzlibrary{fadings}
\usetikzlibrary{decorations.pathmorphing,patterns}
\tikzset{snake it/.style={decorate, decoration=snake}}
\usetikzlibrary{arrows, decorations.markings}
\pgfplotsset{compat=1.14}
\tikzset{
vecArrow/.style={
  thick,
  decoration={markings,mark=at position
   1 with {\arrow[scale=2,thin]{open triangle 60}}},
  double distance=1.4pt, shorten >= 10.5pt,
  preaction = {decorate},
  postaction = {draw,line width=1.4pt, white,shorten >= 4.5pt}
  },
innerWhite/.style={
  semithick,
  white,
  line width=1.4pt,
  shorten >= 4.5pt
  }
}

\makeatletter
\def\@bibdataout@aps{%
\immediate\write\@bibdataout{%
@CONTROL{%
apsrev41Control%
\longbibliography@sw{%
    ,author="08",editor="1",pages="1",title="0",year="1"%
    }{%
    ,author="08",editor="1",pages="1",title="",year="1"%
    }%
  }%
}%
\if@filesw \immediate \write \@auxout {\string \citation {apsrev41Control}}\fi 
}
\makeatother
\definecolor{dgreen}{rgb}{0,0.4,0}
\colorlet{Changes@Color}{red}
\setaddedmarkup{\textcolor{dgreen}{#1}}
\newcommand{\stkout}[1]{\ifmmode\text{\st{\ensuremath{#1}}}\else\st{#1}\fi}
\setdeletedmarkup{}

\definecolor{orange}{rgb}{1,0.5,0}
\definecolor{darkgreen}{rgb}{0,0.4,0.1}
\DeclareUnicodeCharacter{300}{à}

\newcommand*{\state}{\ensuremath{\nu}}
\newcommand*{\statep}{\ensuremath{ {\nu'} }}

\DeclareMathAlphabet{\mathsfit}{\encodingdefault}{\sfdefault}{m}{sl}
\SetMathAlphabet{\mathsfit}{bold}{\encodingdefault}{\sfdefault}{bx}{sl}

\begin{document}


\title{Generalization of Fourier's law into viscous heat equations}

\author{Michele Simoncelli}
\email{michele.simoncelli@epfl.ch}
\affiliation{Theory and Simulation of Materials (THEOS) and National Centre for Computational Design and Discovery of Novel Materials (MARVEL), \'Ecole Polytechnique F\'ed\'erale de Lausanne, Lausanne, Switzerland.}

\author{Nicola Marzari}
\affiliation{Theory and Simulation of Materials (THEOS) and National Centre for Computational Design and Discovery of Novel Materials (MARVEL), \'Ecole Polytechnique F\'ed\'erale de Lausanne, Lausanne, Switzerland.}

\author{Andrea Cepellotti}
\affiliation{Department of Physics, University of California at Berkeley and Materials Sciences Division, Lawrence Berkeley National Laboratory, Berkeley, California 94720, USA}

\begin{abstract}
Heat conduction in dielectric crystals originates from the propagation of atomic vibrations, whose microscopic dynamics is well described by the linearized 
phonon Boltzmann transport equation. 
Recently, it was shown that thermal conductivity can be resolved exactly and in a closed form as a sum over relaxons,
\textit{i.e.} collective phonon excitations that are the eigenvectors of Boltzmann equation's scattering matrix [Cepellotti and Marzari, Phys. Rev. X \textbf{6}, 041013 (2016)]. Relaxons have a well-defined parity, and only odd relaxons contribute to the thermal conductivity. 
Here, we show that the complementary set of even relaxons determines another quantity --- the thermal viscosity --- that enters into the description of heat transport, and is especially relevant in the hydrodynamic regime, where dissipation of crystal momentum by Umklapp scattering phases out. 
We also show how the thermal conductivity and viscosity parametrize two novel viscous heat equations --- two coupled equations for the temperature and drift-velocity fields --- which represent the thermal counterpart of the Navier-Stokes equations of hydrodynamics in the linear, laminar regime. These viscous heat equations are derived from a coarse-graining of the linearized Boltzmann transport equation for phonons, 
\added{and encompass both limits of Fourier's law and of second sound, taking place, respectively, in the regimes of strong or weak momentum dissipation. 
Last, we introduce the Fourier deviation number as a descriptor that captures the deviations from Fourier's law due to hydrodynamic effects.
We showcase these findings in a test case of a complex-shaped device made of graphite, obtaining a remarkable agreement with the very recent experimental demonstration of hydrodynamic transport in this material. 
The present findings also suggest that hydrodynamic behavior can appear at room temperature in micrometer-sized diamond crystals.
The present formulation rigorously generalizes Fourier's heat equation, extending the reach of physical and computational models for heat conduction also to the hydrodynamic regime.
}
\end{abstract}

\maketitle

\section{Introduction} 
\label{sec:introduction}

Thermal transport in insulating crystals takes place through the evolution and dynamics of the vibrations of atoms around their equilibrium positions.
The first predictive theoretical framework to describe thermal transport was developed by Peierls in 1929~\cite{peierls1929kinetischen,peierls1955quantum,ziman1960electrons}, who envisioned a microscopic theory in terms of a Boltzmann transport equation (BTE) for the propagation of vibrational excitations (phonon wavepackets).
In the 1960s significant progress took place in this field, propelled by newly discovered  hydrodynamic phenomena in crystals, with striking signatures such as Poiseuille-like heat flow~\cite{mezhov1966measurement} and second sound~\cite{PhysRevLett.16.789}.
The former manifests itself with a heat flux that is akin to the flow of a fluid in a pipe (\textit{i.e.} showing a parabolic-like profile with a maximum in the center and minimum at the boundaries, due to viscous effects); the latter instead results in heat propagation in the form of a coherent temperature wave, rather than a diffusing heat front. 
Second sound has been observed experimentally in a handful of solids; first, in solid helium~\cite{PhysRevLett.16.789}, followed by sodium fluoride \cite{PhysRevLett.25.26,PhysRevLett.36.480}, bismuth~\cite{PhysRevLett.28.1461}, sapphire~\cite{danil1979observation}, and strontium titanate~\cite{PhysRevLett.75.2416,PhysRevB.76.075207} -- all at cryogenic conditions. Importantly, neither Poiseuille flow nor second sound can be described by
the \added{{macroscopic Fourier's equation, which is limited to a diffusive description of heat propagation.}
}

These experimental observations have been accompanied by several pioneering efforts aimed at providing a quantitative description of heat hydrodynamics \added{at the 
mesoscopic level, \textit{i.e.} in terms of partial differential equations (PDEs) that are simpler than the microscopic integro-differential BTE (we use ``mesoscopic model'' to denote any description requiring more fields or PDEs than the Fourier's PDE for the temperature field~\cite{bergamasco2018mesoscopic}).}
Sussmann and Thellung~\cite{Sussmann_Thellung_1963}, starting from the linearized BTE (LBTE) in the absence of momentum-dissipating (Umklapp) phonon-phonon scattering events, derived mesoscopic equations in terms of the temperature and of the phonon drift velocity, \textit{i.e.} the thermal counterparts of pressure and fluid velocity in liquids.
Further advances came from Gurzhi~\cite{gurzhi1964thermal,gurzhi_hydrodynamic_1968} and Guyer \& Krumhansl~\cite{PhysRev.148.766,PhysRev.148.778} who, including the effect of weak crystal momentum dissipation, obtained equations for damped second sound and for Poiseuille heat flow.
Among early works, we also mention the discussions of phonon hydrodynamics in the framework of many-body theory,  as in Refs.~\cite{griffin1965detection,enz1968one}.
While correctly capturing the qualitative features of phonon hydrodynamics, all the theoretical investigations mentioned above assume simplified phonon dispersion relations (either power-law~\cite{gurzhi1964thermal,gurzhi_hydrodynamic_1968} or linear isotropic~\cite{Sussmann_Thellung_1963, PhysRev.148.766,PhysRev.148.778}), or neglect momentum dissipation~\cite{Sussmann_Thellung_1963}.
A  more rigorous and general formulation --- albeit valid only in the hydrodynamic regime of weak Umklapp scattering --- was introduced by Hardy, who extended the study of second sound~\cite{hardy1970phonon} and, together with Albers, of Poiseuille flow in terms of mesoscopic transport equations~\cite{PhysRevB.10.3546}. 

The turn of the century brought renewed interest in the theory of heat conduction; computational and algorithmic advances now allow to solve exactly the LBTE 
--- employing iterative~\cite{omini1995iterative,broido2007intrinsic,carrete2017almabte}, variational~\cite{fugallo2013ab}, or exact diagonalization~\cite{chaput2013direct,cepellotti2016thermal} methods --- and thus investigate the accuracy of the LBTE and the models derived from it.
In addition, nowadays it is possible to solve the LBTE without any fitting parameter, deriving all quantities from first-principles; this has been shown to describe accurately the thermal properties of bulk crystals~\cite{ broido2007intrinsic, esfarjani2011heat, Chen_Science_12, Nanoscale_Thermal_14, cepellotti2015phonon, paulatto2013anharmonic, paulatto2015first,lindsay_first_2016, cepellotti2017boltzmann,  PhysRevB.53.9064, mcgaughey2019phonon}, provided phonon branches remain well-separated \cite{simoncelli2019unified}. 

Further applications of the LBTE, combined with state-of-the-art first-principles simulations, have also recently predicted the existence of hydrodynamic phenomena at non-cryogenic temperatures in low-dimensional or layered materials such as graphene \cite{ lee_hydrodynamic_2015, cepellotti2015phonon, cepellotti2016second_sound}, other 2D materials \cite{cepellotti2015phonon}, carbon nanotubes~\cite{lee_hydrodynamic_2017} and graphite~\cite{ding_phonon_2018}.
Fittingly, and remarkably, these theoretical suggestions have now been confirmed by the experimental finding of second sound in graphite~\cite{Chen_science_2019} at $\sim 100$ K.

In the hydrodynamic regime, where Poiseuille flow or second sound occur, Fourier's law fails~\cite{martelli_thermal_2018,second_sound_diamond,machida_observation_2018,Chen_science_2019}, depriving us of the most common tool used to predict the temperature profile in a device.
The LBTE, in principle, allows to predict accurately thermal transport under these conditions, but its complexity prevents a straightforward application to materials with complex geometries (used in experiments and relevant for applications), thus posing limitations to the study of how a material's shape alters transport~\cite{cepellotti2017boltzmann}.
Recent research efforts have been directed at developing mesoscopic models that correct the shortcomings of Fourier's law \added{and extend it} at a lower computational cost than the \added{solution of the full} LBTE. \added{Different strategies have been suggested;
some approaches reduce the complexity of the LBTE by neglecting the effects of the phonon modes repopulation due to scattering events (the so-called single-mode relaxation-time approximation (SMA)), thus allowing for analytical~\cite{hua2014analytical, PhysRevB.91.085202, PhysRevB.84.195206, PhysRevB.91.165311} or asymptotic~\cite{Peraud2016} solutions.}
From the LBTE in the SMA, mesoscopic models that generalize Fourier's law accounting for ultrafast thermal processes or  ballistic effects \added{have been derived~\cite{chen2001ballistic,PhysRevLett.96.184301,ordonez2011constitutive,ramu2014enhanced,hua2019experimental}.}
Other works have derived mesoscopic models without relying on the LBTE~\cite{cao2007equation}, 
or have generalized the Guyer-Krumhansl equation to account for the effect of the boundaries on the heat flow~\cite{alvarez2009phonon,guo2015phonon,Guo2018,ziabari_full-field_2018,PhysRevMaterials.2.076001}.
A hydrodynamic transport model has been derived from the LBTE in the Callaway approximation, defining a phonon viscosity which can be computed from atomistic data~\cite{PRB_Viscosity_Callaway_in_s}.

Here, we \added{provide a general and universal solution to the challenge of extending Fourier's law all the way to the hydrodynamic regime, deriving from the LBTE two novel coupled mesoscopic heat transport equations that cover exactly and on equal footing Fourier diffusion, hydrodynamic propagation, and all regimes in between.}

\added{To this aim, we first show that one can define the thermal viscosity of a crystal starting from an exact solution of the LBTE in terms of the eigenvectors of the scattering matrix (\textit{i.e.} the relaxons introduced in Ref.~\cite{cepellotti2016thermal} to determine thermal conductivity), and evaluate from it the crystal-momentum flux generated in response to a drift-velocity gradient.
The relaxons' parity~\cite{cepellotti2016thermal} highlights the complementary character of thermal conductivity and viscosity, with the former being determined by odd relaxons, and the latter by even relaxons.}

Next, we use a coarse-graining procedure to derive two novel ``viscous'' heat equations: these
are two coupled equations for the local temperature and drift-velocity fields, and are parametrized in terms of the thermal conductivity and viscosity.
The viscous heat equations represent the thermal counterpart of the Navier-Stokes equations for fluids in the laminar regime, and, as mentioned, include Fourier's law and second sound in the limits of strong and weak crystal momentum dissipation, respectively.

Last, we introduce the Fourier deviation number (FDN), a dimensionless parameter that quantifies the deviation from Fourier's law due to hydrodynamic effects.
We test this formalism on \added{graphite,} diamond and silicon, 
\added{showing that the FDN predicts a temperature and size for the window of hydrodynamic thermal transport that replicates the explicit solution of the viscous heat equations, but at a negligible computational cost. 
Most importantly, 
the prediction of the present formulation for the hydrodynamic window of graphite is found to be in excellent agreement with recent, pioneering experiments~\cite{Chen_science_2019}. 
%
In passing, we also predict that hydrodynamic behavior can appear in diamond at room temperature for micrometer-sized crystals.}

\section{Thermal viscosity} 
\label{sec:thermal_viscosity}
A microscopic description of thermal transport is given by the LBTE:
\begin{equation}
  \frac{\partial n_{\state} (\bm{r},t)}{\partial t}
  + \bm{v}_{\state}\cdot \nabla { n_{\state} (\bm{r},t)} 
  = -\frac{1}{V} \sum_{\statep} \Omega_{\state \statep} n_{\state}(\bm{r},t) \;,
  \label{eq:LBTE}
\end{equation}
where $\state$ labels a phonon state (\textit{i.e.} an index running on all the phonon wavevectors $\bm{q}$ and phonon branches $s$), $\bm{v}_{\state}$ is the phonon group velocity, $V$ is the crystal volume \footnote{$V=\mathcal{V}N_{\rm c}$ \textit{i.e.}  the unit cell volume $\mathcal{V}$ times the number of unit cells that constitute the crystal $N_{\rm c}$. $N_{\rm c}$ is also the number of wavevectors $\bm{q}$ used to sample the Brillouin zone}, and $\Omega_{\state \statep}$ is the phonon-phonon scattering matrix \cite{cepellotti2016thermal}.
Eq.~(\ref{eq:LBTE}) governs the evolution of the deviation $n_{\state}(\bm{r},t)$ of the phonon populations from equilibrium:
\begin{equation}
  n_{\state}(\bm{r},t)=N_{\state}(\bm{r},t)-\bar{N}_{\state}\;,
  \label{eq:out_of_eq_deviation}
\end{equation}
where $N_{\state}(\bm{r},t)$ are the out-of-equilibrium phonon populations at position $\bm{r}$ and time $t$,  $\bar{N}_{\state}=(e^{{\hbar\omega_{\state}}/{(k_B \bar{T}) }}-1)^{-1}$ is the equilibrium Bose-Einstein distribution at temperature $\bar{T}$, and $\omega_{\state}$ are the phonon frequencies.
From the solution of the LBTE one can derive the local lattice energy $E(\bm{r},t) {=} \frac{1}{V}\sum_{\state} \hbar\omega_\state N_{\state}(\bm{r},t)$ and the total crystal momentum $\bm{P}(\bm{r},t){=} \frac{1}{V}\sum_{\state} \hbar \bm{q} N_{\state}(\bm{r},t)$ \cite{hardy1970phonon}.
The former is often studied in connection with the thermal conductivity~\cite{cepellotti2016thermal}, while the latter becomes relevant in the hydrodynamic regime of thermal transport~\cite{Klemens_1951,Sussmann_Thellung_1963,gurzhi_hydrodynamic_1968}.
\added{We note in passing that this latter emerges only in ``simple'' crystals, \textit{i.e.} those where phonon interbranch spacings are much larger than their linewidths \cite{simoncelli2019unified}.}

The energy flux generated in response to a temperature gradient determines the thermal conductivity; \added{correspondingly,} the  crystal-momentum flux generated in response to a perturbation of the drift velocity \added{determines} the thermal viscosity (for the electronic analogous, see Ref.~\cite{rice1967theory}).
Therefore, we start by considering a crystal in the hydrodynamic regime of thermal transport (\textit{i.e.} carrying a finite amount of crystal momentum);
under the constraint of fixed total energy and crystal momentum, the local equilibrium is given by the phonon drifting distribution $N^D_{\state}(T(\bm{r},t) ,\bm{u}(\bm{r},t))$~\cite{gurzhi_hydrodynamic_1968} 
\begin{equation}
\label{eq:drifting_distribution}
    N^D_{\state} (T(\bm{r},t) ,\bm{u}(\bm{r},t)) 
    = 
    \frac{1}{e^{\frac{1}{k_B T(\bm{r},t)}(\hbar\omega_{\state}- \hbar\bm{q}\cdot\bm{u}(\bm{r},t))}-1} \;.
\end{equation}
\added{The drifting distribution} differs from the Bose-Einstein distribution due to the presence of a drift velocity $\bm{u}$ (a parameter \added{expressing} the amount of local momentum, just \added{as the} temperature \added{does for} the local energy); it depends implicitly on $\bm{r},t$ through $T(\bm{r},t)$ and $\bm{u}(\bm{r},t)$.
Next, we study the effect of small perturbations in the temperature and drift velocity.
To this aim, we expand the out-of-equilibrium drifting distribution~(\ref{eq:out_of_eq_deviation}) in proximity of the local thermal equilibrium \cite{hardy1970phonon}, finding
\begin{equation}
\begin{split}
   n_{{\state}}(\bm{r},t) 
    &=  \frac{\partial {N}^D_{\state} }{\partial T}\bigg|_{\rm eq}\hspace*{-3mm} (T(\bm{r},t) - \bar{T})
    + \frac{\partial N^D_{\state} }{\partial \bm{u}}\bigg|_{\rm eq}\hspace*{-3mm} \cdot \bm{u}(\bm{r},t)  + n^{\delta}_{\state}(\bm{r},t) \\
    &=n^T_{\state}(\bm{r},t) + n^D_{\state}(\bm{r},t) + n^{\delta}_{\state}(\bm{r},t) ,
    \label{eq:hardy_decomposition}
    \end{split}
    \raisetag{4.5mm}
\end{equation}
where $n^T_{\state}$ arises from the change in the local temperature~\cite{Allen_2018_2}, $n^D_{\state}$ from the local drift velocity, and $n^{\delta}_{\state}$ accounts for all the information that cannot be mapped to a local equilibrium state; the derivatives are computed at equilibrium where $T(\bm{r},t)=\bar{T}$ and $\bm{u}(\bm{r},t)=0$.
In analogy with previous work~\cite{fugallo2013ab,cepellotti2016thermal}, \added{we consider the steady-state case and linearize the LBTE around the constant temperature and drift-velocity gradients (\textit{i.e.} $n^{\delta}_{\state}$, $\nabla T$, and $\nabla \bm{u}$ are constant).}
\deleted{of the order of the temperature or drift velocity gradients and  
has to be determined solving the LBTE.}
Then, we substitute Eq.~(\ref{eq:hardy_decomposition}) in Eq.~(\ref{eq:LBTE}) and, keeping only 
terms linear in the temperature and drift-velocity gradients, we obtain 
\begin{equation}
\begin{split}
      &\frac{\partial \bar{N}_{\state}}{\partial T}\bm{v}_{\state} \cdot \nabla T 
    + \bm{v}_{\state} \cdot \bigg(\frac{\partial N^D_{\state}}{\partial \bm{u}} \cdot \nabla \bm{u}\bigg)   \\
    &\hspace{10mm}= -\frac{1}{V} \sum_{\statep} \Omega_{\state \statep} \Big(n^T_{\statep}(\bm{r},t) + n^D_{\statep}(\bm{r},t) + n^{\delta}_{\statep}
     \Big).
\end{split}
\label{eq:LBTE_expanded_gradiend}
\end{equation}
We recast Eq.~(\ref{eq:LBTE_expanded_gradiend}) in the symmetric (thus diagonalizable) form, \textit{i.e.} in terms of $\tilde{\Omega}_{\state \statep} = 
    \Omega_{\state \statep} \sqrt{ \frac {\bar{N}_{\statep}(\bar{N}_{ \statep}+1)} {\bar{N}_{\state}(\bar{N}_{\state}+1)} }$ and $\tilde{n}_{\state}{(\bm{r},t)}={ n_{\state}{(\bm{r},t)}}[{{\bar{N}_{\state}(\bar{N}_{\state}+1)}}]^{-\frac{1}{2}}$ with the goal of using the relaxon picture~\cite{cepellotti2016thermal} to gain insight in the physics underlying transport.
We then simplify the symmetrized Eq.~(\ref{eq:LBTE_expanded_gradiend}) exploiting parity: we recall that a function $f_{\state}$ is even if $f_{\state} = f_{-\state}$ (this is e.g. the case of the phonon energy $\hbar\omega_{\state}=\hbar\omega_{-\state}$), and odd if $f_{\state} = - f_{-\state}$ (e.g. the phonon group velocity $\bm{v}_{\state}=-\bm{v}_{-\state}$), and use the notation $-\nu=(-\bm{q},s)$.
Therefore, $\frac{\partial \bar{N}_{\state}}{\partial T}$ and thus $\tilde{n}^{T}_{\state}(\bm{r},t)$ are even, whereas $\frac{\partial \bar{N}^D_{\state}}{\partial \bm{u}}$ and thus $\tilde{n}^{D}_{\state}(\bm{r},t)$ are odd.

Since the eigenvectors of the scattering matrix have a well-defined parity~\cite{hardy1970phonon}\footnote{This follows from the property $\tilde{\Omega}_{\state \statep} =\tilde{\Omega}_{-\state,-\statep}$}, we can split $\tilde{n}^{\delta}_{\state}$ \added{into} $\tilde{n}^{\delta{\rm E}}_{\state}+\tilde{n}^{\delta{\rm O}}_{\state}$, \added{separating} the even ($\tilde{n}^{\delta{\rm E}}_{\state}$) and odd ($\tilde{n}^{\delta{\rm O}}_{\state}$) components.
At steady state, Eq.~(\ref{eq:LBTE_expanded_gradiend}) \added{decouples into} two equations, one for each parity.
The equation for the odd part is
\begin{equation}
       \frac{\bm{v}_{\state} }{\sqrt{\bar{N}_{\state}(\bar{N}_{\state}+1)}}   \left(\frac{\partial \bar{N}_{\state}}{\partial T} \nabla T\right) = -\frac{1}{V} \sum_{\statep} \tilde{\Omega}_{\state \statep}\tilde{n}^{\delta\rm{O}}_{\statep}
\label{eq:LBTE_odd_nablaT}
\end{equation}
\added{and describes the response to a thermal gradient~\cite{fugallo2013ab,cepellotti2016thermal},}
where $\tilde{n}^{\delta{\rm O} }_{\state}$ is the odd out-of-equilibrium phonon population generated in response to a temperature gradient.
In writing Eq.~(\ref{eq:LBTE_odd_nablaT}) we made the assumption that $\frac{1}{V} \sum_{\statep} \tilde{\Omega}_{\state \statep} \big(\tilde{n}^D_{\statep}(\bm{r},t) + \tilde{n}^{\delta\rm{O}}_{\statep}\big)\simeq \frac{1}{V} \sum_{\statep} \tilde{\Omega}_{\state \statep}  \tilde{n}^{\delta\rm{O}}_{\statep}$, as explained in Appendix~\ref{sec:App1_eigv_scatt_matrix}.
The solution of equation~(\ref{eq:LBTE_odd_nablaT}) can then be used to determine the heat flux and the thermal conductivity (see e.g. Refs.~\cite{fugallo2013ab,cepellotti2016thermal}).
The equation for the even part is
\begin{equation}
\label{eq:LBTE_linear_nabla_u}
    \frac{\bm{v}_{\state} }{\sqrt{\bar{N}_{\state}(\bar{N}_{\state}+1)}}  \bigg(\frac{\partial N^D_{\state}}{\partial \bm{u}} \cdot \nabla \bm{u} \bigg)  
    = -\frac{1}{V} \sum_{\statep} \tilde{\Omega}_{\state \statep} \tilde{n}^{\delta{\rm E} }_{\statep} \;,
\end{equation}
\added{and describes the response to a drift-velocity gradient,}
where $\tilde{n}^{\delta{\rm E} }_{\state}$ is the even  out-of-equilibrium phonon population generated in response to it. In writing Eq.~(\ref{eq:LBTE_linear_nabla_u}) we used the property that $\tilde{n}^T_{\state}(\bm{r},t)$ is an eigenvector of the scattering matrix with zero eigenvalue:
$\frac{1}{V} \sum_{\statep} \tilde{\Omega}_{\state \statep} \tilde{n}^T_{\statep}(\bm{r},t) = 0$ (also detailed in Appendix~\ref{sec:App1_eigv_scatt_matrix}).

The phonon deviation $\tilde{n}^{\delta{\rm E} }_{\state}$ gives rise to a total crystal momentum flux $\Pi^{ij}_{\delta{\rm E} }\deleted{(\bm{r},t)} = \frac{1}{V} \sum_{\state}  \hbar q^i v_{\state}^j \sqrt{\bar{N}_{\state}(\bar{N}_{\state}+1)}\tilde{n}^{\delta{\rm E} }_{\state} \deleted{(\bm{r},t)}$ \cite{hardy1970phonon,gurzhi_hydrodynamic_1968}.
Analogously to the electronic case~\cite{rice1967theory}, the flux of total crystal momentum allows to define the thermal viscosity as the $4^{\rm th}$-rank tensor relating the \added{local} response $\Pi_{\delta{\rm E} }^{ij}(\bm{r},t)$ to the \added{local} perturbation $\nabla \bm{u}(\bm{r},t)$: 
\begin{equation}
    \Pi_{\delta{\rm E} }^{ij}(\bm{r},t)= -\sum_{kl} \replaced{\eta^{ijkl}}{\mu^{ijkl}} \frac{\partial u^{k}(\bm{r},t)}{\partial r^l} \;;
    \label{eq:definition_viscosity}
\end{equation}
\added{the uniform drift-velocity gradient is a special space-independent case of this local relationship.}
Eq.~(\ref{eq:LBTE_linear_nabla_u}) has the same mathematical form of the steady-state LBTE \added{linearized in the temperature gradient and} used to compute the thermal conductivity.
Therefore, we can readily solve it applying the methodology introduced in Ref.~\cite{cepellotti2016thermal}, and based on the eigenvectors of the scattering matrix (relaxons), to find a closed expression for $\tilde{n}^{\delta{\rm E} }_{\state}$.
Combining such solution with Eq.~(\ref{eq:definition_viscosity}) \added{we find, after symmetrizing,} the following expression for the \added{symmetrized} thermal viscosity
\begin{equation}
\mu^{ijkl} =\frac{\eta^{ijkl}+\eta^{ilkj}}{2} =\sqrt{A^i A^k} \sum_{\alpha>0} \frac{w^{j}_{i\alpha} w^{l}_{k\alpha}+w^{l}_{i\alpha} w^{j}_{k\alpha}}{2} \tau_{\alpha} \;,
\label{eq:viscosity_tensor}
\end{equation}
where $A^i = \frac{\partial P^i}{\partial u^i} \big|_{\text{eq}} = \frac{1}{k_B T V} \sum_{\state} \bar{N}_{\state}(\bar{N}_{\state}{+}1) (\hbar q^i)^2$ is referred to as the specific momentum, $\tau_\alpha$ is the relaxation time of relaxon $\alpha$ (\textit{i.e.} the inverse eigenvalue associated to the eigenvector $\theta^{\alpha}_{\state}$ of the symmetrized scattering matrix $\tilde{\Omega}_{\state \statep}$~\cite{cepellotti2016thermal}), and $w^{j}_{i\alpha}$ is the velocity tensor $w^{j}_{i\alpha} = \frac{1}{V} \sum_{\state} \phi^i_{\state} v^j_{\state} \theta^{\alpha}_{\state}$ for relaxon $\theta^{\alpha}_{\state}$ and eigenvector $\phi_{\state}^i$ (see Appendix~\ref{sec:App_3_viscosity} for a full demonstration).
$\phi_\state^i$ are three special eigenvectors linked to the crystal momentum of the system; to see this, we first decompose the scattering matrix as $\tilde{\Omega}_{\state \statep}=\tilde{\Omega}^{N}_{\state \statep}+\tilde{\Omega}^{U}_{\state \statep}$, where $\tilde{\Omega}^{N}_{\state \statep}$ and $\tilde{\Omega}^{U}_{\state \statep}$ contain only momentum-conserving (normal) and momentum-dissipating (Umklapp) processes, respectively.
Since the normal part of the scattering matrix conserves crystal momentum, there exists a set of 3 eigenvectors $\phi_{\state}^i$ ($i=1,\dots,3$ where $3$ is the dimensionality of the system) with zero eigenvalue for $\tilde{\Omega}^{N}_{\state \statep}$, which are associated to the conservation of crystal momentum in the 3 Cartesian directions.
Because the viscosity describes the response of the crystal momentum flux to a change of drift velocity, it is not surprising that the eigenvectors $\phi_\state^i$ appear in its definition.
In fact, the local equilibrium distribution (Eq.~(\ref{eq:drifting_distribution})) is linear in the drift velocity, with the proportionality coefficients being these special eigenvectors  (see Appendix~\ref{sec:App1_eigv_scatt_matrix} for a proof), and therefore they appear in the viscosity as well, to describe a perturbation to this local equilibrium.
We note in passing that the thermal viscosity defined in Eq.~(\ref{eq:viscosity_tensor}) has the \replaced{dimensions}{units} of a dynamic viscosity (Pa$\cdot$s) \added{and satisfies the property $\mu^{ijkl}=\mu^{ilkj}=\mu^{kjil}$}.

\begin{figure}[htp!]
\centering
  \includegraphics[width=\columnwidth]{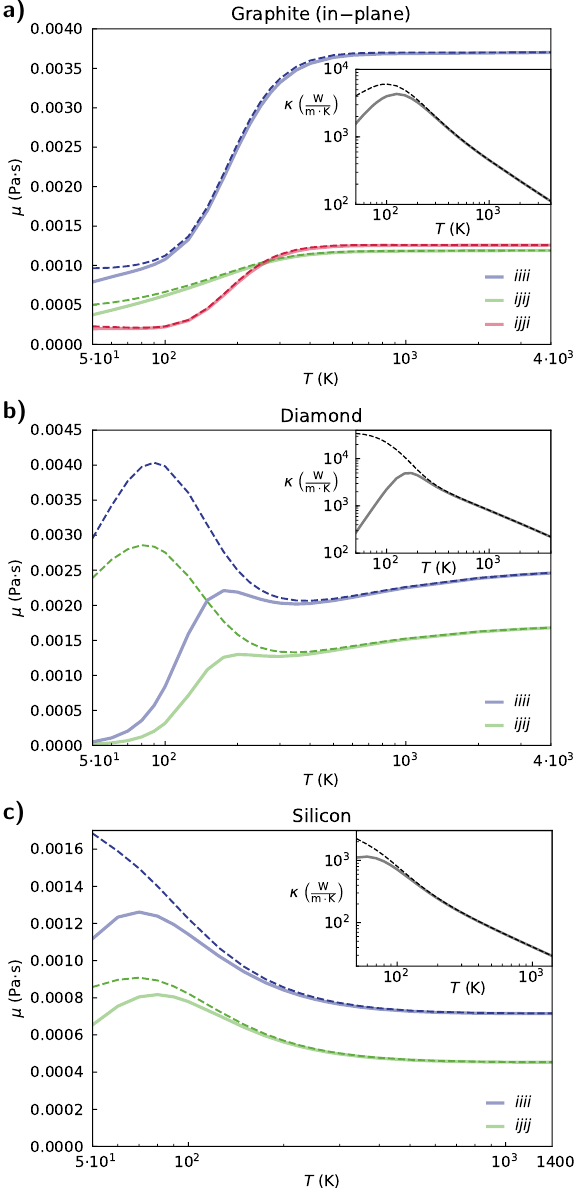}
  \caption{Largest components of the 4$^{\rm th}$-order thermal viscosity tensor for \added{graphite (a),} diamond (b), \added{and silicon (c),} as a function of  temperature \added{(the off-plane tensor components for graphite are reported in Fig.~\ref{fig:graphite_off_plane})}. 
   Insets: total thermal conductivities (\added{in-plane components for graphite and} diagonal components of the isotropic tensor \added{for diamond and silicon}) as a function of  temperature. 
   The dashed lines refer to the bulk materials \added{(Eq.~(\ref{eq:viscosity_tensor}) for viscosity and Eq.~(\ref{eq:thermal_conductivity}) for conductivity)}; the solid lines show finite-size predictions computed \added{using Eq.~(\ref{eq:comb_Mathiessen_1}) and Eq.~(\ref{eq:comb_Mathiessen_2})} for a sample having size \added{(or grain size)} \replaced{$10\;\mu m$}{$5\;\mu m$}  (see main text and Appendix~\ref{sec:App4_ballistic_tc_visc}; \added{for diamond and silicon the component $\mu^{ijji}$ is negligible). 
   } 
  }
  \label{fig:res_viscosity}
\end{figure}

We report in Fig.~\ref{fig:res_viscosity} the first-principles estimates of the thermal viscosity for \replaced{graphite, diamond, and silicon}{diamond and silicon}.
We choose these \replaced{three}{two} materials as prototypes for two different behaviors, the former \added{two} \deleted{a} \added{displaying} hydrodynamic thermal transport \cite{fugallo2013ab,second_sound_diamond,doi:10.1021/nl502059f,ding_phonon_2018,Chen_science_2019,Gandolfi2019}, as opposed to \replaced{the more conventional case of silicon}{the latter being more conventional} \cite{fugallo2013ab,cepellotti2016thermal}.
We account for finite-size effects by combining  the bulk viscosity in Eq.~(\ref{eq:viscosity_tensor}) with its ballistic limit via Matthiessen's rule for a \added{sample having size (or grain size) of 10} $\mu m$; \added{the same renormalization is applied also to the thermal conductivity (see Appendix~\ref{sec:App4_ballistic_tc_visc} for details).}
\added{
This is an approximate treatment of surface scattering effects, which provides an estimate of size effects at a much lower computational complexity compared to more refined models~\cite{doi:10.1021/nl502059f,PhysRevB.95.165407}.
In general, when samples' sizes become comparable or smaller than the carriers' diffusion lengths~\cite{heron2009mesoscopic,siemens2010quasi, PhysRevLett.110.025901, regner2013broadband, Wilson2014, ziabari_full-field_2018}, transport coefficients cannot be rigorously defined and a more accurate treatment is required, solving the much more complex and computationally expensive space-dependent LBTE~\cite{size_dependent_sol_LBTE, hua2014analytical, carrete2017almabte, cepellotti2017boltzmann}.
Matthiessen's approach is a good approximation when the values of thermal conductivity and viscosity do not differ significantly from their bulk counterparts, as is the case for the materials, sizes and temperatures considered in this work (e.g., in Fig.~\ref{fig:res_viscosity} the size renormalization is negligibly small in the temperature range considered for graphite and silicon, and becomes so above 300 K for diamond).
}
\begin{figure}
\centering
  \includegraphics[width=0.9\columnwidth]{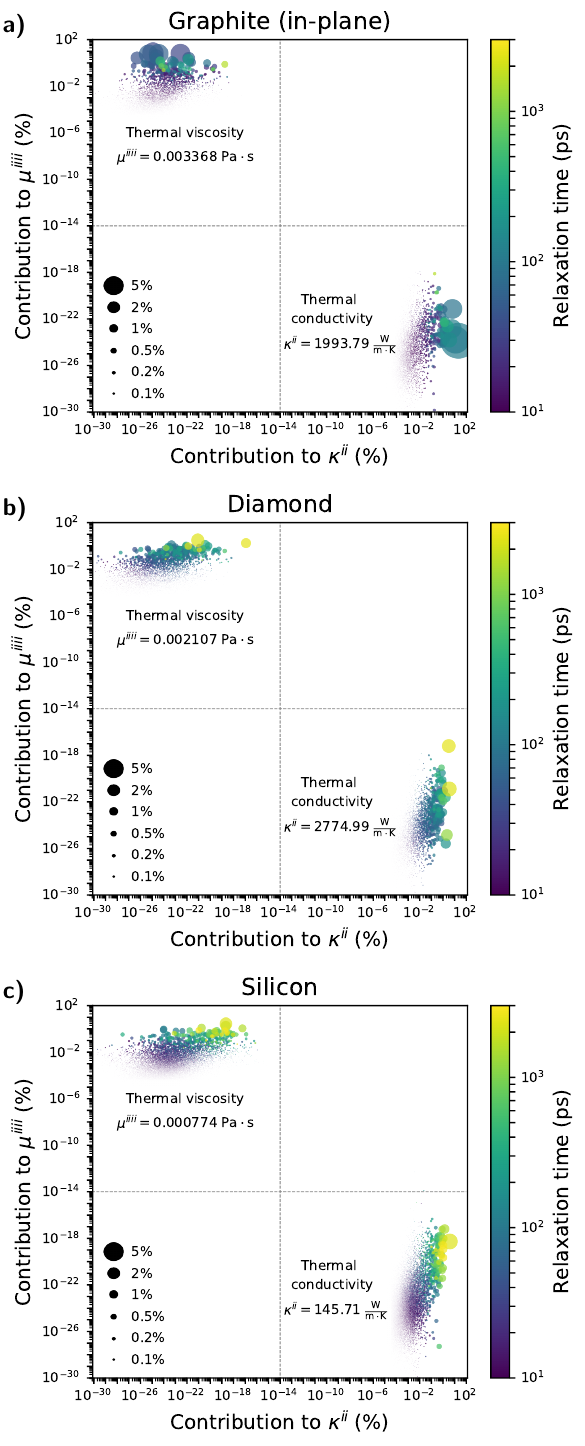}
  \caption{Relaxons' contributions to the bulk thermal conductivity and to the thermal  viscosity at 300 K for \added{ graphite (a, in-plane components),} diamond (b) and silicon (c). 
   Each dot represents a relaxon, with its color labeling its relaxation time, and its area being proportional to the sum of its percentage contributions to the thermal conductivity and viscosity.
   The dashed lines are plotted as a guide to the eye, to underscore how even and odd relaxons are fully decoupled.
   Odd relaxons, which determine the thermal conductivity, yield negligible (zero) contributions to the thermal viscosity; conversely, even relaxons determine the thermal viscosity and yield negligible (zero) contributions to thermal conductivity.}
  \label{fig:relaxon_decomposiion}
\end{figure}

\added{For graphite in the low-temperature ($\lesssim$100 K) regime, the bulk in-plane viscosity components are constant or slowly increasing with temperature;
at higher temperatures, the viscosity components increase with temperature up to reaching their saturation values.
For a graphite polycrystalline sample with a grain size of $10\mu m$, finite-size effects on viscosity (and on conductivity) are small.}
\added{For diamond in the temperature regime below 90 K, the bulk values of the thermal viscosity increase with increasing temperature, then in the intermediate temperature regime ($100 \;{\rm K}\lesssim T\lesssim 300\;{\rm K}$) they decrease with temperature, and finally for increasingly higher temperatures they gently increase up to saturating to the high-temperature limiting values.   
However, size effects renormalize this behavior for temperatures below 300 K, leaving viscosity components that increase with temperature.
In silicon, increasing temperature yields a decrease, up to a high-temperature saturation value, of the bulk thermal viscosity; size effect renormalize the viscosity below $\sim 100$ K yielding viscosity components that increase with temperature (\textit{i.e.} a behavior analogous to that of diamond). 
We note that in the high-temperature limit, \textit{i.e.} when the thermal conductivity decay as $T^{-1}$~\cite{ziman1960electrons,sun2010lattice}, the viscosity components tend to constant values for all these three materials.
The total bulk thermal conductivities are shown in the insets of Fig.~\ref{fig:res_viscosity} for comparison (we only show one component; for diamond and silicon the conductivity tensor is isotropic, for graphite we show the in-plane component). 
The off-plane transport coefficients for graphite are discussed in  Fig.~\ref{fig:graphite_off_plane}a.} 
\added{In graphite, the bulk thermal conductivity below 100 K does not increase monotonically as the temperature is decreased due to the presence of natural-abundance isotopic scattering.
Natural-abundance isotopic scattering has been considered also for diamond and silicon, and the thermal conductivity at low temperatures is much lower than for isotopically-pure diamond~\cite{fugallo2013ab,diamond_0,diamond_1,diamond_2,diamond_3,diamond_4,diamond_5} or silicon~\cite{silicon_isotopes}.
}
We have also verified that the effects of phonon coherences are negligible in these crystals~\cite{simoncelli2019unified}.
We note in passing that, even if the thermal conductivities of \replaced{diamond and silicon}{the two materials} differ by more than one order of magnitude, their largest thermal-viscosity components differ only by a factor of 3.
These results may be compared with the case of water, whose dynamic (shear) viscosity is $8.5{\cdot} 10^{-3}$ Pa$\cdot$s at room temperature, indicating that the thermal viscosity found here is comparable or larger.
To a good approximation, water is an incompressible fluid, and thus its largest viscosity component is $ijij$, also called ``first viscosity'' or ``shear viscosity''.
For compressible fluids, the $iiii$ components of the viscosity tensor --- also called ``second viscosity'' or ``volume viscosity''~\cite{landau1980kinetics} --- are non-negligible.
Here, in contrast with water, the $iiii$ component of the thermal viscosity tensor is the largest, which \added{elicits} an analogy to a compressible fluid.
\replaced{It is important to mention that the present formulation may need further extension to deal with unstrained  2D materials, since their flexural phonon modes have a quadratic dispersion $\omega_\state\propto q^2$ at $T=0K$ \cite{bonini2012acoustic}. 
Under these conditions, long-wavelength phonons lead to $\bm{q}\cdot \bm{u}$ being larger than $\omega_\state$, causing negative values of the phonon drifting distribution which are not compatible with a semiclassical description of transport.
Further work is needed to address this issue, for example considering the phonon renormalization due to coupling between bending and stretching degrees of freedom of the monolayer~\cite{PRL_renormalization_graphene} or the presence of a substrate~\cite{Pop_substrate}, or introducing the Wigner-function formalism~\cite{simoncelli2019unified,weinbub2018recent}.
}{It is worth mentioning that the present formulation may need to be extended 
in order to describe 2D materials, in which thermal transport is often hydrodynamic~\cite{cepellotti2015phonon}.
In fact, the presence of quadratic phonon modes in 2D materials makes the drifting distribution~(\ref{eq:drifting_distribution}) negative for small phonon wavevectors, when $\bm{q}\cdot \bm{u}>\omega_\state$.
To the best of our knowledge, negative-valued phonon distributions require a treatment based on the Wigner function extension of the LBTE~\cite{simoncelli2019unified}, and this will be the subject of future work.}

Finally, it is worth checking the complementary character of the thermal conductivity and viscosity that arises from their decoupled relaxons' contributions.
As commented above, decomposing the thermal conductivity~\cite{cepellotti2016thermal} and viscosity~(\ref{eq:viscosity_tensor}) in terms of single relaxons it is possible to show 
that the thermal viscosity is uniquely determined by the even part of the relaxon spectrum, while the  thermal conductivity is determined uniquely by the odd part of the relaxon spectrum~\cite{cepellotti2016thermal}.
In Fig.~\ref{fig:relaxon_decomposiion} we highlight the contributions of each relaxon to the total thermal conductivity and viscosity, confirming numerically this picture.

\section{Viscous heat equations} 
\label{sec:viscous_heat_equations}
We show here that heat conduction can be described by \added{two novel} viscous heat equations that cover \added{on the same footing} both the Fourier and hydrodynamic \added{limits, and all intermediate regimes}.
These are \added{two} coupled equations in the temperature $T(\bm{r},t)$ and drift-velocity $\bm{u}(\bm{r},t)$ fields, which are parametrized by the thermal viscosity and conductivity.
These equations represent the thermal counterpart of the Stokes equations of fluid dynamics --- \textit{i.e.} the Navier-Stokes equations in the linear regime, whose solution yields the laminar flow --- where temperature takes the role of pressure and the phonon drift velocity that of the fluid velocity.
In the kinetic regime, when momentum-dissipating (Umklapp) scattering processes dominate~\cite{cepellotti2015phonon}, these viscous heat equations \added{become equivalent} to Fourier's heat equation.  

As underscored before, hydrodynamic thermal transport is characterized by energy conservation and crystal momentum quasi-conservation (the latter being exactly conserved only in absence of Umklapp processes~\cite{Sussmann_Thellung_1963}).
Conserved quantities in the LBTE dynamics can be related to the eigenvectors of the full or normal scattering matrix  
with zero eigenvalues~\cite{hardy1970phonon} (see also Appendix~\ref{sec:App1_eigv_scatt_matrix}).  
Four of these eigenvectors (\textit{i.e.} phonon distribution functions) can be identified.
The first one is the Bose-Einstein eigenvector $\phi^0_{\state} \propto \hbar \omega_{\state} \propto  \frac{\partial \bar{N}_{\state}}{\partial T}\propto \tilde{n}^T_{\state}$, which is the eigenvector of zero eigenvalue for the symmetrized full scattering matrix $\tilde{\Omega}_{\state\statep}$; its zero eigenvalue is associated to energy conservation in scattering events (both normal and Umklapp).
The other three eigenvectors are the drift eigenvectors $\phi_{\state}^i\propto \hbar q^i\propto \frac{\partial N^D_{\state}}{\partial u^i}\propto n^D_{\state}$, $i=1,2,3$ (where $3$ is the dimensionality of the system) already introduced for the evaluation of the viscosity; these $\phi_{\state}^i$  are eigenvectors with zero eigenvalue for the normal part of the scattering matrix $\tilde{\Omega}^N_{\state\statep}$ and are associated to the conservation of crystal momentum by normal scattering events~\cite{Sussmann_Thellung_1963,PhysRev.148.766, hardy1970phonon}.
We note that these four eigenvectors constitute the first two terms of the phonon distribution expansion in Eq.~(\ref{eq:hardy_decomposition}).
We can thus derive the mesoscopic equations that govern the evolution of the temperature ($T(\bm{r},t)$) and drift velocity ($\bm{u}(\bm{r},t)$) fields projecting the microscopic LBTE in the subspaces spanned by $\phi^0_{\state}$ and by $\phi_{\state}^i$ ($i=1,\dots,3$). 
In order to derive a closed-form equation for the drift velocity, when projecting in the subspace spanned by $\phi_{\state}^i$ we consider the effects of momentum dissipation only within that subspace.
The result is the following set of equations (see Appendix~\ref{sec:App5_mesoscopic_equations} for a detailed derivation):
\begin{widetext}
  \begin{equation}
  C \frac{\partial T(\bm{r},t)}{\partial t} 
  + \sum_{i,j=1}^3 W_{0j}^i \sqrt{\bar{T}A^jC} \frac{\partial u^{j}(\bm{r},t)}{\partial r^i} 
  - \sum_{i,j=1}^3\kappa^{ij} \frac{\partial^2 T(\bm{r},t)}{\partial r^i\partial r^j} = 0 \;,
  \label{eq:macro_diff_eq1}
\end{equation}  
\begin{equation}
    \label{eq:macro_diff_eq2}
    A^i  \frac{\partial u^i(\bm{r},t)}{\partial t}
    +\sqrt{\frac{C A^i}{\bar{T}}} \sum_{j=1}^3 {W}_{i0}^j \frac{\partial T(\bm{r},t)}{\partial r^j}  - \sum_{j,k,l=1}^3 \mu^{ijkl} \frac{\partial^2 u^k(\bm{r},t)}{\partial r^j \partial r^l} 
    =   - \sum_{j=1}^3 \sqrt{A^i A^j} {D}_U^{ij} u^j(\bm{r},t) \;,
\end{equation}
\end{widetext}
where $C=\frac{1}{k_B \bar{T}^2 V}\sum_{\state} \bar{N}_{\state}\big(\bar{N}_{\state}+1\big)(\hbar \omega_{\state})^2$ is the specific heat, 
${W}_{0 j}^i = \frac{1}{V} \sum_{\state} \phi_{\state}^0{v}^i_{\state} \phi_{\state}^j$ is a velocity tensor
\added{that arises from the non-diagonal form of the diffusion operator in the basis of the eigenvectors of the normal part of the scattering matrix (see Appendix~\ref{sec:App5_mesoscopic_equations})},  
\added{$\bar{T}$ is the reference (equilibrium) temperature on which a perturbation is applied  (see Sec.~\ref{sec:case_study} and Appendix~\ref{sub:the_bose_eigenvector}),}
\added{$A^i$ is the specific momentum in direction $i$ (defined in Sec.~\ref{sec:thermal_viscosity}),
$\mu^{ijkl}$ is the thermal viscosity tensor, }
$\kappa^{ij}$ is the thermal conductivity tensor \cite{cepellotti2016thermal}, 
and ${D}_U^{ij} = \frac{1}{V} \sum_{\state\statep} \phi^i_\state \tilde{\Omega}_{\state\statep}^U \phi^j_{\statep}$ is the momentum dissipation rate\added{; the latter is caused both by the presence of Umklapp processes as well as boundary scattering (\textit{i.e.} ${D}_U^{ij}$ is sensitive to size effects like the thermal conductivity and viscosity).
Treating boundary scattering as in Ref.~\cite{doi:10.1021/nl502059f}, we compute ${D}_U^{ij}$ as ${D}_U^{ij}={D}_{U,{\rm bulk}}^{ij}+{D}_{U,{\rm boundary}}^{ij}(L_S)$, where ${D}_{U,{\rm bulk}}^{ij}=\frac{1}{V} \sum_{\state\statep} \phi^i_\state \tilde{\Omega}_{\state\statep}^U \phi^j_{\statep}$ is the momentum dissipation caused by Umklapp scattering in the bulk and ${D}_{U,{\rm boundary}}^{ij}(L_S)=\frac{1}{V} \sum_{\state\statep} \phi^i_{\state} \frac{|\bm{v}_{\state}| \delta_{\state,\statep}}{L_S} \phi^j_{\statep}$ is the momentum dissipation caused by scattering with boundaries at a distance $L_S$.
}
\added{
The scalar equation~(\ref{eq:macro_diff_eq1}) and the vectorial (3-components) equation~(\ref{eq:macro_diff_eq2}) rule the coupled evolution of the scalar temperature field and of the vector drift-velocity field; they constitute the main result of this work and we name these ``viscous heat equations''.}
These transport equations are reminiscent of the linearized Stokes equations for fluids: to see this more clearly, we note that \added{local} energy $E(\bm{r},t)$ and crystal momentum $\bm{P}(\bm{r},t)$ are proportional to temperature and drift velocity respectively ($E(\bm{r},t)=C\;T(\bm{r},t)$ and ${P}^i(\bm{r},t) = A^i\; {u}^i(\bm{r},t)$, where $C$ is the specific heat and $\bm{A}$ is the specific momentum).
Exploiting these relationships, it is possible to rewrite the viscous heat equations in a more familiar form, namely as energy $E$ and momentum $P^i$ balance equations:
\begin{gather}
\label{eq:stokes_1}
  \frac{\partial E(\bm{r},t)}{\partial t} + \nabla \cdot \Big(\bm{Q}^{\delta}(\bm{r},t)+\bm{Q}^{D}(\bm{r},t)\Big)=0  \;, \\
   \frac{\partial P^i(\bm{r},t)}{\partial t} 
    {+}\hspace*{-1.mm} \sum_j \hspace*{-1mm}\left(\hspace*{-1mm}\frac{\partial \Pi_T^{ij}(\bm{r},t)}{\partial r^j}
    + \frac{\partial \Pi_{\delta{\rm E} }^{ij}(\bm{r},t)}{\partial r^j}\hspace*{-1mm}\right)\hspace*{-1mm}
    = - \frac{\partial P^i(\bm{r},t)}{\partial t} \bigg|_{\text{Umkl}},
\label{eq:stokes_2}
\end{gather}
where, on the basis of the phonon population expansion in Eq.~(\ref{eq:hardy_decomposition}), we distinguish the drifting heat flux into the contributions from the \added{local} temperature gradient ${Q}^{\delta,i}(\bm{r},t) = - \sum_{j}\kappa^{ij} \nabla^j T(\bm{r},t)$ and from the drift velocity field ${Q}^{D,i}(\bm{r},t)=\sum_jW_{0j}^i\sqrt{\bar{T}A^jC} {u}^j(\bm{r},t)$.
Similarly, the momentum flux receives separate contributions from the \added{local} temperature $\Pi^{ij}_{T}(\bm{r},t) = \sqrt{C A^i/\bar{T}} \;{W}_{i0}^j  T(\bm{r},t)$ and from variations of the drift velocity through Eq.~(\ref{eq:definition_viscosity}).
Finally, $\frac{\partial P^i}{\partial t} \big|_{\text{Umkl}}$ accounts for the dissipation of crystal momentum by Umklapp processes \added{or scattering with boundaries}; further details can be found in Appendix~\ref{sec:App5_mesoscopic_equations}.
The distinction between temperature-driven  and drifting components of the heat flux \added{($\bm{Q}^\delta(\bm{r},t)$ and $\bm{Q}^D(\bm{r},t)$, respectively)} is essential for hydrodynamic transport.
\added{
We will show in Sec.~\ref{sec:deviations_from_fourier_s_law} that Fourier's law is recovered in the limiting 
case where crystal momentum dissipation dominates over viscous effects and is the fastest timescale of the phonon dynamics. }
\deleted{
In fact, Fourier's law is recovered in the limiting case of $\bm{u}=0$, since Eq.~(\ref{eq:stokes_1}) reduces to $C \frac{\partial T}{\partial t} - \kappa^{ij} \frac{\partial^2 T}{\partial r^i\partial r^j} = 0 $; so it is the drifting flux that introduces a correction to Fourier's law.}

It is worth mentioning that the viscous heat equations \added{introduced here} differ from the Stokes equations for fluids in two major ways.
First, there is no analogous to the mass conservation satisfied by Stokes equations, since the total phonon number is not a constant of motion (e.g. a phonon coalescence event decreases the number of phonons in the system).
Second, while collisions between molecules in the fluid conserve momentum, scattering among phonons does not conserve crystal momentum in  presence of Umklapp processes.

The most relevant feature of the viscous heat equations is their capability to describe hydrodynamic thermal transport in terms of mesoscopic quantities, \textit{i.e.} temperature and drift velocity, resulting in a much simpler and computationally less expensive approach than the microscopic LBTE.
The parameters entering Eqs.~(\ref{eq:macro_diff_eq1},~\ref{eq:macro_diff_eq2}) can be determined from first-principles calculations or, possibly less accurately, from classical potentials, and are tabulated in Appendix \ref{sec:App6_parameter_sAppendix} for \added{graphite,} diamond and silicon.

In order to be solved, the viscous heat equations require appropriate boundary conditions on the temperature and drift velocity. 
Boundary conditions on temperature have been widely studied in conjunction with Fourier's heat equation~\cite{logan2014applied}: typically, one makes assumptions on the system's capability to exchange heat at the boundaries, and on the temperature at those boundaries (Neumann and Dirichlet boundary conditions, respectively~\cite{logan2014applied}). 
In the next section~\ref{sec:case_study} we consider a system in which the temperature is fixed at some boundaries, while the others 
are assumed to be adiabatic (that is, the heat flux across these boundaries is zero).
In contrast, boundary conditions on the drift velocity, \textit{i.e.} on crystal momentum at the sample's borders, have not been studied as extensively. 
Since crystal momentum is not conserved at boundaries~\cite{Sussmann_Thellung_1963}, we impose a no-slip condition of zero drift velocity $\bm{u}(\bm{r},t)$  on all boundaries, ensuring thus zero drifting heat $\bm{Q}^D(\bm{r},t)\propto \bm{u}(\bm{r},t)$.
As discussed in past work~\cite{ziman1960electrons,cepellotti2017boltzmann}, more comprehensive boundary conditions require quantifying phonon reflection at surfaces, and are beyond the scope of this work.

\added{
The viscous heat equations~(\ref{eq:macro_diff_eq1},\ref{eq:macro_diff_eq2}) improve upon past work on different levels.
First, they are valid for general dispersion relations for phonons, and thus overcome the limitations --- pointed out by Hardy and Albers \cite{PhysRevB.10.3546} already in 1974 --- of the pioneering mesoscopic models developed in the 1960s by Guyer-Krumhansl~\cite{PhysRev.148.766, PhysRev.148.778} or Gurzhi~\cite{gurzhi_hydrodynamic_1968}, which assumed linear-isotropic or power-law phonon dispersion relations, respectively (assumptions which arose from the hypothesis that hydrodynamic phenomena would occur at cryogenic temperatures only).
The limitations of the Guyer-Krumhansl~\cite{PhysRev.148.766, PhysRev.148.778} and Gurzhi~\cite{gurzhi_hydrodynamic_1968} models were overcome in 1974 by Hardy and Albers~\cite{PhysRevB.10.3546}, who derived from the LBTE a set of mesoscopic equations for energy and crystal momentum valid for a general phonon dispersion relation (\textit{i.e.} not necessarily linear-isotropic). 
Hardy and Albers' formulation relies on the hydrodynamic approximation --- \textit{i.e.} that the fastest timescale of phonon dynamics is that of normal processes --- that is valid only within the hydrodynamic regime, where Umklapp collisions are rare events. 
In fact, Hardy and Albers' equations are limited just to the hydrodynamic regime and do not incorporate Fourier's law as a limit (and thus any intermediate regimes).
The viscous heat equations address exactly this issue, 
and Fourier's law emerges 
when crystal momentum dissipation (due to Umklapp processes or scattering with boundaries) dominates over viscous effects and is the fastest timescale of the phonon dynamics (see Sec.~\ref{sec:deviations_from_fourier_s_law}). 
Finally, the viscous heat equations take into account the entire collision matrix, at variance with recent mesoscopic models derived from the LBTE either in the SMA~\cite{Guo2018} or in the  Callaway approximation~\cite{PRB_Viscosity_Callaway_in_s}.}

\section{Second sound} 
\label{sec:second_sound}

Second sound is the coherent propagation of a temperature wave \cite{Sussmann_Thellung_1963, PhysRev.148.778,kwok1967dispersion, enz1968one,ackerman1968temperature,hardy1970phonon, maris1981second,cepellotti2016second_sound,  Chen_science_2019,PhysRev.131.2013}, and it is an effect properly described by the viscous heat equations.
From a phenomenological point of view, second sound appears when the temperature field satisfies the following damped wave equation~\cite{hardy1970phonon} (we define $x$ as the second sound propagation direction):
\begin{equation}
  \frac{\partial^2 T(x,t)}{\partial t^2} + \frac{1}{\tau_{ss}}\frac{\partial T(x,t)}{\partial t} 
  - v_{ss}^2 \frac{\partial^2 T(x,t)}{\partial x^2} = 0 \;,
  \label{eq:second_sound_damped_wave}
\end{equation}  
where $\tau_{ss}$ and $v_{ss}$ are the second-sound relaxation time and undamped propagation velocity, yet to be determined.
\added{In contrast to Fourier's law, which has the form of a classical (non-relativistic) diffusion equation and states that a temperature-gradient variation causes an instantaneous variation of the heat flux, Eq.~(\ref{eq:second_sound_damped_wave}) has the physical property that a sudden localized change of temperature propagates in space with a finite speed (\textit{i.e.} it is not felt instantly everywhere in space)~\cite{cattaneo1948sulla,RevModPhys.61.41,tzou1995generalized,bergamasco2018mesoscopic}.}
The temperature profile that solves Eq.~(\ref{eq:second_sound_damped_wave}) has the form of a damped wave:
\begin{equation}
  T(x,t) = \bar{T} + \delta T\;  e^{i( k x-\bar{\omega}(k) t)} e^{- t / \tau_{ss}}\;, 
  \label{eq:temperature_wave}
\end{equation}
where the second sound frequency $\bar{\omega}(k)$ 
depends on the second-sound wavevector $k$.
In Appendix~\ref{sec:second_sound_appendix} \added{we derive the second-sound equation from the viscous heat equations~(\ref{eq:macro_diff_eq1},\ref{eq:macro_diff_eq2}) following two different approaches (bottom-up and top-down).}
In the first \added{bottom-up} approach we find the conditions for which the damped wave equation~(\ref{eq:second_sound_damped_wave}) emerges form the viscous heat equations Eq.~(\ref{eq:macro_diff_eq1},\ref{eq:macro_diff_eq2}).
When this happens, the solution of Eq.~(\ref{eq:second_sound_damped_wave}) is the damped wave equation for temperature~(\ref{eq:temperature_wave}) shown above, with the second-sound dispersion relation given by $\bar{\omega}(k)=\sqrt{v_{ss}^2k^2-(2\tau_{ss})^{-2}}$ (this can be easily verified substituting Eq.~(\ref{eq:temperature_wave}) into Eq.~(\ref{eq:second_sound_damped_wave})).
This allows to express $\tau_{ss}$ and $v_{ss}$ in terms of the parameters appearing in the viscous heat equations;
in particular, $\tau_{ss}=\frac{C({W}_{x0}^x)^2 }{\kappa ({D}^{xx}_U)^2+ {D}^{xx}_UC({W}_{x0}^x)^2}$ and $v_{ss}=\frac{\kappa{D}^{xx}_U+C({W}_{x0}^x)^2}{C{W}_{x0}^x}$.
The propagation velocity of second sound is affected by damping and depends on the wavevector $k$: it is given by the group velocity $v_{g}(k) = \frac{\partial \bar{\omega}(k)}{\partial k}=k v_{ss}[k^2+(2\tau_{ss}v_{ss})^{-2}]^{-\frac{1}{2}}$, and we note that it reduces to the undamped propagation velocity $v_{ss}$ in the undamped limit $\tau_{ss}\to 0$.

These results are consistent with empirical expectations on second sound: in the limit of weak crystal momentum dissipation the second-sound relaxation time increases, while the velocity becomes smaller, making second sound more likely to be observed in the hydrodynamic regime~\cite{PhysRev.148.778,hardy1970phonon}.
In fact, when $D^{xx}_U \to 0$ we find that $\tau_{ss}\to ({D}^{xx}_U)^{-1}$ and $v_{ss}\to W^x_{x0}$.
We note that the viscous heat equations describe not only the propagation of the temperature field, but also that of the drift velocity.
In Appendix~\ref{sec:second_sound_appendix} we show that when second sound emerges the drift-velocity field propagates as a damped wave as well (\textit{i.e.} similar to Eq.~(\ref{eq:temperature_wave})), with the same relaxation time and velocity of temperature, but with a phase shift of $\pi/2$. 
\added{
Phase shifts between hydrodynamic and resistive components of the frequency-dependent heat flux have also been discussed recently in the context of an improved Callaway approximation for the LBTE~\cite{PhysRevB.88.144302}.
}

\added{
As an alternative, we took inspiration from Ref.~\cite{cepellotti2016second_sound}, which derived the second-sound dispersion relations by taking advantage of the Laplace transform of the LBTE, to identify solutions in the form of a damped wave following a top-down approach.
In particular, we take a damped-wave ansatz for temperature, Eq.~(\ref{eq:temperature_wave}), and a similar one for the drift-velocity field (with the same frequency and decay time of temperature); we substitute these into the viscous heat equations~(\ref{eq:macro_diff_eq1},\ref{eq:macro_diff_eq2}) and determine the conditions under which these are acceptable solutions.}
As detailed in Appendix~\ref{sec:second_sound_appendix}, we find that the dispersion relations of second sound in the long-wavelength limit reduce to $\omega(k) - \frac{i}{\tau_{ss}} \approx - \frac{i D_U^{xx}}{2} + \sqrt{ - \bigg(\frac{D_U^{xx}}{2}\bigg)^2 +  k^2 \bigg(W^2 - \frac{D_U^{xx} \kappa}{C} - \frac{2 \mu D_U^{xx}}{A} \bigg) }$.
In this long wavelength limit and in the hydrodynamic regime $D_U\to0$ we have $\tau_{ss} \approx \frac{1}{D_U^{xx}}$ and $v_{g}(k) \approx W$, which is consistent with the first bottom-up approach presented to derive the equations for second sound.
We further stress that the LBTE can only rigorously describe second sound in the long-wavelength limit $k\to0$ in order for the temperature to be slowly varying (for which the two approaches shown in this section provide the same result); for smaller wavelengths, the definition of temperature itself becomes questionable~\cite{lepri2003thermal}.

We also recall that Enz~\cite{enz1968one} and Hardy~\cite{hardy1970phonon} distinguished between ``drifting''  and ``driftless'' second sound.
The former emerges when normal processes dominate and is described in terms of balance equations for energy and momentum; the latter is determined by a uniform energy flux that decays exponentially. 
The second sound discussed here is of the drifting kind, as it emerges from a set of balance equations for energy and crystal momentum derived from the LBTE.

\begin{figure*}
\includegraphics[width=\textwidth]{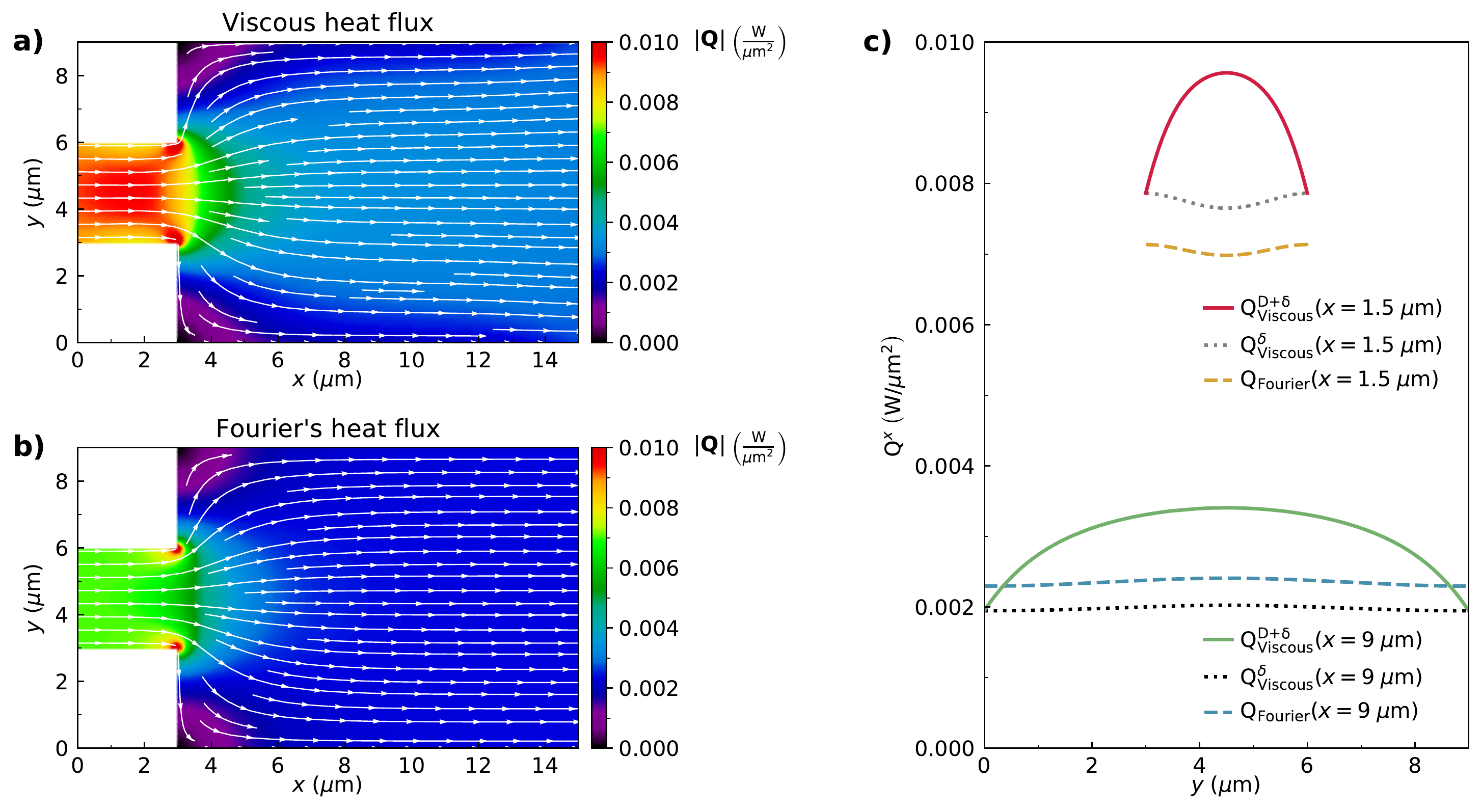}
  \caption{ 
  \added{In-plane ($x-y$) heat fluxes in graphite around $\bar{T}=70$ K, for a sample infinite in the $z$ direction (the $z$ direction of the sample coincides with the off-plane direction of graphite). Panel a) shows the total heat flux (${\bm{Q}}^{D}+{\bm{Q}}^{\delta}$) obtained from the viscous heat equations~(\ref{eq:macro_diff_eq1},~\ref{eq:macro_diff_eq2}).
  Panel b) shows instead the Fourier heat flux (${\rm Q}^{i}_{\rm Fourier}(\bm{r},t){=}-\sum_j\kappa^{ij} \nabla^j T(\bm{r},t)$) obtained solving the steady-state Fourier equation  $\sum_{ij}\kappa^{ij}\frac{\partial^2 T(\bm{r},t)}{\partial r^i\partial r^j}=0$. 
  Boundary conditions are as follows: the temperature is set at 80K/60K on the left/right boundaries, a zero total heat flux is imposed at the other boundaries, and zero drift velocity is imposed on all boundaries (no-slip boundary condition). 
  With this choice of boundary conditions, the phonon distribution reduces to the  Bose-Einstein distribution on the left/right boundaries, where the temperature is fixed.
  Despite having local equilibrium at the left/right boundaries, a non-zero drift velocity is present in the material as a consequence of the coupling in Eq.~(\ref{eq:macro_diff_eq1}) between drift velocity and temperature.
 Panel c) shows the $x$-component of the heat flux along sections taken at $x=1.5\;\mu m$ (orange, gray, red) and $x=9.0\;\mu m$ (black, blue, green); solid lines are the results obtained from the viscous heat equations, dashed lines are the results from Fourier's law, and dotted lines are the heat fluxes due to the temperature gradient within the viscous heat equations ($\bm{Q}^{\delta}$).} }
  \label{fig:2heat_flux}
\end{figure*}
\section{Case study} 
\label{sec:case_study}
We showcase the solution for the viscous heat equations for \replaced{graphite}{diamond} around the equilibrium temperature \added{$\bar{T}=70$ K} in the geometry shown in Fig.~\ref{fig:2heat_flux}, often used as an illustrative example in textbooks on fluid dynamics.
The equations are solved numerically using a finite-element solver implemented in Mathematica~\cite{Mathematica}, imposing a temperature of \added{80 K} on the left side ($x=0\;\mu m$) and of \added{60 K} on the right side ($x=\added{15}\;\mu m$), assuming all boundaries at $x\neq0$ and $x\neq\added{15}\;\mu m$ to be adiabatic, and imposing a no-slip condition on $\bm{u}$ at all boundaries.

We \replaced{show}{plot} in Fig.~\ref{fig:2heat_flux} the \replaced{viscous}{drifting} heat flux \replaced{$\bm{Q}^{\delta}+\bm{Q}^{D}$}{$\bm{Q}^{D}$},  and \deleted{, the flux due to changes of local temperature $\bm{Q}^{\delta}$ (panel b), and, for comparison,}  the flux predicted by Fourier's law $\bm{Q}_{\rm Fourier}$ \replaced{(left panels)}{c} \added{and contrast the Fourier and viscous heat flux components across two different sections (right panel).}
\deleted{The total heat flux resulting from the viscous equations is given by the sum $\bm{Q}^{D}{+}\bm{Q}^{\delta}{=}\bm{Q}^{D+\delta}$ (panel~\ref{fig:2heat_flux}d).}

We stress that Fourier's law lacks a description of the contribution to the heat flux derived from the local drift velocity \cite{Sussmann_Thellung_1963, Maris_1981_key, simons1983relation}.
As a result, Fourier's law misses qualitative and quantitative properties of the heat flux profile.
The largest differences are observed in proximity of spatial inhomogeneities, such as boundaries or corners.
For example, $\bm{Q}^{D}$ quickly increases (decreases) in proximity of the thermal reservoir on the hot-left (cold-right) side of the sample; these changes in $\bm{Q}^{D}$ determine opposite changes in $\bm{Q}^{\delta}$. 
Microscopically, these variations are caused by the rapid transition of the phonon distribution from the Bose-Einstein equilibrium distribution imposed at the boundaries to an out-of-equilibrium distribution carrying non-zero total crystal momentum (\textit{i.e.} a drift velocity) inside the sample.

\begin{figure}
  \centering
  \includegraphics[width=\columnwidth]{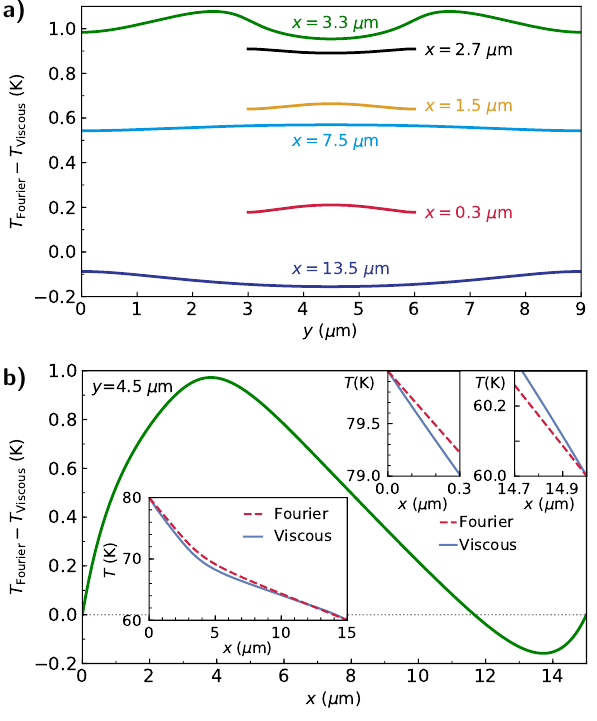}
  \caption{Difference between the temperature profiles predicted by Fourier's law and by the viscous heat equations, plotted in the top panel a) along $y$ for various values of $x$ and in the bottom panel b) along $x$ for $y=\replaced{4.5}{0.55}\;\mu m$. The insets of panel b) show the temperature profile along $x$ for $y=\replaced{4.5}{0.55}\;\mu m$; the result of the viscous heat equations~(\ref{eq:macro_diff_eq1},~\ref{eq:macro_diff_eq2}) is solid/blue, Fourier's law prediction is dashed/red.}
  \label{fig:temp_difference}
\end{figure}
We report in Fig.~\ref{fig:2heat_flux}\replaced{c}{d} the total heat flux profiles along two transversal sections of the sample, contrasting the prediction from the viscous heat equations (solid lines) with that of Fourier's law (dashed lines).
Along these sections, Fourier's law predicts a flat heat flux profile, while the viscous heat equations yield a Poiseuille-like profile. 
The results from the viscous heat equations are thus substantially different from Fourier's predictions, and
the behavior of the heat flux can be understood from a simple analytical 1D solution of the viscous heat equations in the absence of Umklapp processes~\cite{Sussmann_Thellung_1963}: as discussed in Appendix~\ref{sec:App_Sussman_Tellung}, the flux is described by hyperbolic functions with a characteristic length scale $b=\sqrt{\frac{\mu \kappa}{A C (W)^2}}$ (an estimate of the friction lengths, see Ref.~\cite{cepellotti2017boltzmann}).
At distances from the surface larger than $b$ we recover the flat behavior typical of the bulk.
We also note that these results mimic those from the (computationally very expensive) space-dependent solution of the LBTE \added{(either in the frequency-dependent SMA approximation~\cite{size_dependent_sol_LBTE} or considering the full scattering operator~\cite{cepellotti2017boltzmann})}, which generates a minimum flux on surfaces and maximum at the sample's center. 
We further note that, at variance with classical fluid dynamics and as pointed out in Ref.~\cite{cepellotti2017boltzmann}, the total heat flux does not drop to zero at the boundaries: the no-slip condition sets the drifting heat flux $\bm{Q}^D$ to zero, but the temperature-driven component $\bm{Q}^\delta$ is still allowed to be nonzero.

In Fig.~\ref{fig:temp_difference} we plot the difference between the temperature profile predicted by Fourier's law and the viscous heat equations along longitudinal (top panel) and transversal (bottom panel) directions. 
The insets in Fig.~\ref{fig:temp_difference}b show the results of the viscous equations (solid blue line) and Fourier's law (dashed red line) along the section $y=\replaced{4.5}{0.55}\; \mu m$.
Along the transversal direction (Fig.~\ref{fig:temp_difference}a), Fourier's law and Eqs.~(\ref{eq:macro_diff_eq1},~\ref{eq:macro_diff_eq2}) predict temperature profiles which are substantially different in the presence of variations of the sample's shape (green line corresponding to $x=\replaced{3.3}{1.01}\;\mu m$), while they are merely shifted by a positive or negative offset away from these; the precise amount depends on the distance from the fixed-temperature boundaries.
These differences become more clear by inspecting the longitudinal direction (Fig.~\ref{fig:temp_difference}b), where the discrepancy between the temperature predicted by Fourier's law and Eqs.~(\ref{eq:macro_diff_eq1},~\ref{eq:macro_diff_eq2}) is largest at \replaced{$x\simeq3\;\mu m$}{$x=1\;\mu m$}, \textit{i.e.} where the sample of Fig.~\ref{fig:2heat_flux} changes geometry. 
We show in the inset of Fig.~\ref{fig:temp_difference}b that also the longitudinal temperature \added{profile} (for $y=\replaced{4.5}{0.55}\; \mu m$) changes when going from Fourier's law (dashed red line) to the viscous heat equations (solid blue line).
The difference is maximized close to variations of the sample's shape at $x=\replaced{3}{1}\;\mu m$ and at the boundaries $x=0$ and \replaced{$x=15$}{$x=5$} $\mu m$; in this latter case, the viscous heat equations predict steeper-than-Fourier's law or non-linear temperature gradients that are reminiscent of those obtained in molecular dynamics simulations~\cite{nonlinear_profile_NEMD_1,Zhang2017,Khadem2013,puligheddu2017first,PhysRevE.97.032102,he2012lattice,PhysRevB.79.115201,PhysRevB.95.144309,Allen_2018_1,li2019influence} and in explicit solutions of the LBTE~\cite{cepellotti2017boltzmann}. 

\section{Fourier Deviation Number} 
\label{sec:deviations_from_fourier_s_law}

In this section, we introduce a \replaced{descriptor}{scalar} that parametrizes the conditions under which hydrodynamic heat conduction is observed; we will refer to this as the ``Fourier deviation number'' (FDN).
In particular, we aim at distinguishing the diffusive regime from the hydrodynamic regime: in the former case the viscous heat equations become equivalent to Fourier's law, while in the latter case Fourier's law no longer holds and the viscous heat equations are required.

\begin{figure*}[t]
  \centering
    \includegraphics[width=\textwidth]{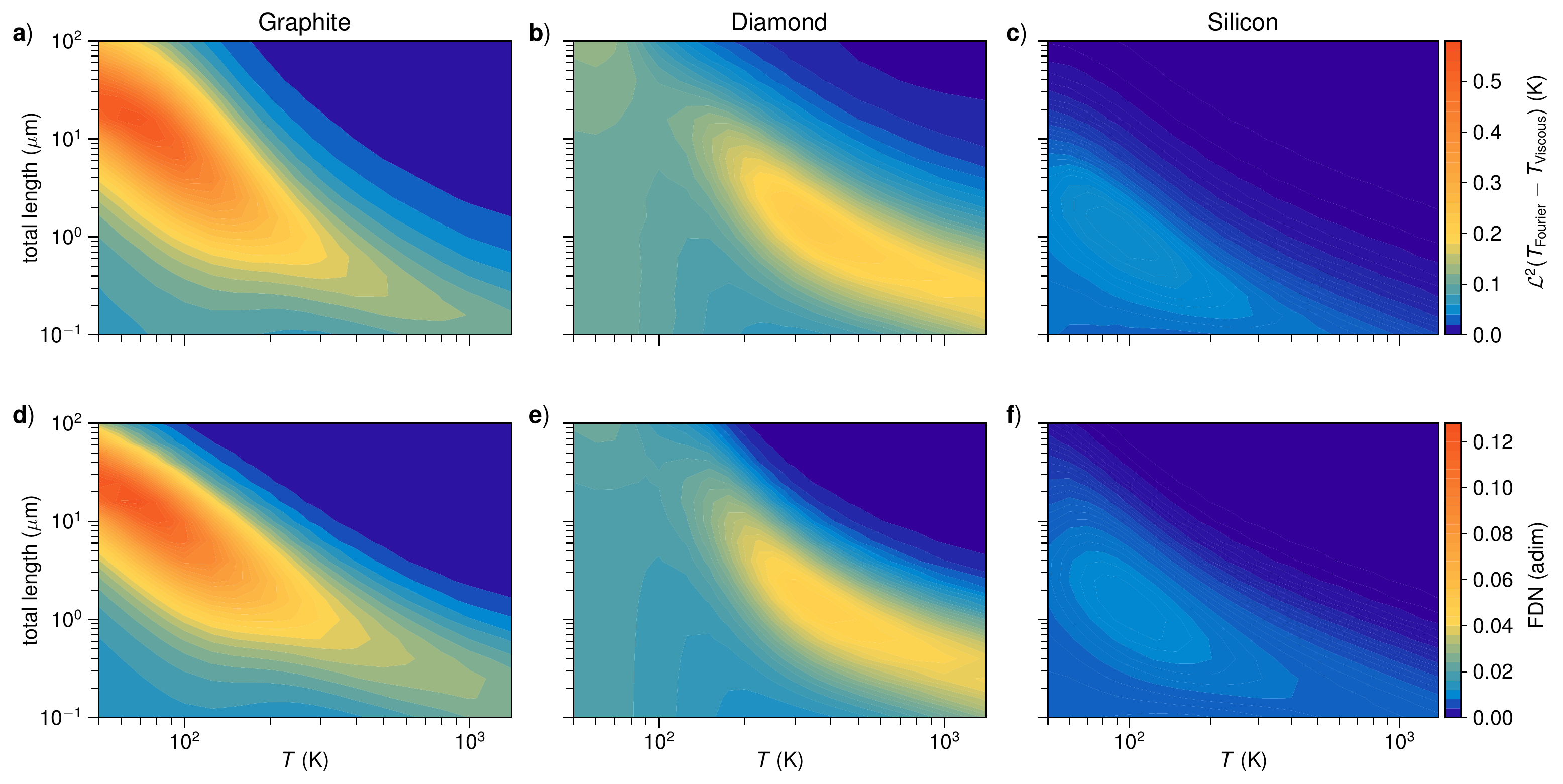}

    \caption{
    \added{Top row, $\mathcal{L}^2$ distance (see text) between the temperature profile predicted by Fourier's law and that of the viscous heat equations~(\ref{eq:macro_diff_eq1},~\ref{eq:macro_diff_eq2}), shown as a function of temperature and total length of a sample similar (in the geometric sense) to the shape of Fig.~\ref{fig:2heat_flux} for the cases of graphite (a), diamond (b) and silicon (c). 
    The color quantifies the  $\mathcal{L}^2$  distance between the temperature profiles predicted by Fourier's law ($T_{\rm Fourier}(x,y)$) and the viscous heat equations~(\ref{eq:macro_diff_eq1},~\ref{eq:macro_diff_eq2}) ($T_{\rm Viscous}(x,y)$), computed over the region $G$ corresponding to points with $x>\frac{1}{5}l_{\rm TOT}$ where $l_{\rm TOT}$ is the total length of the sample geometrically similar to that in Fig.~\ref{fig:2heat_flux} (see main text for details).  Bottom row, Fourier Deviation Number (FDN, as defined in Eq.~(\ref{eq:FDN_def})) for the same materials as a function of temperature and total length of the sample (for the geometry of Fig.~\ref{fig:2heat_flux}, the characteristic size $L$ appearing in FDN is the shortest length-scale, \textit{i.e.} 1/5 of the total length $l_{\rm TOT}$). It is apparent that the FDN correctly identifies the deviations between the solutions of the Fourier and of the viscous heat equations solutions.
    }
    }
  \label{fig:deviation_Fourier}
\end{figure*}

\added{{In order to investigate how the hydrodynamic deviations from Fourier's law depend on the sample's sizes and reference temperature $\bar{T}$, 
we solve the viscous heat equations and Fourier's equation for several samples having different dimensions, and at various reference temperatures $\bar{T}$. We perform this analysis for graphite (Fig.~\ref{fig:deviation_Fourier}a), diamond (Fig.~\ref{fig:deviation_Fourier}b) and silicon (Fig.~\ref{fig:deviation_Fourier}c). 
In particular, we consider samples that are similar --- in the geometric sense --- to the reference sample shown in Fig.~\ref{fig:2heat_flux}, \textit{i.e.} we generate samples of different dimensions starting from the reference sample of Fig.~\ref{fig:2heat_flux} and varying the sizes via a uniform scaling. 
Following this protocol, we vary the sample's length $l_{\rm TOT}$, reported on the $y-$axis of Fig.~\ref{fig:deviation_Fourier}a-c, from $0.1$ to $100$ $\mu$m 
(e.g. the reference sample of Fig.~\ref{fig:2heat_flux} corresponds to $y=15\;\mu m$, $y=100\;\mu m$ corresponds to a sample obtained magnifying uniformly by a factor of $100/15\simeq 6.66$ the sizes of the reference sample).
The color in Fig.~\ref{fig:deviation_Fourier}a-c represents the normalized $\mathcal{L}^2$ distance between the temperature profile predicted by Fourier’s law and the viscous heat equations for a given sample length $l_{\rm TOT}$ and reference temperature $\bar{T}$:
\begin{equation}
\begin{split}
 	{\mathcal{L}^2}[&{T}_{\rm Fourier}-{T}_{\rm Viscous}](l_{\rm TOT}, \bar{T})=\\
 	&\sqrt{\frac{\int_{G}[ T_{\rm Fourier}(x,y)-T_{\rm Viscous}(x,y)]^2 dx dy }{\int_{G} dx dy}}\;,
\end{split}
\end{equation}
and in order to ease the qualitative interpretation of these results later, we evaluate $\mathcal{L}^2$ in the spatially-homogeneous region $G$ defined by $x>\frac{1}{5}l_{\rm TOT}$. 
We also inspected the effects of shape on the deviation $\mathcal{L}^2$, finding that the magnitude of $\mathcal{L}^2$ is larger for geometries with spatial non-homogeneities that imply larger values of the drift velocity's second derivative; nevertheless, the dependence of $\mathcal{L}^2$ on  $l_{\rm TOT}$ and $\bar{T}$ is qualitatively unchanged for different shapes. 
}}

\added{Results in Figs.~\ref{fig:deviation_Fourier}a-c have been computed accounting for finite-size effects as discussed in Sec.~\ref{sec:thermal_viscosity} and Appendix~\ref{sec:App4_ballistic_tc_visc}. 
Additionally, to better compare with the experimental results of Ref.~\cite{Chen_science_2019}, we make the hypothesis of working with polycrystalline graphite with an average crystal grain size of $10\;\mu m$ as in Ref.~\cite{Chen_science_2019}: as a result, the transport coefficients of graphite are renormalized by the system size for $l_{\rm TOT}<10\;\mu m$, and by the grain size for $l_{\rm TOT}>10\;\mu m$ (therefore, for $l_{\rm TOT}>10\;\mu m$ size effects are given only by the boundary conditions effects on the temperature and drift-velocity fields).
In graphite, the difference between Fourier's law and the viscous heat equations is largest in the temperature range 60-80 K and for sizes $10\mu m\lesssim l_{\rm TOT}\lesssim 20\mu m$.
}

\added{Turning our attention to diamond, we first recall that its thermal properties are characterized by a very large thermal conductivity, which originates from having large group velocities and weak Umklapp scattering~\cite{diamond_0, diamond_1, diamond_2, diamond_3, diamond_4, diamond_5}; the latter condition being favorable to the emergence of hydrodynamic effects.
This is confirmed by the results shown in Fig.~\ref{fig:deviation_Fourier}b,
with the hydrodynamic deviation being largest around room temperature and dimension $l_{\rm TOT}\gtrsim 1\;\mu m$ (we note in passing that, at this temperature and size, the surface-renormalization of transport coefficients is small and well compatible with the mesoscopic approach described in this work).
These results suggest that a hydrodynamic window exists also for diamond, and that therefore hydrodynamic behavior (e.g. second sound) might be measurable in this material at temperatures even larger than graphite.
}

\added{For silicon, instead, the deviation from Fourier's law is smaller and takes place at lower temperatures.
Silicon is an example of a material for which the thermal conductivity computed within the SMA approximation is very close to the conductivity computed from the exact solution of the LBTE~\cite{broido2007intrinsic,fugallo2013ab,li2014shengbte,cepellotti2016thermal}, and it is known that the SMA works reasonably well in systems where momentum-dissipating (Umklapp) scattering events dominate over those conserving crystal momentum (normal)~\cite{li2014shengbte,lindsay_first_2016}. Therefore, the negligible magnitude of hydrodynamic effects predicted by the viscous heat equations for silicon is consistent with the predominance of  Umklapp over normal scattering events known to occur in this material.
We last remark that, in contrast to graphite, results for diamond and silicon have been computed for single crystals, \textit{i.e.} transport coefficients are not limited by grains' size.
}

In order to capture intuitively and inexpensively all these trends we introduce an approach inspired by the definition of the Reynolds number, and we rewrite the viscous heat equations~(\ref{eq:macro_diff_eq1}, \ref{eq:macro_diff_eq2}) in adimensional form (we follow the standard procedure used e.g. in fluid dynamics, which is also called ``Buckingham Pi theorem''~\cite{PhysRev.4.345,white2003fluid}).
First, we extract the magnitudes of the tensors appearing in the viscous heat equations, factorizing the largest component: $A^i{=}A_{\rm m}\mathsf{a}^{i}$, where $A_{\rm m}{=}\max\limits_{i}(|A^i|)$ and $\mathsf{a}^{i}=A^i/A_{\rm m}$ is an adimensional tensor with the largest component having modulus equal to $1$. 
\added{Similarly, we factorize the largest component of all other tensors: $\kappa^{ij}{=}\kappa_{\rm m}\mathsf{k}^{ij}$ (with $\kappa_{\rm m}{=}\max\limits_{i,j}(|\kappa^{ij}|)$ and $\mathsf{k}^{ij}{=}\kappa^{ij}/\kappa_{\rm m}$); $W^j_{i0}{=}W_{\rm m}\mathsf{w}^{j}_{i}$ (with $W_{\rm m}{=}\max\limits_{i,j}(|W^j_{i0}|)$ and $\mathsf{w}^{j}_{i}{=}W^j_{i0}/W_{\rm m}$); $D^{ij}_{U}{=}D_{U,{\rm m}}\mathsf{d}^{ij}$ (with $D_{U,{\rm m}}{=}\max\limits_{i,j}(|D^{ij}_{U}|)$ and $\mathsf{d}^{ij}{=}D^{ij}_{U}/D_{U,{\rm m}}$); $\mu^{ijkl}{=}\mu_{\rm m}\mathsf{m}^{ijkl}$ (with $\mu_{\rm m}{=}\max\limits_{i,j,k,l}(|\mu^{ijkl}|)$ and $\mathsf{m}^{ijkl}{=}\mu^{ijkl}/\mu_{\rm m}$).}
Then, we define a set of dimensionless variables $\bm{r}^*=\bm{r}/L$, $\bm{u}^* = \bm{u} / u_0$, and $T^*=T/{\delta T}$, where $L$, $u_0$ and $\delta T$ are a characteristic size, drift velocity and temperature perturbation (more on these later).
Substituting these variables in Eqs.~(\ref{eq:macro_diff_eq1},~\ref{eq:macro_diff_eq2}), and limiting ourselves to the steady-state regime, we obtain:
\begin{align}
  &\pi_1\sum_{i,j=1}^3\frac{\mathsf{w}^{i}_{j}\sqrt{\mathsf{a}^j}}{g(\mathsf{w}{\cdot}\sqrt{\mathsf{a}})_{2}}\frac{\partial{u}^{*j}(\bm{r}^*)}{\partial r^{*i}} -\sum_{i,j=1}^3 \frac{\mathsf{k}^{ij}}{g(\mathsf{k})_2}\frac{\partial^2 T^{*}(\bm{r}^*)}{\partial r^{*i}\partial r^{*j} } = 0\;,\label{eq:fou_analog} 
\end{align}
\begin{align} 
  &\pi_2\sqrt{\mathsf{a}^i}\sum_{j=1}^3 \frac{\mathsf{w}^j_i}{g^i(\mathsf{w})_1}\frac{\partial T^{*}(\bm{r}^*)}{\partial r^{*j}}
  -\pi_{3}\sum_{j,k,l=1}^{3}\frac{\mathsf{m}^{ijkl}}{g^i(\mathsf{m})_3}\frac{\partial^2 u^{*k}(\bm{r}^*)}{\partial r^{*j}\partial r^{*l} }=\nonumber \\
  &\hspace*{3.5cm}-\sum_{j=1}^{3}\frac{\sqrt{\mathsf{a}^i\mathsf{a}^j}\mathsf{d}^{ij}}{g^i(\sqrt{\mathsf{a}{\otimes}\mathsf{a}}{\cdot}\mathsf{d})_1}{u}^{*j}(\bm{r}^*)\label{eq:drift_eq} \;,
  \raisetag{8mm}
\end{align} 
\added{{where we have introduced the quantities\\ 
$g(\mathsf{w}{\cdot}\sqrt{\mathsf{a}})_{2}=({\max_{ij}[\mathsf{w}^i_j\sqrt{\mathsf{a}^j}]})^{-1}\sum_{ij} {\mathsf{w}^i_j\sqrt{\mathsf{a}^j}}$,\\ 
$g(\mathsf{k})_2=({\max_{ij}[\mathsf{k}^{ij}]})^{-1}\sum_{ij} {\mathsf{k}^{ij}}$,\\ 
$g^i(\mathsf{w})_1=({\max_{j}[\mathsf{w}^{i}_j]})^{-1}\sum_{j} {\mathsf{w}^{i}_j}$,\\ 
$g^i(\sqrt{\mathsf{a}{\otimes}\mathsf{a}}{\cdot}\mathsf{d})_1=({\max_{j}[\sqrt{\mathsf{a}^{i}\mathsf{a}^{j}}\mathsf{d}^{ij}]})^{-1}\sum_{j} \sqrt{\mathsf{a}^{i}\mathsf{a}^{j}}\mathsf{d}^{ij}$,\\ 
$g^i(\mathsf{m})_3=({\max_{jkl}[\mathsf{m}^{ijkl}]})^{-1}\sum_{jkl}{\mathsf{m}^{ijkl}}$.
The role of these parameters is to account for the correct order of magnitude of the five summations appearing in Eqs.~(\ref{eq:fou_analog},\ref{eq:drift_eq}); they carry a Cartesian superscript $i$ whenever they may depend on direction, and their subscripts indicate the number of indexes summed in their definition. For example, for a $3{\times}3$ isotropic tensor $\chi^{ij}=\chi_{\rm m} \delta^{ij}$ (where $\chi_{\rm m}$ is a constant and $\delta^{ij}$ is the Kronecker tensor), $g(\chi)_2=(\chi_{\rm m})^{-1} \sum_{ij} \chi^{ij}=3$ and $g^i(\chi)_1=(\chi_{\rm m})^{-1} \sum_{j} \chi^{ij}=1\;\forall\; i$.
Because we are computing the FDN for the planar geometry discussed in Fig.~\ref{fig:2heat_flux}, we can discard the $z$ component (\textit{i.e.} the Cartesian direction indexed by $3$) from Eqs.~(\ref{eq:fou_analog},\ref{eq:drift_eq}).
Then, for diamond, silicon and in-plane graphite, transport properties are isotropic (\textit{i.e.} all the 2$^{\rm nd}$-rank tensors appearing in Eq.~(\ref{eq:fou_analog}) and Eq.~(\ref{eq:drift_eq}) are diagonal).
From this it follows that $g(\mathsf{w}{\cdot}\sqrt{\mathsf{a}})_{2}=2$;
$g(\mathsf{k})_{2}=2$;
$g^i(\mathsf{w})_1=1\;\forall \;i=1,2$; and
$g^i(\sqrt{\mathsf{a}{\otimes}\mathsf{a}}{\cdot}\mathsf{d})_1=1\;\forall \;i=1,2$.
Finally, the tensor $\mathsf{m}^{ijkl}$ is not isotropic (because the thermal viscosity tensor from which it is derived is not isotropic), and from the values in Fig.~\ref{fig:res_viscosity} it follows that $g^i(\mathsf{m})_1=1+\frac{\mu^{xyxy}}{\mu^{xxxx}}\forall i=1,2,3$ for diamond and silicon, and $g^i(\mathsf{m})_1=1+\frac{\mu^{xyxy}}{\mu^{xxxx}}+2\frac{\mu^{xyyx}}{\mu^{xxxx}}\;\forall\;i=1,2$ for graphite.}}

The final expressions for the dimensionless parameters $\pi_1,\;\pi_2,\;\pi_3$ are thus:
\begin{equation}
\pi_1 = \frac{\sqrt{\bar{T}A_{\rm m}C}W_{\rm m}u_0 L}{\kappa\cdot\delta T}\frac{g(\mathsf{w}{\cdot}\sqrt{\mathsf{a}})_{2}}{g(\mathsf{k})_2 }  \;,\label{eq:pi_1} 
\end{equation}
\begin{equation}
\pi_2 =\sqrt{\frac{C}{A_{\rm m}\bar{T}}}  \frac{W_{\rm m} \delta T}{L{D}_{U,{\rm m}} u_0} \frac{g^i(\mathsf{w})_1}{g^i(\sqrt{\mathsf{a}{\otimes}\mathsf{a}}{\cdot}\mathsf{d})_1} \;,\label{eq:pi_2}\end{equation}
\begin{equation}
\pi_3 = \frac{\mu_{\rm m}}{{D}_{U,{\rm m}} L^2 A_{\rm m} } \frac{g^i(\mathsf{m})_3}{g^i(\sqrt{\mathsf{a}{\otimes}\mathsf{a}}{\cdot}\mathsf{d})_1}
 \;.\label{eq:pi_3}
\end{equation}
\added{From the derivation of these parameters, it is clear that they can be interpreted in terms of average values (we indicate with $\big< \dots \big>$ the average over space) of physical quantities: $\pi_1{\sim {\big<{Q}^D\big>}/{\big<Q^\delta\big>}}$, $\pi_2\sim {\big< \frac{\partial \Pi_T}{\partial r} \big>}/{\big< \frac{\partial P}{\partial t} \big|_{\text{Umkl}}\big>}$ and $\pi_3\sim{\big< \frac{\partial \Pi_{\delta{\rm E} }}{\partial r} \big>}/{\big< \frac{\partial P}{\partial t} \big|_{\text{Umkl}} \big>}$.}
Still, to evaluate these parameters we need to know the characteristic size $L$ and temperature perturbation $\delta T$, and estimate the characteristic values of the drift velocity $u_0$. 
Focusing on the setup discussed in the previous section (Fig.~\ref{fig:2heat_flux}), we have \replaced{$L=\frac{1}{5}l_{\rm TOT}$}{L=$1\mu$m} \added{(corresponding to the shortest length scale of the geometry considered, which has total length $l_{\rm TOT}$ along $x$)} and \replaced{$\delta T$=10K.}{$\delta T$=20K.}
As shown in Appendix~\ref{sec:App8_estimation_of_the_typical_velocity}, the characteristic drift velocity $u_0$ is found by interpolating the asymptotic behavior at low ($u_L$) and high ($u_H$) temperatures $u_0^{-1} = u_L^{-1} + u_H^{-1}$.
In the low-temperature limit, where Umklapp scattering is frozen, viscous effects determine the drift velocity, and one can show that \replaced{
  $u_L=\frac{3}{7}\sqrt{\frac{C A}{\bar{T}}} \frac{{W}^x_{x0}\;\delta T\; L }{\mu^{xyxy}\left(1+\frac{\mu^{xxxx}}{\mu^{xyxy}}\right)}$
}{$u_L=
\frac{\delta T}{36} 
{\cdot}
\sqrt{\frac{CA^x}{\bar{T}}}
\frac{W^x_{x0}}{2\mu^{xyxy}}L^2$.}.
In the high temperature limit, the drift velocity is mainly determined by the crystal momentum dissipation rate and 
\replaced{$u_H=\sqrt{\frac{C}{\bar{T}A^x}}\frac{{W}_{x0}^x }{{D}_U^{xx}}\frac{2}{7}\frac{\delta T}{L}$}{
$u_H=\sqrt{ \frac{C}{\bar{T}A^x} }\frac{{W}_{x0}^x\delta T }{{3\; D_U^{xx}}L}$}.
We can therefore estimate the values of all the $\pi_1$, $\pi_2$, and $\pi_3$ factors. 

\begin{figure}[b]
  \centering
    \includegraphics[width=\columnwidth]{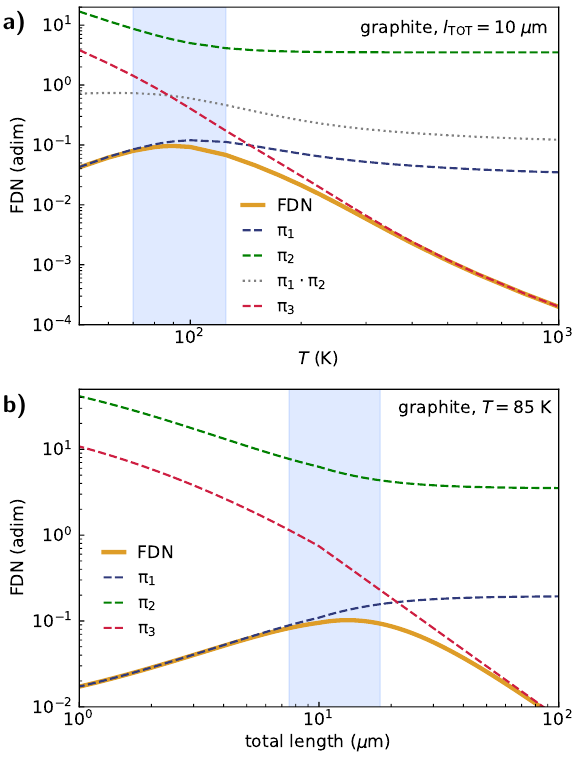}
  \caption{\added{Fourier Deviation Number (FDN, solid line) and its different contributions $\pi_1$ and $\pi_3$ defined in Eq.~(\ref{eq:FDN_def}) for a graphite sample of fixed length $l_{\rm TOT}=10\mu m$ as a function of temperature (panel a), and at fixed $T$=85 K as a function of sample's length $l_{\rm TOT}$ (panel b).
  The shaded areas are the hydrodynamic (second-sound) windows measured at $10\mu$m (panel a) or 85 K (panel b) by \citet{Chen_science_2019}.
  The terms $\pi_2$ and $\pi_1\pi_2$ are also reported in panel a) because in the high-temperature limit Fourier's law is recovered when $\pi_1\pi_2\ll 1$ (see main text for details).
  Notably, FDN predicts hydrodynamic effects to be largest at temperatures and sizes that are in excellent agreement with these observed in the experiments.
  }}
  \label{fig:HFDN_vs_T}
\end{figure}

\added{
Looking at the definition of $\pi_1$, $\pi_2$ and $\pi_3$, it is straightforward to identify the conditions for which the viscous heat equations reduce to Fourier's law.
When  $\pi_3{\ll}1$, \textit{i.e.} for crystal momentum dissipation dominating over viscous effects and/or very large characteristic size $L$, the temperature gradient  in Eq.~(\ref{eq:drift_eq}) is proportional to the drift velocity:
$\pi_2\sum_{j} \frac{\mathsf{w}^j_i}{g^i(\mathsf{w})_1}\frac{\partial T^{*}(\bm{r}^*)}{\partial r^{*j}}
{\simeq} -\sum_{j}\frac{\sqrt{\mathsf{a}^j}\mathsf{d}^{ij}}{g^i(\sqrt{\mathsf{a}{\otimes}\mathsf{a}}{\cdot}\mathsf{d})_1}{u}^{*j}(\bm{r}^*)$. 
While it is intuitive to understand the emergence of Fourier's law when crystal momentum dissipation (e.g. due to Umklapp scattering) dominates over viscous effects, the Fourier-like behavior of large-size samples can be rationalized qualitatively recalling the simple analytical 1D solution of the viscous heat equations in the absence of Umklapp processes discussed in Appendix~\ref{sec:App_Sussman_Tellung}: the heat flux is described by hyperbolic functions with a characteristic length scale $b=\sqrt{\frac{\mu_{\rm m} \kappa_{\rm m}}{A_{\rm m} C (W_{\rm m})^2}}$, and for a characteristic size $L\gg b$, a flat Fourier-like heat flux profile is recovered.
It is worth mentioning that $\pi_3\ll 1$ --- and thus Fourier-like behavior --- can also emerge for very small (sub-micrometer) length scales, where ballistic effects are relevant and strongly renormalize the transport coefficients, as discussed in Appendix~\ref{sec:App4_ballistic_tc_visc}. 
The emergence of Fourier-like behavior in the ballistic regime (with a size-dependent thermal conductivity alike to that employed here) has been explained in a recent work \cite{maassen2015steady}, confirming that the results for sub-micrometer sizes obtained with the aforementioned approximated treatment of boundary scattering are qualitatively correct.
If we insert the condition $\pi_3\ll 1$ in Eq.~(\ref{eq:fou_analog}), we obtain a Fourier-like equation for the temperature:
$ \sum\limits_{i,j} \hspace*{-1mm}\Big[\hspace*{-1mm}\Big(\hspace*{-1mm}\pi_1\pi_2 
\frac{g(\mathsf{k})_{2} g^i(\sqrt{\mathsf{a}{\otimes}\mathsf{a}}{\cdot}\mathsf{d})_1}{g(\mathsf{w}{\cdot}\sqrt{\mathsf{a}})_{2} g^i(\mathsf{w})_1}  
\hspace*{-1mm}\sum\limits_{k,l}\hspace*{-0.3mm}  (\mathsf{w}^{i}_{k}\mathsf{d}^{-1})^{kl}\mathsf{w}^{j}_{l}\hspace*{-0.3mm}\Big){+} {\mathsf{k}^{ij}}\hspace*{-0.3mm}\Big]\hspace*{-1mm}\frac{\partial^2 T^{*}(\bm{r}^*)}{\partial r^{*i}\partial r^{*j} } {=} 0$. 
So, the thermal conductivity is corrected by a term proportional to $\pi_1 \pi_2$ {(since in all the setups considered here, $\frac{g(\mathsf{k})_{2} g^i(\sqrt{\mathsf{a}{\otimes}\mathsf{a}}{\cdot}\mathsf{d})_1}{g(\mathsf{w}{\cdot}\sqrt{\mathsf{a}})_{2} g^i(\mathsf{w})_1}=1$)}, 
but the qualitative behavior is that of Fourier's law.
Numerically, we find that $\pi_1\pi_2{\ll}1$ for graphite in the high-temperature limit (see Fig.~\ref{fig:HFDN_vs_T}a), so that the viscous heat flux is well approximated by Fourier's flux; {the same limiting behavior is observed also in  diamond and silicon.}
On the other hand, deviations from the temperature profile predicted by Fourier's law appear when both $\pi_1$ and $\pi_3$ are large.
In fact, $\pi_3\gtrsim 1$ implies that the temperature gradient in Eq.~(\ref{eq:drift_eq}) is coupled to the second derivative of the drift velocity, and one can not obtain a Fourier-like equation as before; large values of $\pi_1$ imply that the drift velocity affects the evolution of temperature in Eq.~(\ref{eq:fou_analog}).
It follows that Fourier's law is mathematically recovered for crystal momentum dissipation dominating over viscous effects and/or large characteristic size ($\pi_3{\ll}1$), but deviations between the viscous and Fourier temperature profile can be small even when viscous effects dominate over crystal momentum dissipation ($\pi_3{\gg}1$) if the coupling between drift velocity and temperature in Eq.~(\ref{eq:fou_analog}) is small ($\pi_1{\ll}1$).
We note in passing that while this reasoning has been made here at steady state, it can be straightforwardly extended to the time-dependent case, considering time variations of the drifting velocity slow compared to the crystal momentum dissipation timescale.
}

\added{On the basis of this reasoning, it is convenient to} introduce a Fourier deviation number (FDN) as:
\begin{equation}
  {\rm FDN} =\left( \frac{1}{\pi_1}+\frac{1}{\pi_3}\right)^{-1} \;,
 \label{eq:FDN_def}
\end{equation}
which provides a simple estimate of the deviations from Fourier's law: the larger the FDN, the larger the deviations from Fourier's law.
%
\added{In Fig.~\ref{fig:deviation_Fourier}d-f, we plot the FDN for graphite, diamond and silicon as a function of sample's total length $l_{\rm TOT}$ and reference temperature $\bar{T}$.
Remarkably, the FDN captures the trends of the exact solution of the viscous heat equations (Fig.~\ref{fig:deviation_Fourier}a-c), thus identifying accurately the regime where hydrodynamic behavior emerges. }
\added{The magnitude of FDN is directly proportional to the magnitude of hydrodynamic effects: a larger FDN implies a larger difference between the Fourier and viscous temperature profiles.}
\added{
A detailed analysis of the $\pi$ terms as a function of temperature for a graphite sample geometrically similar to that in Fig.~\ref{fig:2heat_flux} and with a fixed total length $l_{\rm TOT}=10\mu$m is shown in Fig.~\ref{fig:HFDN_vs_T}a, where $l_{\rm TOT}=10 \mu$m has been chosen to match the length fixed in the experiments by \citet{Chen_science_2019} to investigate the magnitude of hydrodynamic effects as a function of temperature.
Importantly, the FDN at fixed $l_{\rm TOT}=10 \mu$m predicts hydrodynamic effects to be largest at temperatures that match very well those measured in recent experiments for second sound in graphite~\cite{Chen_science_2019}.
Then, we rationalize the trend of FDN as a function of temperature in terms of the $\pi$ terms entering in Eq.~(\ref{eq:FDN_def}): $\pi_1$ increases with temperature for T$\lesssim100$ K and saturates to a constant value at high temperature and
$\pi_3$ decreases asymptotically like $T^{-2}$.
Fig.~\ref{fig:HFDN_vs_T}b shows the analysis of FDN as a function of the sample's size $l_{\rm TOT}$ (varied according to the aforementioned protocol) at fixed temperature 85 K, \textit{i.e.} the temperature fixed in the experiments by \citet{Chen_science_2019} to investigate the magnitude of hydrodynamic effects as a function of size.
Also in this case, FDN predicts hydrodynamic effects to be largest in a size range that matches very well the sizes at which hydrodynamic effects have been measured.
We thus conclude that the present model is capable of describing quantitatively the hydrodynamic window that has been recently measured in graphite.
}
\added{In Appendix~\ref{sec:second_sound_in_very_large_diamond_samples}, we also show that for diamond extending the sizes in Fig.~\ref{fig:deviation_Fourier}e up to $10$ mm yields an increase of FDN for temperatures around 50-60 K and dimensions $\sim 1$ mm.
This result is in qualitative agreement with the predictions for diamond's second sound window performed using the reduced isotropic crystal model or the Callaway model~\cite{second_sound_diamond}.
Finally, the parameters entering in the $\pi_i\;(i=1,2,3)$ are exactly the same used for the numerical solutions of the viscous heat equations in Fig.~\ref{fig:deviation_Fourier}a-c.}

\section{Conclusions} 
\label{sec:conclusions}

\added{
We have introduced a framework to describe heat conduction beyond Fourier's law and to capture in the process the hydrodynamic transport regime where the phonon gas assumes a drift velocity and heat propagation resembles fluid dynamics.
Just as a perturbation of the temperature generates an energy flux that is proportional to thermal conductivity, a perturbation of the drift velocity generates a crystal momentum flux, with the proportionality tensor coefficient between the two being a thermal viscosity.
We have shown that thermal conductivity and viscosity can be determined exactly and in a closed form as a sum over relaxons (\textit{i.e.} the eigenvectors of the phonon scattering matrix).
Relaxons have a well-defined parity and even relaxons contribute exclusively to the thermal viscosity (odd relaxons contribute exclusively to the thermal conductivity \cite{cepellotti2016thermal}).}

\added{
Most importantly, the microscopic LBTE has been coarse-grained into two novel mesoscopic viscous heat equations, which are coupled equations parametrized by the thermal conductivity and viscosity that govern the evolution of the temperature and drift-velocity fields.
The viscous heat equations provide a general description of heat transfer
encompassing both the Fourier and the hydrodynamic regimes, and all intermediate cases, 
allowing for the emergence of second sound, and of Poiseuille-like heat flow associated to a temperature profile deviating from Fourier's law.
In addition, these viscous heat equations show that 
the thermal conductivity is not sufficient to describe thermal transport in general terms, but also its complementary quantity, the thermal viscosity, must be taken into account.
}

\added{
We have characterized the hydrodynamic behavior in terms of the Fourier deviation number (FDN), a dimensionless parameter that quantifies hydrodynamic deviations from Fourier's law. 
At a negligible computational cost, the FDN enables to study the temperature-and-size window at which hydrodynamic phenomena such as second sound emerge.}

\added{
The full solution of the viscous heat equations allows to predict measurable temperature and heat flux profiles in complex shaped devices in the mesoscopic heat transport regime at a much reduced cost, and more transparently, than solving the full LBTE, therefore paving the way towards understanding shape and size effects in complex and microscopic devices, and especially when novel physics arises, such is the case in phononic devices \cite{PhysRevLett.88.094302,PhysRevLett.93.184301,Chang1121,hu2009thermal,wang2014phonon,PhysRevB.91.245417,volz2016nanophononics,wang2017experimental,liu2017model,chen2018efficient,felix2018thermal,sood2018electrochemical,de2019phonon}.
These results are particularly relevant for the emerging field of materials that display hydrodynamic thermal transport,  often characterized by large thermal conductivities~\cite{lee_hydrodynamic_2015,cepellotti2015phonon,lee_hydrodynamic_2017,ding_phonon_2018,Chen_science_2019,cepellotti2016second_sound,martelli_thermal_2018,second_sound_diamond,machida_observation_2018}.
Such behavior has been investigated in devices made either of graphite, diamond or silicon.
For graphite, the present formulation predicts the hydrodynamic windows (as a function of temperature for a fixed size, or as a function of size for a fixed temperature) in excellent agreement with recent experimental findings~\cite{Chen_science_2019}, while suggesting that hydrodynamic behavior can also appear in diamond at room temperature for micrometer-sized crystals.}

\added{
Finally, we remark that the present methodology can be adapted to describe viscous phenomena for electronic conduction~\cite{levitov2016electron,Bandurin1055,moll2016evidence,PhysRevLett.118.226601}, or any other transport phenomena described by a linearized Boltzmann equation.}

\section{Acknowledgements} 
\label{sec:acknowledgements}
We acknowledge support from the Swiss National Science Foundation 
under project ID P2ELP2\_168546 
and the MARVEL NCCR. Computational resources have been provided by the Swiss National Supercomputing Center (CSCS) and PRACE.

\appendix

\section{Eigenvectors of the scattering matrix}
\label{sec:App1_eigv_scatt_matrix}
The scattering matrix $\Omega_{\state,\statep}$ appearing in Eq.~(\ref{eq:LBTE}) is not symmetric, but it can be recast in a symmetric (and thus diagonalizable) form by means of the the following transformation~\cite{hardy1970phonon,chaput2013direct,cepellotti2016thermal}:
\begin{gather}
    \tilde{\Omega}_{\state \statep} = 
    \Omega_{\state \statep} \sqrt{ \frac {\bar{N}_{\statep}(\bar{N}_{ \statep}+1)} {\bar{N}_{\state}(\bar{N}_{\state}+1)} } \;, \\
     \tilde{n}_{\state}(\bm{r},t)=\frac{ n_{\state}(\bm{r},t)}{\sqrt{\bar{N}_{\state}(\bar{N}_{\state}+1)}} \;,
  \label{eq:symmetrization_LBTE}
\end{gather}
where also the distribution $n_{\state}(\bm{r},t)$ appearing in Eqs.~(\ref{eq:LBTE_odd_nablaT},\ref{eq:LBTE_linear_nabla_u}) has to be transformed for consistency.
The symmetrized scattering operator $\tilde{\Omega}_{\state,\statep}$ is real and symmetric, and can thus be diagonalized~\cite{cepellotti2016thermal,hardy1970phonon,chaput2013direct}:
\begin{equation}
  \frac{1}{V} \sum_{\statep} \tilde{\Omega}_{\state \statep} \theta^{\alpha}_{\statep}
  =\frac{1}{\tau_\alpha} \theta^{\alpha}_{\state} \;,
\end{equation}
where $\theta^{\alpha}_{{\state}}$ denotes a relaxon (\textit{i.e.} an eigenvector), $\alpha$ is the relaxon index and the inverse eigenvalue ${\tau_{\alpha}}$ is the relaxon lifetime.
We also define a scalar product \cite{cepellotti2016thermal}:
\begin{equation}
\left<\alpha|\alpha'\right>
= \frac{1}{V} \sum_{\state} \theta^{\alpha}_{\state}\theta^{\alpha'}_{\state} 
= \delta_{\alpha,\alpha'} \;,
\end{equation}
used to orthonormalize eigenvectors.

In order to show that eigenvectors with zero eigenvalues are related to conserved quantities in the LBTE dynamics, we rewrite the scattering operator distinguishing scattering events that conserve crystal momentum -- normal (N) -- from those that do not -- Ukmlapp (U):
\begin{equation}
    \tilde{\Omega}_{\state \statep} = \tilde{\Omega}^N_{\state \statep} + \tilde{\Omega}^U_{\state \statep} \;.
    \label{eq:separation_N_U}
\end{equation}
As stated in the main text, there are four eigenvectors 
with zero eigenvalues, 
that we will discuss in the next sections.

\subsection{The Bose-Einstein eigenvector: local temperature} 
\label{sub:the_bose_eigenvector}
Applying the transformation~(\ref{eq:symmetrization_LBTE}) to Eq.~(\ref{eq:LBTE_expanded_gradiend}) and considering the steady-state conditions, one obtains the following equation for the even part \added{(see below for the odd part)}:
\begin{equation}
\begin{split}
    \frac{\bm{v}_{\state} }{\sqrt{\bar{N}_{\state}(\bar{N}_{\state}+1)}}  
&\cdot \bigg(\frac{\partial N^D_{\state}}{\partial \bm{u}} \cdot \nabla \bm{u}\bigg)\\
&= -\frac{1}{V} \sum_{\statep} \tilde{\Omega}_{\state,\statep} \big(\tilde{n}^{T}_{\statep}(\bm{r},t)+\tilde{n}^{\delta{\rm E} }_{\statep} \big)\;.
\end{split}
\label{eq:LBTE_viscosity_symm}
\end{equation}
The distribution $\tilde{n}^T_{\state}(\bm{r},t)$ is obtained applying the symmetrization~(\ref{eq:symmetrization_LBTE}) to the distribution ${n}^T_{\state}(\bm{r},t)$ appearing in equation~(\ref{eq:hardy_decomposition}) and it is thus evident that $\tilde{n}^T_{\state}(\bm{r},t)\propto \omega_{\state}(T(\bm{r},t)-\bar T)$.
From the energy conservation in scattering events (both normal and Umklapp), it follows that~\cite{spohn2006phonon,hardy1970phonon}
\begin{equation}
\begin{split}
    &\frac{1}{V} \sum_{\statep} \tilde{\Omega}_{\state \statep}\tilde{n}^T_{\state}(\bm{r},t)=\\
    &  \frac{1}{V} \sum_{\statep} \tilde{\Omega}_{\state \statep}  \frac{1}{\sqrt{\bar{N}_{\state}(\bar{N}_{\state}+1)}}  \frac{\partial \bar{N}_{\state}}{\partial T} (T(\bm{r},t)-\bar T)=\\
   & \frac{1}{V} \sum_{\statep} \tilde{\Omega}_{\state \statep} {\sqrt{\bar{N}_{\statep}(\bar{N}_{\statep}+1)}}  \frac{\hbar\omega_\statep}{k_B\bar{T}^2} (T(\bm{r},t)-\bar T)
  = 0 \;.
  \label{eq:demo_Bose_eigen}
\end{split}
\raisetag{18mm}
\end{equation}
As a result, we identify $\frac{\partial \bar{N}_{\state}}{\partial T}$ as an eigenvector with zero eigenvalue $\theta_{\state}^0$, that, after normalization, is
\begin{equation}
\label{eq:energy_eigenvector}
    \theta_{\state}^0 = \sqrt{\frac{\bar{N}_{\state}(\bar{N}_{\state}+1)}{k_BT^2C}} \hbar \omega_{\state} 
    = \sqrt{\frac{k_BT^2}{C \bar{N}_{\state}(\bar{N}_{\state}+1)}} \frac{\partial \bar{N}_{\state}}{\partial T}  \;,
\end{equation}
where the specific heat $C$ is
\begin{equation}
    C = \frac{\partial E}{\partial T}\bigg|_{NV} 
    = \frac{1}{k_B \bar{T}^2 V}\sum_{\state} \bar{N}_{\state}\big(\bar{N}_{\state}+1\big)(\hbar \omega_{\state})^2 \;.
    \label{eq:specific_heat}
\end{equation}
From equation~(\ref{eq:demo_Bose_eigen}), it follows that $\tilde{n}^T_{\state}(\bm{r},t)$ disappears from Eq.~(\ref{eq:LBTE_viscosity_symm}). Therefore, by removing the symmetrization~(\ref{eq:symmetrization_LBTE}), Eq.~(\ref{eq:LBTE_viscosity_symm}) gives Eq.~(\ref{eq:LBTE_linear_nabla_u}) in the main text.
In the context of the decomposition~(\ref{eq:separation_N_U}), the Bose-Einstein eigenvector~(\ref{eq:energy_eigenvector}) is an eigenvector to both the normal and Umklapp scattering operator, and will be denoted as $\phi_{\state}^0$ when we will later consider the basis of eigenvectors of the normal scattering operator in Appendix~\ref{sec:App5_mesoscopic_equations}.

\subsection{The drift eigenvectors: local drift velocity}
\label{sub:the_momentum_eigenvector}
Starting from Eq.~(\ref{eq:LBTE_expanded_gradiend}) at the steady-state, one obtains the following equation for the odd part:
\begin{equation}
\begin{split}
\label{eq:LBTE_LBTE_odd_nablaT_symmetrized}
  \frac{\bm{v}_{\state} }{\sqrt{\bar{N}_{\state}(\bar{N}_{\state}+1)}}  
&\cdot \bigg(\frac{\partial \bar{N}_{\state}}{\partial T} \cdot \nabla T\bigg)\\
&= -\frac{1}{V} \sum_{\statep} \tilde{\Omega}_{\state,\statep}\big(\tilde{n}^{D }_{\statep}(\bm{r},t)+\tilde{n}^{\delta{\rm O} }_{\statep}\big) \;.
\raisetag{18mm}
\end{split}
\end{equation}
The distribution $\tilde{n}^D_{\state}(\bm{r},t)$ is obtained applying the symmetrization~(\ref{eq:symmetrization_LBTE}) to the distribution ${n}^D_{\state}(\bm{r},t)$ appearing in equation~(\ref{eq:hardy_decomposition}).
We note in particular that $\tilde{n}^D_{\state}(\bm{r},t)\propto \sum_{i=1}^3 q^i \cdot u^i(\bm{r},t)$.

The drifting distribution $\tilde{n}^D_{\state}(\bm{r},t)$ is the stationary distribution for a system conserving crystal momentum. 
Therefore, recalling the decomposition~(\ref{eq:separation_N_U}), we have that 
\begin{align}
   &\frac{1}{V} \sum_{\statep} \tilde{\Omega}^N_{\state,\statep} \tilde{n}^{D }_{\statep}(\bm{r},t)  \\
   & = \frac{1}{V} \sum_{\statep}  \tilde{\Omega}^N_{\state,\statep}  {\sqrt{\bar{N}_{\statep}(\bar{N}_{\statep}+1)}}  \frac{\hbar}{k_B\bar{T}}\bm{q'}\cdot \bm{u}(\bm{r},t) = 0 \;, \nonumber
   \label{eq:n_D_is_eigenvector}
\end{align}
since $\tilde{\Omega}^N_{\state,\statep}$ accounts only for Normal scattering events that conserve crystal momentum~\cite{Sussmann_Thellung_1963,PhysRev.148.766, hardy1970phonon}.
From Eq.~(\ref{eq:n_D_is_eigenvector}) it is possible to identify three eigenvectors of $\tilde{\Omega}^N_{\state,\statep}$ with zero eigenvalues~\cite{hardy1970phonon}:
\begin{equation}
  \label{eq:momentum_eigen}
    \phi_{\state}^i 
  = \frac{1}{\sqrt{\bar{N}_{\state}(\bar{N}_{\state}+1)}} \sqrt{\frac{k_BT}{A^i}} \frac{\partial N^D_{\state}}{\partial u^i} = \sqrt{\frac{\bar{N}_{\state}(\bar{N}_{\state}+1)}{k_B T A^i}} \hbar q^i \;,
\end{equation}
where $i=1,2,3$ and $A^i$ is a normalization constant. 
The drift eigenvectors~(\ref{eq:momentum_eigen}) are, in general, not orthogonal~\cite{PhysRevB.10.3546}. 
Nevertheless, 
we work in a Cartesian coordinate system 
for $\bm{q'}$ and $\bm{u}(\bm{r},t)$, so that these 3 eigenvectors are orthogonal and can also be normalized, choosing $A^i$ from the condition $\left<\phi^i|\phi^i\right>=1$.
In computing the normalization constants $A^i$, we note that they can be expressed in terms of physically meaningful quantities. 
In particular, we note that the crystal momentum density associated to the drifting distribution is:
\begin{equation}
    \bm{P} = \frac{1}{V} \sum_{\state} N^D_{\state} \hbar \bm{q} ,
\end{equation}
and its derivative with respect to the drift velocity is:
\begin{align}
    \frac{\partial P^i}{\partial u^j} \bigg|_{\text{eq}} 
    &= \frac{1}{V} \sum_{\state} \frac{\partial N^D_{\state}}{\partial u^j} \bigg|_{\text{eq}} \hbar q^i \nonumber \\
    &= \frac{1}{k_B T V} \sum_{\state} \bar{N}_{\state}(\bar{N}_{\state}+1) \hbar q^i \hbar q^j \;.
    \label{eq:specific_momentum}
\end{align}
Comparing Eq.~(\ref{eq:momentum_eigen}) with Eq.~(\ref{eq:specific_momentum}) we note that 
$A^i = \frac{\partial P^i}{\partial u^i} \big|_{\text{eq}}$ and $\frac{\partial P^i}{\partial u^j} \big|_{\text{eq}}{=}0\iff i\neq j$.
Therefore, we will refer to $A^i$ as the specific momentum, due to its formal similarity with specific heat.
It can be shown that in the high temperature limit, $A^i(T\to\infty) \propto T$ (see also Appendix~\ref{sec:App6_parameter_sAppendix}).

Finally, it is worth mentioning that in going from equation~(\ref{eq:LBTE_expanded_gradiend}) to equation~(\ref{eq:LBTE_odd_nablaT}), the term $n^D_{\state}(\bm{r},t)$ has disappeared due to the following approximation:
\begin{equation}
\begin{split}
   &\frac{1}{V} \sum_{\statep} \tilde{\Omega}_{\state,\statep} (\tilde{n}^{D }_{\statep}(\bm{r},t)+\tilde{n}^{\delta{\rm O}}_{\statep})=\\
   &\frac{1}{V} \sum_{\statep} (\tilde{\Omega}^N_{\state,\statep}+\tilde{\Omega}^U_{\state,\statep}) (\tilde{n}^{D }_{\statep}(\bm{r},t)+\tilde{n}^{\delta{\rm O}}_{\statep})=\\
    &\frac{1}{V} \sum_{\statep} \tilde{\Omega}^U_{\state,\statep}\tilde{n}^{D }_{\statep}(\bm{r},t)+ 
    \frac{1}{V} \sum_{\statep} \tilde{\Omega}_{\state,\statep}
    \tilde{n}^{\delta{\rm O}}_{\statep}\simeq\\
   &\frac{1}{V} \sum_{\statep} \tilde{\Omega}_{\state,\statep} \tilde{n}^{\delta{\rm O}}_{\statep} \;.
   \raisetag{7mm}
   \label{eq:n_D_is_eigenvector}
\end{split}
\end{equation}
This is correct both \added{at high and low} temperatures, because at high temperatures the strong crystal momentum dissipation ensures $\tilde{n}^D_{\statep}(\bm{r},t)\propto \bm{u}(\bm{r},t) \approx \bm{0}$ \added{(see also Fig.~\ref{fig:estimate_u})}, and at low temperatures Umklapp processes are frozen ($\tilde{\Omega}^U_{\state \statep}\approx 0$).

\subsection{Local equilibrium} 
\label{sub:local_equilibrium}
From Eq.~(\ref{eq:demo_Bose_eigen}) and Eq.~(\ref{eq:n_D_is_eigenvector}) it follows that, in the hydrodynamic regime, the distributions $n^T_{\state}(\bm{r},t)$ and $n^D_{\state}(\bm{r},t)$ are left unchanged by the dynamics described by the LBTE; therefore, these are local equilibrium distributions. 
It follows that $n^T_{\state}(\bm{r},t)$ and $n^D_{\state}(\bm{r},t)$ do not appear in Eq.~(\ref{eq:LBTE_odd_nablaT}) and Eq.~(\ref{eq:LBTE_linear_nabla_u}) and thus do not contribute to the thermal conductivity and viscosity, which respectively describe the response to a perturbation of the local temperature and drift velocity. 
It is worth mentioning that, in the kinetic regime, $n^D_{\state}(\bm{r},t)$ vanishes and $n^T_{\state}(\bm{r},t)$ is still a stationary distribution for the LBTE.

\section{Thermal viscosity}
\label{sec:App_3_viscosity}
The total crystal momentum flux tensor $\Pi^{ij}_{\rm tot}$\cite{hardy1970phonon} is defined as
\begin{equation}
  \Pi^{ij}_{\rm tot}(\bm{r},t) 
  = \frac{1}{V} \sum_{\state} \hbar q^i v^j_{\state} N_{\state} (\bm{r},t) \;.
\end{equation}
Due to the odd parity of $q^i$ and $v^j_{\state}$, only the even part of the phonon distribution contributes to the crystal momentum flux. 
Using the decomposition~(\ref{eq:hardy_decomposition}) introduced in the main text, we identify three contributions to the crystal momentum flux:
\begin{align}
  \Pi^{ij}_{\rm tot}(\bm{r},t) &=\frac{1}{(2\pi)^3}\hspace{-1mm}\sum_s\hspace{-1.5mm}\int\limits_{BZ}\hspace{-1.5mm}\hbar q^i{v}_{\state}^j\hspace{-.4mm} \Big(\hspace{-.7mm}\bar{N}_{\state} {+} {n}^T_{\state}(\bm{r},t){+}{n}^{\delta{\rm E} }_{\state}(\bm{r},t)
\hspace{-.7mm}\Big)\hspace{-.4mm}d^3\hspace{-.5mm}{q} \nonumber \\
  &= \bar{\Pi}^{ij} + \Pi_{T}^{ij}(\bm{r},t) + \Pi^{ij}_{\delta {\rm E}}(\bm{r},t) \;,
\label{eq:def_momentum_flux_tensor}
\end{align}
where $\bar{\Pi}^{ij}$ is the equilibrium (constant) crystal momentum flux, which is not affected by the LBTE's dynamics; ${\Pi}^{ij}_T(\bm{r},t)$ is the momentum flux related to the local equilibrium temperature and ${\Pi}^{ij}_{\delta {\rm E}}(\bm{r},t)$ is the out-of-equilibrium momentum flux generated in response to deviations from local equilibrium conditions and is further discussed below.

\added{As stated by Eq.~(\ref{eq:definition_viscosity}), the thermal viscosity tensor $\eta^{ijkl}$ relates a drift velocity perturbation to the momentum flux generated as a response to that perturbation.}
\added{Eq.~(\ref{eq:definition_viscosity}) is valid in the mesoscopic regime mentioned in Sec.~\ref{sec:thermal_viscosity} and \ref{sec:case_study}, 
where the thermal viscosity is space-independent and non-homogeneities arise mainly from variations of the local temperature and local drift velocity.}
In particular, $\eta^{ijkl}$ is determined by deviations from local equilibrium and thus depends only on $n^{\delta{\rm E} }$.
To determine the distribution $n^{\delta{\rm E} }$, we must solve the LBTE linearized in the drift velocity gradient Eq.~(\ref{eq:LBTE_linear_nabla_u}).
To this aim, we first symmetrize the LBTE using the transformations~(\ref{eq:symmetrization_LBTE}), finding
\begin{equation}
\label{eq:LBTE_viscosity}
  \frac{\bm{v}_{\state} }{\sqrt{\bar{N}_{\state}(\bar{N}_{\state}+1)}}  
\cdot \bigg(\frac{\partial N^D_{\state}}{\partial \bm{u}} \cdot \nabla \bm{u}\bigg)
= -\frac{1}{V} \sum_{\statep} \tilde{\Omega}_{\state,\statep} \tilde{n}^{\delta{\rm E} }_{\statep} \;.
\end{equation}
Next, using the relaxon approach discussed in Ref.~\cite{cepellotti2016thermal} for thermal conductivity, we write the response to the perturbation $\nabla \bm{u}$ as a linear combination of even eigenvectors:
\begin{equation}
  \tilde{n}_{\state}^{\delta{\rm E} }=\sum_{\beta i j} f^{ij}_\beta \theta_{\state}^{\beta} \frac{\partial u^i}{\partial r^j} 
\;.
  \label{eq:response_relaxons}
\end{equation}
Substituting this relation in the LBTE, and noting that the left term is related to the eigenvector $\phi^i$ of the normal scattering matrix (Eq.~(\ref{eq:momentum_eigen})), we obtain
\begin{equation}
v^j_{\state} \sqrt{\frac{A^i}{k_BT}} \phi^i_{\state}
= - \sum_{\alpha} \frac{1}{\tau_\alpha} f_\alpha^{ij}  \theta^\alpha_{\state} \;.
\label{eq:LBTE_in_terms_delta_5}
\end{equation}
Taking the scalar product with a generic eigenvector $\theta_{\state}^\alpha$, we find
\begin{equation}
 \sqrt{\frac{A^i}{k_B T }} w^{j}_{i\alpha} = - \frac{f_\alpha^{ij}}{\tau_\alpha}  \;,
\label{eq:LBTE_in_terms_delta_6}
\end{equation}
where $w^{j}_{i\alpha}$ is a velocity tensor given by
\begin{equation}
  w^{j}_{i\alpha} = 
  \frac{1}{V} \sum_{\state} \phi^i_{\state} v^j_{\state} \theta^{\alpha}_{\state} \;.
\end{equation}
Thanks to the odd parity of $\phi^i_{\state}$ and $v^j_{\state}$, the velocity $w_{i\alpha}^j$ is different from zero only for even eigenvectors $\alpha$; Eq.~(\ref{eq:LBTE_in_terms_delta_6}) can thus be trivially solved  for $f^{ij}_{\alpha}$.

With the knowledge of the LBTE solution $f^{ij}_{\alpha}$ at hand, the crystal momentum flux tensor is readily computed.
We thus express $\Pi^{ij}_{\delta {\rm E}}$ in the relaxon basis, finding
\begin{align}
\label{eq:momentum_flux_sol}
\Pi^{ij}_{\delta {\rm E}}
&=\frac{1}{V} \sum_{\state} {n}^{\delta{\rm E} }_{\state} \hbar q^i {v}^j_{\state}  \\
  &= \frac{1}{V} \sum_{\state} \tilde{n}^{\delta{\rm E} }_{\state}\sqrt{\bar{N}_{\state}(\bar{N}_{\state}+1)} \hbar q^i v^j_{\state}  \nonumber \\
  &= \frac{1}{V} \sum_{\state\alpha k l} f_\alpha^{kl} \theta_{\state}^\alpha \frac{\partial u^k}{\partial r^l}  \sqrt{\bar{N}_{\state}(\bar{N}_{\state}+1)} \hbar q^i \; v^j_{\state} \nonumber \\
  &= \sqrt{k_BT A^i} \sum_{\alpha k l} f_{\alpha}^{kl} w_{i\alpha}^j \frac{\partial u^k}{\partial r^l} \nonumber\;. 
\end{align}
Substituting Eq.~(\ref{eq:LBTE_in_terms_delta_6}) in Eq.~(\ref{eq:momentum_flux_sol}), we obtain the expression for the \added{asymmetric} thermal viscosity tensor:
\begin{equation}
\eta^{ijkl} 
= \sqrt{A^i A^k} \sum_\alpha w^{j}_{i\alpha} w^{l}_{k\alpha} \tau_{\alpha} .
  \label{eq:viscosity}
\end{equation}
\added{The tensor $\eta^{ijkl}$ is, in general, not symmetric under exchange of indexes $j\leftrightarrow l$. We will show later that, in a way analogous to fluids, only the sum over the spatial derivatives of the momentum flux tensor is relevant for the mesoscopic description of hydrodynamic thermal transport:
\begin{equation}
\begin{split}
    \sum_j \frac{\partial \Pi^{ij}_{\delta {\rm E}}(\bm{r},t)}{\partial r^j}&=\sum_{jkl} \eta^{ijkl}  \frac{\partial^2 u^k(\bm{r},t)}{\partial r^j\partial r^l} \\
   &= \sum_{jkl} \mu^{ijkl}  \frac{\partial^2 u^k(\bm{r},t)}{\partial r^j\partial r^l} \;,
\end{split}
  \label{eq:divergence_of_pi}
\end{equation}
where in the last term of Eq.~(\ref{eq:divergence_of_pi}) we have rewritten the divergence of the momentum flux tensor using the symmetrized viscosity tensor $\mu^{ijkl}$:
\begin{equation}
  \mu^{ijkl}=\frac{\eta^{ijkl}+\eta^{ilkj}}{2}\;.
\end{equation}
}

It is worth drawing a parallel between thermal viscosity and conductivity, where the latter can be written as \cite{cepellotti2016thermal}
\begin{equation}
    \kappa^{ij} = C \sum_{\alpha} w^i_{0\alpha} w^j_{0\alpha} \tau_\alpha \;.
    \label{eq:thermal_conductivity}
\end{equation}
Notably, $w_{0\alpha}^i$ is different from zero only for odd eigenvectors.
As a result, thermal conductivity and viscosity are two quantities describing the transport due to the odd and even part of the spectrum respectively, \textit{i.e.} energy and crystal momentum.


\subsection{Single-mode relaxation-time approximation} 
\label{sub:single_mode_relaxation_time_approximation}
\begin{figure}[t]
  \centering
  \includegraphics[width=\columnwidth]{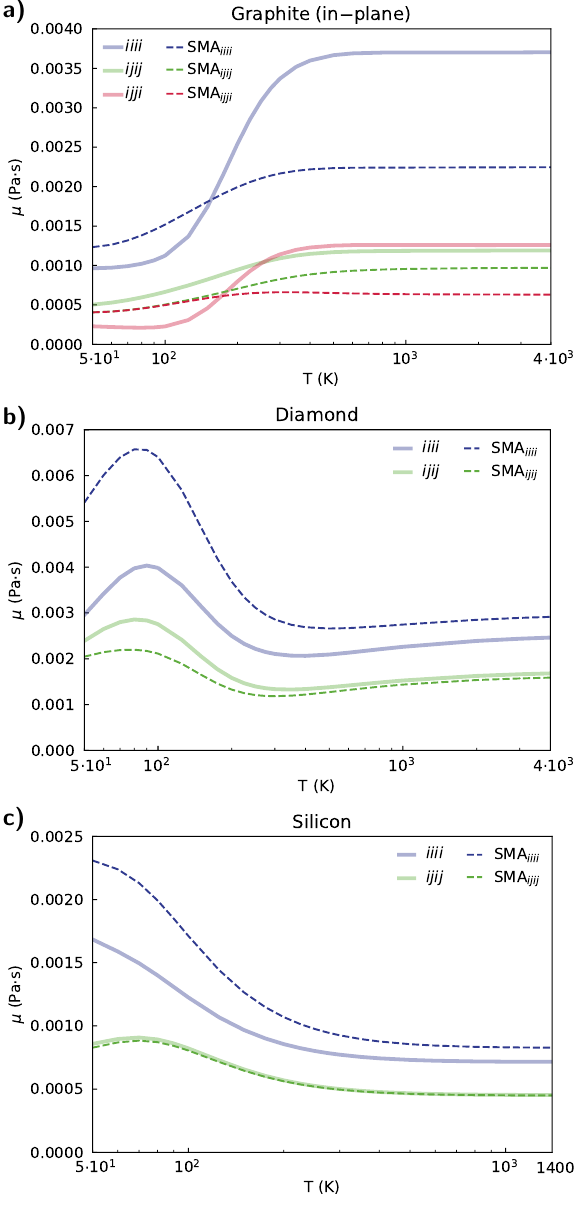}
  \caption{Comparison between the exact bulk thermal viscosity (solid lines) and the SMA bulk thermal viscosity (dashed line) for \added{in-plane graphite (a),} diamond (b) and silicon (c).}
  \label{fig:compare_SMA_vs_exact_viscosity}
\end{figure}
In this section we derive the expression for thermal viscosity within the single-mode relaxation-time approximation (SMA).
Using the SMA, the LBTE at Eq.~(\ref{eq:LBTE_viscosity}) becomes
\begin{equation}
\label{eq:LBTE_viscosity_SMA}
  \frac{\bm{v}_{\state} }{\sqrt{\bar{N}_{\state}(\bar{N}_{\state}+1)}}  
\cdot \bigg(\frac{\partial N^D_{\state}}{\partial \bm{u}} \cdot \nabla \bm{u}\bigg)
= -\frac{ \tilde{n}^{\delta{\rm E} }_{\state}}{\tau^{\rm ^{SMA}}_{\state}} \;.
\end{equation}
The deviation from equilibrium $\tilde{n}^{\delta{\rm E} }_{\state}$ is readily found as
\begin{equation}
 \tilde{n}^{\delta{\rm E} }_{\state}=-\frac{\bm{v}_{\state} }{\sqrt{\bar{N}_{\state}(\bar{N}_{\state}+1)}}  
\cdot \bigg(\frac{\partial N^D_{\state}}{\partial \bm{u}} \cdot \nabla \bm{u}\bigg)\tau^{\rm ^{SMA}}_{\state} \;.
\end{equation}
We insert this result in the definition of momentum flux, obtaining
\begin{align}
\label{eq:momentum_flux_sol_SMA}
\Pi^{ij}_{\delta \rm E,  SMA}
&=\frac{1}{V} \sum_{\state} {n}^{\delta{\rm E} }_{\state} \hbar q^i {v}^j_{\state}  \\
  &=- \frac{1}{V} \sum_{\state} \hbar^2  q^i v^j_{\state} {q}^k  {v}^l_{\state}   
\frac{\bar{N}_{\state}\big(\bar{N}_{\state}+1\big) }{k_B \bar{T}} \tau^{\rm ^{SMA}}_{\state} \frac{\partial u^k}{\partial r^l}\nonumber\;.
\end{align}
From the definition of the thermal viscosity~(\ref{eq:definition_viscosity}), the SMA thermal viscosity is therefore
\begin{equation}
  \mu^{ijkl}_{\rm {\tiny SMA}}=\frac{\eta^{ijkl}_{\rm \tiny  SMA}{+}\eta^{ilkj}_{\rm \tiny SMA}}{2}= \frac{1}{V} \sum_{\state} \hbar^2  q^i v^j_{\state} {q}^k  {v}^l_{\state}   
\frac{\bar{N}_{\state}\big(\bar{N}_{\state}+1\big) }{k_B \bar{T}} \tau^{\rm ^{SMA}}_{\state} \;,
\label{eq:SMA_thermal_visc}
\end{equation}
\added{where we highlight the fact that in the SMA approximation $\eta^{ijkl}_{\rm SMA}=\mu^{ijkl}_{\rm SMA}$, since in the SMA  the viscosity tensor $\eta^{ijkl}_{\rm SMA}$ has the $j{\leftrightarrow} l$ symmetry.}
A comparison between the exact bulk thermal viscosity~(\ref{eq:viscosity_tensor}) and the SMA bulk thermal viscosity~(\ref{eq:SMA_thermal_visc}) is shown in Fig.~\ref{fig:compare_SMA_vs_exact_viscosity}.
We highlight how the SMA approximation --- which neglects the off-diagonal elements of the scattering operator and works well when the Umklapp processes dominate over normal processes~\cite{PhysRev.148.766,fugallo2013ab,lindsay_first_2016} --- overestimates the largest component of the thermal viscosity, especially at low temperatures. This overestimation is more pronounced in diamond compared to silicon, in agreement with results from previous works where the SMA approximation was reported to yield quite accurate results for the thermal conductivity of silicon~\cite{cepellotti2016thermal} but not for diamond~\cite{fugallo2013ab}.

\section{Size effects on the thermal conductivity and viscosity} 
\label{sec:App4_ballistic_tc_visc}

In order to obtain a simple estimate of size effects, we compute the total effective thermal conductivity and viscosity using a Matthiessen sum of the diffusive and ballistic limit:
\begin{align}
  \frac{1}{\kappa^{ij}}&=\frac{1}{\kappa^{ij}_{\rm bulk}}{+}\frac{1}{\kappa^{ij}_{\rm ballistic}} \;,\label{eq:comb_Mathiessen_1} \\
\frac{1}{\mu^{ijkl}}&=\frac{1}{\mu^{ijkl}_{\rm bulk}}{+}\frac{1}{\mu^{ijkl}_{\rm ballistic}} \label{eq:comb_Mathiessen_2}\;.
\end{align}
The ballistic conductivity and viscosity are computed for a sample of size $L_S$ 
as $\kappa^{\added{ij}}_{\rm ballistic}=K^{\added{ij}}_S\cdot L\deleted{^x}_S$ and $\mu^{ijkl}_{\rm ballistic}=M^{ijkl}\cdot L\deleted{^x}_S$, where the prefactors $K_S$ and $M^{ijkl}$ are 
\begin{align}
  K^{\added{ij}}_S &= \frac{1}{V}\sum_\state (\hbar \omega_\state)^2 \frac{\bar{N}_\state(\bar{N}_\state+1)}{k_B \bar{T}^2}v^{\replaced{i}{x}}_\state\added{v^j_\state \frac{1}{|\bm{v}_{\state}|}}  \;, \\
  M^{ijkl} &= \frac{1}{V} \sum_{\state} \hbar^2  q^i v^j_{\state} {q}^k  {v}^l_{\state}   
\frac{\bar{N}_{\state}\big(\bar{N}_{\state}+1\big) }{k_B \bar{T}} \frac{1}{\added{|\bm{v}_{\state}|}} \;.
\end{align}
These prefactors are obtained after setting $\tau_\state = \frac{L_S}{|\bm{v}_\state|}$ in the SMA expressions of $k$ and $\mu$.
\added{$L_S$ is the length that determines boundary effects, \textit{i.e.} the sample's size for a single crystal or the grain size for a polycrystalline sample.}
The numerical values of $K^{\added{ij}}_S$ and $M^{ijkl}$  can be computed from first-principles and are tabulated in Tabs.~\ref{tab:param_graphite},~\ref{tab:param_C},~\ref{tab:param_Si}.


\section{Viscous heat equations}
\label{sec:App5_mesoscopic_equations}

In this section we derive an extension to Fourier's law from the LBTE, which describes \added{mesoscopic} hydrodynamic thermal transport in terms of the temperature $T(\bm{r},t)$ and drift velocity $\bm{u}(\bm{r},t)$ fields. 
\added{We focus on the mesoscopic regime where transport coefficients are well-defined and non-homogeneities arise from variations of the temperature and drift velocity. In contrast to Sec.~\ref{sec:thermal_viscosity}, here the temperature and drift velocity gradients are space-dependent, and consequently the LBTE is not linearized in the temperature and drift velocity gradients.}
We start recalling that hydrodynamic thermal transport emerges when most collisions between phonon wavepackets conserve the crystal momentum.
This can happen, for example, when the mean free path for normal collisions $\Lambda^N$ is much smaller than the boundary scattering's length $L_S$ (for a single crystal $L_S$ is the sample's size, in the case of a polycrystalline sample $L_S$ is the size of the grains)  or the mean free path for Umklapp collisions $\Lambda^U$: $\Lambda^N \ll L_S,\Lambda^U$~\cite{gurzhi_hydrodynamic_1968, cepellotti2015phonon}.
Under these conditions, the local equilibrium is expressed in terms of the four special eigenvectors $\phi^0_{\state}$ (also denoted $\theta^0_{\state}$, since this is a common eigenvector for the full and normal scattering operator, see Appendix~\ref{sec:App1_eigv_scatt_matrix}), $\phi^i_{\state}$ ($i=1,2,3$) described in Sec.~(\ref{sub:the_bose_eigenvector},\ref{sub:the_momentum_eigenvector}), and of the local temperature $T(\bm{r},t)$ and drift velocity $\bm{u}(\bm{r},t)$ fields. 
As explained in Sec.~(\ref{sub:the_bose_eigenvector},\ref{sub:the_momentum_eigenvector}), these four special eigenvectors do not contribute to thermal conductivity and viscosity (\textit{i.e.} they do not appear in Eqs.~(\ref{eq:LBTE_odd_nablaT}) and ~(\ref{eq:LBTE_linear_nabla_u})), since these latter coefficients only describe the response to a perturbation of the local equilibrium.

In order to exploit the relationship between the drift velocity $\bm{u}(\bm{r},t)$ and the drift eigenvectors $\phi^i_\state$, we choose to work with the basis of eigenvectors of the normal scattering matrix $\tilde{\Omega}^N_{\state \statep}$. 
To this aim, we diagonalize $\tilde{\Omega}^N_{\state \statep}$ as
\begin{equation}
  \frac{1}{V} \sum_{\statep} \tilde{\Omega}^N_{\state \statep} \phi^{\beta}_{\statep} = \frac{1}{\tau^N_{\beta}}\phi^{\beta}_{\state} \;,
  \label{eq:relaxons_normal}
\end{equation}
where $\beta$ is an eigenvalue index, $\phi^{\beta}_{\state}$ an eigenvector, and $\frac{1}{\tau^N_{\beta}}$ is an eigenvalue.
Among these eigenvectors, we know the analytic expression for the 4 of them associated with energy and momentum conservation, which we label with $\beta=0$ for the energy eigenvector (in this section we will use $\phi^0_\state$ to label the Bose-Einstein eigenvector~(\ref{eq:energy_eigenvector})), and $\beta=1,\dots,3$ for the momentum eigenvectors, Eq.~(\ref{eq:momentum_eigen}).

Noting that the set of ``normal'' eigenvectors $\{\phi^{\beta}_{\state}\}$ (``normal'' in the sense that they diagonalize \added{$\tilde{\Omega}^N_{\state \statep}$, the part of the scattering matrix that accounts for normal processes only)} is a complete basis set~\cite{PhysRev.148.766}, we write the deviation from equilibrium $\tilde{n}_{\state}(\bm{r},t)$ as a linear combination of these eigenvectors:
\begin{equation}
  \tilde{n}_{\state}(\bm{r},t) = \sum_\beta z_{\beta}(\bm{r},t) \phi_{\state}^{\beta} \;.
  \label{eq:response_relaxons_normal}
\end{equation}
After inserting Eq.~(\ref{eq:response_relaxons_normal}) in (\ref{eq:LBTE}), we write the LBTE in the basis of the eigenvectors of the normal scattering operator
\begin{align}
     \sum_\beta & \frac{\partial  z_\beta(\bm{r},t) }{\partial t}\phi_{\state}^\beta +\bm{v}_{\state}\cdot \bigg(\sum_\beta \nabla z_\beta(\bm{r},t) \phi_{\state}^\beta \bigg) = \nonumber \\
     =& - \sum_{\beta>3} \frac{z_\beta(\bm{r},t)}{\tau^N_\beta} \phi_{\state}^\beta -\frac{1}{V} \sum_{\statep, \beta>0} z_\beta(\bm{r},t) \tilde{\Omega}^U_{\state \statep}\phi_{ \statep}^\beta \;.
     \label{eq:LBTE_normal_eigenvectors}
\end{align}
This equation is formally equivalent to the LBTE, but allows us to take advantage of the knowledge of the first $4$ eigenvectors to derive mesoscopic equations.

\subsection{The projection of the LBTE on the $1^{\rm st}$ (Bose-Einstein) eigenvector: energy moment}
\label{ssub:the_bose_eigenvector}
Here we show how to obtain an energy balance equation. 
First, we note from Eq.~(\ref{eq:hardy_decomposition}) that the phonon population expansion of Eq.~(\ref{eq:response_relaxons_normal}) can be recast as
\begin{align}
\label{eq:fundamental_projections_2}
  \tilde{n}_{\state}&(\bm{r},t)
  = \tilde{n}^T_{\state}(\bm{r},t) + \tilde{n}^D_{\state}(\bm{r},t) + \tilde{n}^{\delta}_{\state}(\bm{r},t) \nonumber \\
  =& \sqrt{\frac{C}{k_B \bar{T}^2}} \phi^0_{\state} (T(\bm{r},t)-\bar{T}) \nonumber
  + \sum_{i=1}^3 \sqrt{\frac{A^i}{k_B\bar{T}}} \phi^i_{\state} u^i (\bm{r},t)  \\
  &+ \sum_{\beta>3} \phi^\beta_{\state} z_\beta (\bm{r},t)  \;.
\end{align}
The coefficients $z_\beta$ in front of the $4$ special eigenvectors of $\tilde{\Omega}^N_{\state \statep}$ are associated to temperature and drift velocity, which fully determine local equilibrium; in detail
\begin{gather}
\label{eq:known_terms_normal}
    z_0 (\bm{r},t) = \sqrt{\frac{C}{k_B\bar{T}^2}} (T(\bm{r},t)-\bar{T}) \;, \\
    z_i (\bm{r},t) = \sqrt{\frac{A^i}{k_B\bar{T}}} u^i(\bm{r},t), \; i=1,2,3 \;.
    \label{eq:known_terms_normal2}
\end{gather}
We now project the LBTE~(\ref{eq:LBTE_normal_eigenvectors}) in the subspace spanned by the Bose-Einstein eigenvector $\phi^0_{\state}$, \textit{i.e.} 
we take the scalar product of Eq.~(\ref{eq:LBTE_normal_eigenvectors}) with $\phi^0_{\state}$, finding
\begin{equation}
      \frac{\partial z_0(\bm{r},t)}{\partial t} +\sum_{\beta>0}\bm{W}_{0\beta}\cdot \nabla z_\beta(\bm{r},t)= 0 \;, 
    \label{eq:LBTE_relaxon_z_zero}
\end{equation}
where we used the fact that $\phi^0_{\state}$ is an eigenvector of zero eigenvalue to $\tilde{\Omega}^U_{\state \statep}$ (see Sec.~(\ref{sub:the_bose_eigenvector})) and we defined the velocity tensor 
\begin{equation}
  W^j_{\alpha \beta} = \frac{1}{V} \sum_{\state} \phi_{\state}^\alpha v^j_{\state} \phi_{\state}^\beta \;.
  \label{eq:def_W}
\end{equation}
Note that the velocity $W^j_{\alpha \beta}$ differs from the velocity $w^j_{\alpha \beta}$ introduced in Sec.~(\ref{sec:thermal_viscosity}) for thermal viscosity; the difference arises from the use in Eq.~(\ref{eq:def_W}) of the ``normal'' eigenvectors (of $\tilde\Omega^N_{\state \statep}$) rather than the general eigenvectors of $\tilde\Omega_{\state \statep}$ ($W^j_{\alpha \beta} \neq w^j_{\alpha \beta}$ in presence of Umklapp processes).

Substituting Eqs.~(\ref{eq:known_terms_normal},~\ref{eq:known_terms_normal2}) in (\ref{eq:LBTE_relaxon_z_zero}) we obtain
\begin{align}
  \sqrt{\frac{C}{k_B\bar{T}^2}} \frac{\partial T(\bm{r},t)}{\partial t} 
  &+ 
  \sum_{i,j=1}^3 \sqrt{\frac{A^i}{k_B\bar{T}}} W_{0i}^j \frac{\partial u^i(\bm{r},t)}{\partial r^j} + \nonumber \\
  &+ 
  \sum_{\beta>3} \bm{W}_{0 \beta}\cdot \nabla z_\beta(\bm{r},t) = 0 \;.
\label{eq:zero_eigenvector_proj}
\end{align}
To elucidate the meaning of this equation, we note that the harmonic heat flux can be written as~\cite{PhysRev.132.168}:
\begin{align}
  \label{eq:formula_Q}
  \bm{Q} (\bm{r},t) &= \frac{1}{V} \sum_{\state} \bm{v}_{\state} \hbar \omega_{\state}
  N_{\state}(\bm{r},t)  \nonumber \\
  &= \frac{1}{V} \sum_{\state} \bm{v}_{\state} \hbar \omega_{\state} 
  \big(n^D_{\state}\added{(\bm{r},t)} +n^{\delta {\rm O} }_{\state}\added{(\bm{r},t)} \big) \;,
\end{align}
where we used the fact that only odd components of the phonon distribution contribute to the heat flux.
Therefore, the heat flux receives contributions from both the drift velocity~\cite{simons1983relation} and the temperature gradient~\cite{cepellotti2016thermal}.
In the basis of ``normal'' eigenvectors, the drifting contribution can be written as
\begin{align}
   Q^{D,i}(\bm{r},t)
   &= \frac{1}{V} \sum_{\state} v^i_{\state} \hbar \omega_{\state}
 n^D_{\state}\added{(\bm{r},t)} = \nonumber  \\
 &= \frac{1}{V} \sum_{\state} v^i_{\state} \hbar \omega_{\state}
 \frac{\partial N^D_{\state}}{\partial \bm{u}} \cdot \bm{u}\added{(\bm{r},t)}\nonumber \\
 &= \frac{1}{V} \sum_{\state,j} v^i_{\state} \theta^0_{\state} \sqrt{\bar{T}A^jC} \phi^j \bm{u}^j\added{(\bm{r},t)} \nonumber\\
 &= \sum_{j} \sqrt{\bar{T}A^jC} W^i_{0j} \bm{u}^j\added{(\bm{r},t)} \;,
\label{eq:drift_en_flux}
\end{align}
while for the contribution from the deviation from local equilibrium $n_{\state}^{\delta{\rm O} }(\bm{r},t)$, we find
\begin{align}
  Q^{\delta {\rm O},i}\added{(\bm{r},t)}
  &= \frac{1}{V} \sum_{\state} {v}_{\state}^i \hbar \omega_{\state} n^{\delta {\rm O} }_{\state}\added{(\bm{r},t)}\nonumber \\
  &= \frac{1}{V} \sum_{\state} {v}_{\state}^i \hbar \omega_{\state} \sqrt{\bar{N}_{\state}(\bar{N}_{\state}+1)} \tilde{n}^{\delta {\rm O} }_{\state}\added{(\bm{r},t)} \nonumber \\
  &= \frac{\sqrt{{Ck_B \bar{T}^2 }}}{V} \sum_{\beta>3} \sum_{\state} z_{\beta}(\bm{r},t)\phi^0_{\state} v^i_{\state} \phi_{\state}^{\beta} \nonumber \\
  &= \sqrt{Ck_B\bar{T}^2} \sum_{\beta>3} W^i_{0\beta} z_{\beta}(\bm{r},t) \;,
  \label{eq:Fourier_en_flux}
\end{align}
where only odd eigenvectors contribute ($W^i_{0\beta}=0$ for even $\beta$ eigenvectors).
As explained in Sec.~(\ref{sub:local_equilibrium}), the thermal conductivity is determined only from odd eigenvectors that are not related to local equilibrium (that is, all the odd eigenvectors minus the three drift eigenvectors).
\added{
At this point it is crucial to recall that the relaxons have a well defined parity (even or odd), deriving from the symmetries of the full scattering operator $\tilde{\Omega}_{\state \statep} =\tilde{\Omega}_{-\state,-\statep}$~\cite{hardy1965lowest,hardy1970phonon}. 
Because the same symmetries apply to $\tilde{\Omega}^N_{\state,\statep}$, also the eigenvectors of the normal scattering operator have a well defined parity~\cite{hardy1965lowest,hardy1970phonon,PhysRevB.10.3546}.
The odd component of the heat flux (and hence the thermal conductivity) is determined only from the odd part of the LBTE's solution $\tilde{n}^{\delta{\rm O}}_{\state}(\bm{r},t)$~\cite{hardy1965lowest,cepellotti2016thermal}, and this can be written either as a linear combination of odd relaxons (as done in Ref.~\cite{cepellotti2016thermal}) or, equivalently, as a linear combination of odd eigenvectors of the normal scattering matrix, as done here. 
}
\added{From the completeness of these basis sets it follows that the heat flux~(\ref{eq:Fourier_en_flux}) arising from the odd out-of-equilibrium phonon distribution determined from equation~(\ref{eq:LBTE_odd_nablaT}),
and described as a linear combination of odd eigenvectors of the normal scattering matrix, is equivalent to the heat flux of Ref.~\cite{cepellotti2016thermal} that is written as a linear combination of odd relaxons (eigenvectors of the full scattering matrix). In an equivalent way, the equation for the odd part of the LBTE~(\ref{eq:LBTE_odd_nablaT}) is equivalent to the equation used to determine the thermal conductivity in Ref.~\cite{cepellotti2016thermal}, the only difference being the basis chosen for the decomposition of the solution (the relaxons in Ref.~\cite{cepellotti2016thermal} and in the computation of the thermal viscosity reported in Fig.~\ref{fig:res_viscosity}, and the eigenvectors of the  normal scattering matrix here). 
The choice of the basis set does not affect the resulting local heat flux, which can therefore be} \deleted{is} related to the \added{local} temperature gradient via the thermal conductivity~\cite{cepellotti2016thermal}: 
\begin{equation}
  Q^{\delta {\rm O},i}\added{(\bm{r},t)} = - k^{ij} \nabla^j T\added{(\bm{r},t)}\;.
  \label{eq:Q_delta_origin}
\end{equation}
\added{In Eq.~(\ref{eq:Q_delta_origin}) the thermal conductivity is space-independent, \textit{i.e.} Eq.~(\ref{eq:Q_delta_origin}) is valid in the mesoscopic regime where the thermal conductivity and viscosity are space-independent
and non-homogeneities arise solely from variations of the local temperature and local drift velocity.}
Eq.~(\ref{eq:Q_delta_origin}) can be used to rewrite Eq.~(\ref{eq:zero_eigenvector_proj}) in terms of the local temperature and drift velocity fields, \added{obtaining Eq.~(\ref{eq:macro_diff_eq1}) of the main text:}
\begin{align}
C \frac{\partial T(\bm{r},t)}{\partial t} 
  &+ \sum_{i,j=1}^3 \sqrt{\bar{T}A^iC} W_{0i}^j \frac{\partial u^i(\bm{r},t)}{\partial r^j}  \nonumber \\
  & - \sum_{i,j=1}^3 k^{ij} \frac{\partial^2 T(\bm{r},t)}{\partial r^i\partial r^j} = 0 \;.
\label{eq:mesoscopic_energy}
\end{align}
\deleted{If the drift velocity is set to zero, we find the usual Fourier's equation for temperature.
However, in presence of non-zero total crystal momentum, the second term introduces a correction to Fourier's law.}
Eq.~(\ref{eq:mesoscopic_energy}) is clearly not sufficient to fully describe the hydrodynamic heat conduction problem in which both $T(\bm{r},t)$ and $\bm{u}(\bm{r},t)$ are nonzero.
In the next section we will derive a complementary set of equations that completes the formulation.
\deleted{
Before moving to the next section, we note that Eq.~(\ref{eq:mesoscopic_energy}) has a simple physical interpretation.
Using Eq.~(\ref{eq:drift_en_flux}) and Eq.~(\ref{eq:Fourier_en_flux}), Eq.~(\ref{eq:mesoscopic_energy}) can be rewritten as
\begin{equation}
  C\frac{\partial T(\bm{r},t)}{\partial t} + \nabla\cdot \Big(\bm{Q}^{\delta{\rm O}}\added{(\bm{r},t)}+\bm{Q}^{D}\added{(\bm{r},t)}\Big)=0 \;,
\end{equation}
which is the familiar energy conservation equation.}


\subsection{The projection of the LBTE on the $2^{\rm nd}/3^{\rm rd}/4^{\rm th}$ eigenvectors (the momentum eigenspace)}
In this section, we derive a set of balance equations for crystal momentum.
We start by recalling from Eq.~(\ref{eq:def_momentum_flux_tensor}) that the momentum flux receives contributions from three different terms. 
Of these three, the first term is a constant related to the equilibrium temperature, and thus is not changed by the LBTE.
Therefore, we focus only on the momentum flux related to the local equilibrium temperature ${\Pi}^{ij}_T(\bm{r},t)$  and the out-of-equilibrium momentum flux generated in response to a drift velocity gradient ${\Pi}^{ij}_{\delta {\rm E}}(\bm{r},t)$. 
Using the expression of the four special eigenvectors discussed in Sec.~(\ref{sub:the_bose_eigenvector}, \ref{sub:the_momentum_eigenvector}), we rewrite these two momentum fluxes in the basis of eigenvectors of the normal scattering matrix, finding:
\begin{align}
    \Pi_T^{ij}(\bm{r},t) 
  &= \frac{1}{V} \sum_{\state}  \hbar q^i v^j n^{T}_{\state}\added{(\bm{r},t)}   \nonumber \\
  &= \frac{1}{V} \sum_{\state} \frac{\partial \bar{N}_{\state}}{\partial T} \hbar q^i v^j (T(\mathbf{r},t)-\bar{T}) \nonumber \\
  &= \frac{1}{V} \sqrt{\frac{CA^i}{\bar{T}}} \sum_{\state} \phi_{\state}^0 v^j \phi^i_{\state}  (T(\mathbf{r},t)-\bar{T}) \nonumber \\
    &= \sqrt{\frac{C A^i}{\bar{T}}} \sum_{i=3}^3 W^j_{i0} (T(\mathbf{r},t)-\bar{T}) \;,
    \label{eq:Pi_T}
\end{align}
and
\begin{align}
    \Pi_{\delta{\rm E} }^{ij}(\bm{r},t) 
    &= \frac{1}{V} \sum_{\state}  \hbar q^i v^j n^{\delta{\rm E}}_{\state}n^{T}_{\state}\added{(\bm{r},t)}    \nonumber \\
    &= \frac{\sqrt{k_B\bar{T}A^i}}{V} \sum_{\state,\beta>3}  \phi_{\state}^{\beta} v^j_{\state} \phi^i_{\state}    z_\beta\added{(\bm{r},t)} \nonumber \\
  &= \sqrt{k_B \bar{T} A^i} \sum_{\beta>3} W^j_{i\beta} z_{\beta}(\bm{r},t) \;,
  \label{eq:Pi_deltaE}
\end{align}
where we used the velocity tensor defined in Eq.~(\ref{eq:def_W}).

Next, 
as in the previous section, we take the scalar product of Eq.~(\ref{eq:LBTE_normal_eigenvectors}) with $\phi^{i}_\state\;(i=1,2,3)$, obtaining the following three equations indexed by $i=1,2,3$
\begin{align}
     &\frac{ \partial z_i(\bm{r},t) }{\partial t} 
     + \bm{W}_{i0} \cdot \nabla z_0(\bm{r},t)
     + \sum_{\beta>3} \bm{W}_{i\beta} \cdot \nabla z_\beta(\bm{r},t) = \nonumber \\
     &= - \sum_{j=1}^3 z_j(\bm{r},t) {D}_U^{ij} 
     - \sum_{\beta>3} z_\beta(\bm{r},t) {D}_U^{i\beta} \;,
    \label{eq:LBTE_relaxon_expanded_momentum}
\end{align}
where we used the fact that $\phi^i_\state$ are  eigenvectors of $\tilde{\Omega}^N_{\state\statep}$ with zero eigenvalues and we defined
\begin{equation}
  {D}_U^{i\beta} 
  = \frac{1}{V^2} \sum_{\state,\statep} \phi_{\state}^i \tilde{\Omega}^U_{\state,\statep} \phi_{\statep}^\beta \;.
\end{equation}
From the property $\tilde{\Omega}^U_{\state,\statep} = \tilde{\Omega}^U_{-\state,-\statep}$ it can be shown that ${D}_U^{i\beta}$ vanishes when $\beta$ indexes an even eigenvector. 
Since the coefficients $z_\beta(\bm{r},t)$ for the first four  eigenvectors ($\beta=0,1,2,3$) are known, it is convenient to rewrite  Eq.~(\ref{eq:LBTE_relaxon_expanded_momentum}) as:
\begin{align}
     &\frac{ \partial z_i(\bm{r},t) }{\partial t} 
     + \bm{W}_{i0} \cdot \nabla z_0(\bm{r},t)\nonumber \\
     &+ \sum_{\beta>3} \big(\bm{W}_{i\beta} \cdot \nabla + {D}_U^{i\beta}\big) z_\beta(\bm{r},t) = 
      - \sum_{j=1}^3 z_j(\bm{r},t) {D}_U^{ij}  \;,
    \label{eq:LBTE_relaxon_expanded_momentum_2}
    \raisetag{18mm}
\end{align}
In the hydrodynamic regime, Umklapp momentum dissipation is weak and thus ${D}_U^{i\beta}\to 0$ ($\phi^i_{\state}$ is approximately an eigenvector with a vanishing eigenvalue for $\tilde{\Omega}^U_{\state\statep}$). 
Therefore, we simplify Eq.~(\ref{eq:LBTE_relaxon_expanded_momentum_2}) noting that
\begin{equation}
   \sum_{\beta>3} \bm{W}_{i\beta} \cdot \nabla z_\beta(\bm{r},t) \gg  \sum_{\beta>3}  { D}_U^{i\beta} z_\beta(\bm{r},t) \;.
   \label{eq:simplification}
\end{equation}
Then, we use the expression of the coefficients $z_\beta(\bm{r},t)$  ($\beta=0,1,2,3$), and substitute Eqs.~(\ref{eq:known_terms_normal},\ref{eq:known_terms_normal2}) in the simplified Eq.~(\ref{eq:LBTE_relaxon_expanded_momentum_2}), obtaining:
\begin{align}
    &\sqrt{\frac{A^i}{k_B\bar{T}}} \frac{\partial u^i(\bm{r},t)}{\partial t}
    + \sqrt{\frac{C}{k_B\bar{T}^2}} \sum_{j=1}^3 {W}_{i0}^j \frac{\partial T\added{(\bm{r},t)}}{\partial r^j} \nonumber \\
    &+ \frac{1}{\sqrt{k_B\bar{T}A^i}} \sum_j \frac{\partial {\Pi}^{ij}_{\delta{\rm E}}(\bm{r},t)}{\partial r^j}   
    = - \sum_{j=1}^3 \sqrt{\frac{A^j}{k_B\bar{T} }} {D}_U^{ij} u^j(\bm{r},t) \;.
    \raisetag{21mm}
    \label{eq:LBTE_momentum_tmp}
\end{align}
Next, we note that only even eigenvectors different from the Bose-Einstein eigenvector determine the even distribution $n_\state^{\delta{\rm E}}(\bm{r},t)$ appearing in the expression for the out-of-equilibrium momentum flux tensor~(\ref{eq:Pi_deltaE}).
As shown in \replaced{Eq.~(\ref{eq:divergence_of_pi}), in the mesoscopic regime}{the main text (Eq.~(\ref{eq:definition_viscosity})) and in  Appendix~\ref{sec:App_3_viscosity}} the sum over the spatial derivatives of $\Pi_{\delta\rm{E}}^{ij}(\bm{r},t)$ can be expressed in terms of the viscosity and second derivative of the drift velocity.
We thus find
\begin{align}
   &A^i  \frac{\partial u^i\added{(\bm{r},t)}}{\partial t}
    +\sqrt{\frac{C A^i}{\bar{T}}} \sum_{j=1}^3 {W}_{i0}^j \frac{\partial T\added{(\bm{r},t)}}{\partial r^j} \nonumber \\
    &  - \sum_{j,k,l=1}^3 \mu^{ijkl} \frac{\partial^2 u^k\added{(\bm{r},t)}}{\partial r^j \partial r^l} 
    =   - \sum_{j=1}^3 \sqrt{A^i A^j} {D}_U^{ij} u^j\added{(\bm{r},t)} \;.
    \raisetag{18mm}
    \label{eq:LBTE_relaxon_expanded_momentum_macro}
\end{align}
Combining this with Eq.~(\ref{eq:mesoscopic_energy}) we obtain $4$ equations to be solved in terms of temperature $T(\bm{r},t)$ and drift velocity $\bm{u}(\bm{r},t)$; \added{these are the viscous heat equations at the core of this work and that} are further discussed in the main text.

As a final remark it is worth mentioning that in the kinetic regime, characterized by strong crystal momentum dissipation,
the inequality~(\ref{eq:simplification}) may not be valid.
\added{Nevertheless, in such regime the strong crystal momentum dissipation ($\max_{ij}[ D^{ij}_U]\to \infty$) yields the following (stronger) inequality
\begin{align}
     &\frac{ \partial z_i(\bm{r},t) }{\partial t}{+} \sum_{\beta>3} \big(\bm{W}_{i\beta} {\cdot} \nabla {+} {D}_U^{i\beta}\big) z_\beta(\bm{r},t) \ll 
      - \sum_{j=1}^3 z_j(\bm{r},t) {D}_U^{ij},
    \label{eq:condition_2}
\end{align}
implying that the viscous heat equations reduce to Fourier's law, as discussed in Sec.~\ref{sec:deviations_from_fourier_s_law}.}
\deleted{
Eq.~(\ref{eq:LBTE_relaxon_expanded_momentum_macro}) can be written as a balance equation for momentum. 
Recalling that $A^i = \frac{\partial P^i}{\partial u^i}$ and using Eq.~(\ref{eq:Pi_T}) and Eq.~(\ref{eq:Pi_deltaE}), we obtain
\begin{equation}
   \frac{\partial P^i}{\partial t}
    + \sum_j \frac{\partial \Pi_T^{ij}}{\partial r^j}
    + \sum_j \frac{\partial \Pi_{\delta{\rm E} }^{ij}}{\partial r^j}
    = - \frac{\partial P}{\partial t} \bigg|_{\text{Umkl}} \;.
\end{equation}
Here, in contrast with the conservation equation for energy, we readily see that crystal momentum is dissipated by the presence of Umklapp processes.}

From a mathematical point of view, the projection of the LBTE in the Bose-Einstein subspace, performed computing the scalar product between the  LBTE in the normal eigenvectors basis~(\ref{eq:LBTE_normal_eigenvectors}) and the Bose-Einstein eigenvector~(\ref{eq:energy_eigenvector})  $\propto \omega_\nu$, is equivalent to calculating the energy moment of the LBTE. 
Analogously, the projection in the momentum subspace, performed calculating the scalar product between the LBTE~(\ref{eq:LBTE_normal_eigenvectors}) and the momentum eigenvectors~(\ref{eq:momentum_eigen}) $\propto q^i$, is equivalent to computing the momentum moment of the LBTE.

\section{Parameters entering the viscous heat equations.}
\label{sec:App6_parameter_sAppendix}
\begin{figure*}[t]
  \centering
      \includegraphics[width=\textwidth]{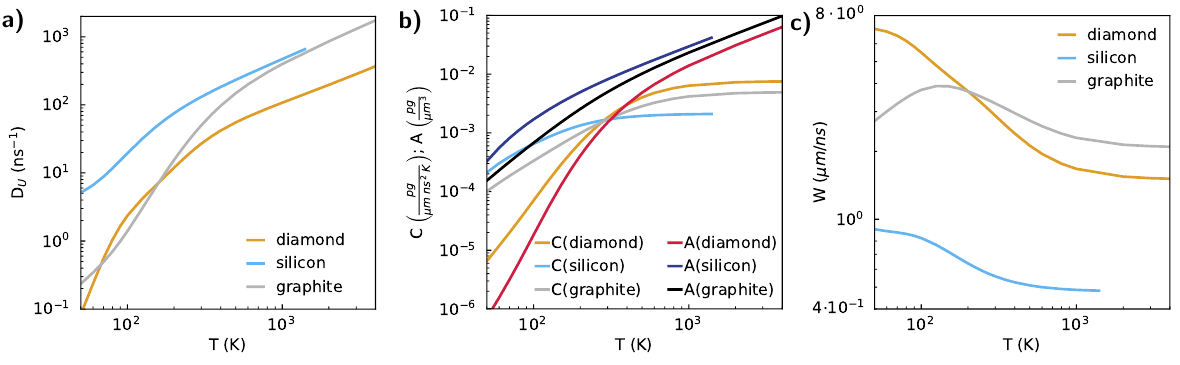}

  \caption{Trends as a function of the temperature of various quantities reported in Tabs.~\ref{tab:param_graphite}, \ref{tab:param_C}, \ref{tab:param_Si}. Panel a), values of the  momentum-dissipation inverse timescale $D_U(T)$; panel b), specific heat $C(T)$ and specific momentum $A(T)$; panel c), values of the velocity $W$. }
  \label{fig:trends}
\end{figure*}
\begin{figure}[hb!]
  \centering
  \includegraphics[width=\columnwidth]{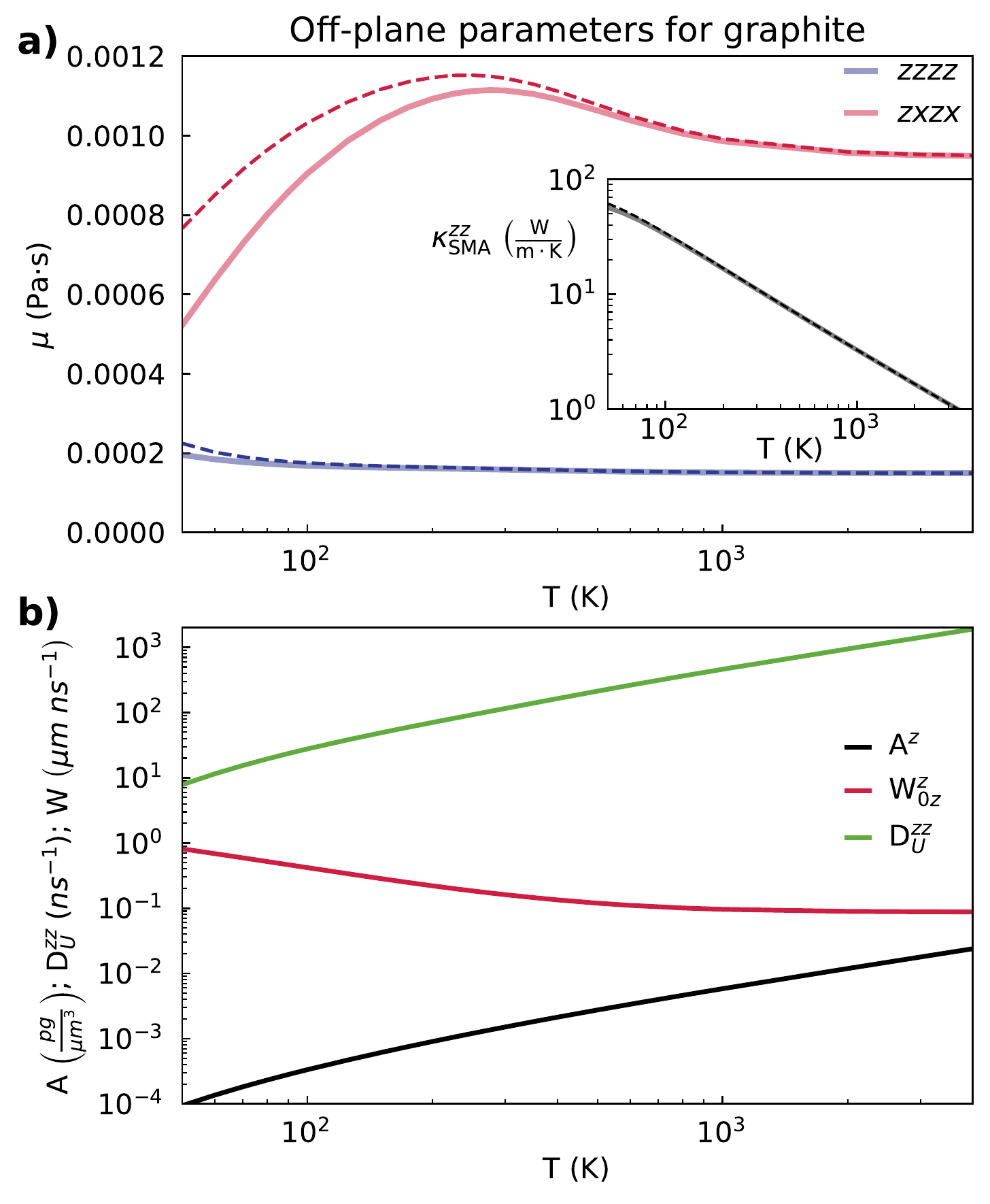}
  \caption{\added{Panel a), off-plane thermal viscosity and off-plane SMA thermal conductivity (inset) of graphite for a sample having a grain size of 10 $\mu m$ (solid line) or for the bulk case (dashed line). The other components of the viscosity tensor involving off-plane directions are negligible (at least one order of magnitude smaller than the largest component reported here).
  Panel b), specific momentum $A^z$, component $W^z_{0z}$ of the velocity tensor, and component $D_U^{zz}$ of the momentum dissipation tensor of graphite.}}
  \label{fig:graphite_off_plane}
\end{figure}
\added{
We report in Fig.~\ref{fig:trends} the trends as a function of temperature of all the parameters needed to solve the viscous heat equations on the geometry of Fig.~\ref{fig:2heat_flux}. Due to the crystal symmetries of the materials considered, several components of vectors and tensors are equivalent.
2$^{\rm nd}$-rank tensors such as thermal conductivity ($\kappa^{ij}$), momentum dissipation (${D}_U^{ij}$) and velocity ($W^j_{0i}$) are isotropic and diagonal for diamond and silicon; for graphite they are isotropic and diagonal for in-plane ($x,y$) directions (as used in the geometry of Fig.~\ref{fig:2heat_flux}).
Similarly, the specific momentum $A^i$ is isotropic for diamond and silicon, and independent from the in-plane direction in graphite ($A^x=A^y$).
The values of ${D}_U^{ij}$, $A^i$, $C$ and $W^j_{0i}$ as a function of temperature are plotted in Fig.~\ref{fig:trends} (we only report the in-plane or isotropic components, indexes are omitted). 
The numerical values of these parameters can be found in Tabs.~\ref{tab:param_graphite}, \ref{tab:param_C}, \ref{tab:param_Si};
in addition, these tables contain also the prefactors discussed in Sec.~\ref{sec:thermal_viscosity},~\ref{sec:viscous_heat_equations} and Appendix~\ref{sec:App4_ballistic_tc_visc} needed to compute the thermal conductivity in the ballistic limit $\kappa^{ij}_{\rm ballistic}$. In particular, the parameter $K^{ij}_S$ (defined for silicon, diamond and in-plane graphite), is defined as $K_S^{ij}=\kappa^{ij}_{\rm ballistic}/L_S$, where $L_S$ is the length that determines surface effects, \textit{i.e.} the sample’s
size for a single crystal or the grain size for a polycrystalline sample.
In the tables, we also report the non-zero irreducible components of the bulk heat flux viscosity tensor, limiting graphite to in-plane components. 
The parameters $M^{ijkl}$ are the prefactors (defined in Appendix~\ref{sec:App4_ballistic_tc_visc}) used to compute the ballistic thermal viscosity as $\mu^{ijkl}_{\rm ballistic}=M^{ijkl}\cdot L_S$.
Similarly, the parameter $F_U^{ij}$ are defined as $F_U^{ij}={D}_{U,{\rm boundary}}^{ij}(L_S)\cdot L_S$. 
}

\added{
Additionally, Fig.~\ref{fig:graphite_off_plane} shows the off-plane transport coefficients (panel a) and the off-plane non-negligible values of $A^i$, ${D}_U^{ij}$, and $W^j_{0i}$ (panel b) for graphite; these parameters are not relevant for the geometry studied in Sec.~\ref{sec:case_study} and are reported for completeness.}

\begin{table*}[h]
  \caption{Parameters entering the viscous heat equations for graphite. We report here only the in-plane components of the tensors needed to perform the calculation of Fig.~\ref{fig:2heat_flux}: 
 $\kappa^{ij}_{\rm P}=\kappa_{\rm P}\delta^{ij}$, $\kappa^{ij}_{\rm C}=\kappa_{\rm C}\delta^{ij}$, $K^{ij}_{S}=K_{S}\delta^{ij}$, ${D}_U^{ij}={D}_U\delta^{ij}$, ${F}_U^{ij}={F}_U\delta^{ij}$, $A=A^i\;\forall \;i$, $W^j_{0i}=W^j_{i0}= W\delta^{ij}$, where the indexes $i,j$ represent the in-plane directions $x,y$ only ($i,j=1,2$).}
  \label{tab:param_graphite}
  \centering
\resizebox{\textwidth}{!}{%
  \begin{tabular}{ccccccccccccccc}
  \hline
  \hline
  {T [K]} & 
  {$\kappa_{\rm P}\;\big[\rm\frac{W}{m{\cdot}K}\big]$} & {$\kappa_{\rm C}\;\big[\rm\frac{W}{m{\cdot}K}\big]$} & ${K_S\;\big[\rm\frac{W}{m^2{\cdot}K}\big]}$ & 
  $\mu^{iiii}\;[\rm Pa{\cdot}s]$ & $\mu^{ijij}\;[\rm Pa{\cdot}s]$ & $\mu^{iijj}\;[\rm Pa{\cdot}s]$ &
    $M^{iiii}\;[\rm\frac{ Pa{\cdot}s}{m}]$ & $M^{ijij}\;[\rm\frac{ Pa{\cdot}s}{m}]$& $M^{ijji}\;[\rm\frac{ Pa{\cdot}s}{m}]$ &
    ${D}_U\;[\rm ns^{-1}]$  &  ${F}_U\;[\rm \frac{m}{s}]$ &
    A $\big[\rm\frac{ pg}{\mu m^3}\big]$ &  $C$ $\big[\rm\frac{ pg}{\mu m{\cdot}ns^2{\cdot}K}\big]$ & W $\big[\rm \frac{\mu m}{ns}\big]$\\
  \hline
50 &          4.05937e+03 & 1.68603e-04 & 2.57075e+08   &      9.64173e-04 & 5.02673e-04 & 2.28503e-04    &      4.39176e+02 & 1.46638e+02 & 1.46269e+02    &      2.35602e-01 & 4.26877e+03 &       1.53433e-04 & 1.00900e-04 & 2.72761e+00 \\
60 &          4.71660e+03 & 2.57901e-04 & 3.95197e+08   &      9.71961e-04 & 5.28685e-04 & 2.19176e-04    &      7.15578e+02 & 2.39244e+02 & 2.38166e+02    &      3.43827e-01 & 4.59801e+03 &       2.29498e-04 & 1.41101e-04 & 3.00647e+00 \\
70 &          5.30227e+03 & 3.86737e-04 & 5.59073e+08   &      9.91550e-04 & 5.60889e-04 & 2.12711e-04    &      1.07264e+03 & 3.59308e+02 & 3.56660e+02    &      4.90888e-01 & 4.88642e+03 &       3.20256e-04 & 1.84524e-04 & 3.25249e+00 \\
80 &          5.73508e+03 & 5.99494e-04 & 7.45726e+08   &      1.02183e-03 & 5.95233e-04 & 2.10580e-04    &      1.51011e+03 & 5.07190e+02 & 5.01440e+02    &      6.99024e-01 & 5.13608e+03 &       4.25102e-04 & 2.30997e-04 & 3.45924e+00 \\
85 &          5.88316e+03 & 7.64389e-04 & 8.46709e+08 &         1.04140e-03 & 6.12652e-04 & 2.11623e-04   &      1.75824e+03 & 5.91502e+02 & 5.83334e+02    &      8.33461e-01 & 5.24694e+03 &       4.82650e-04 & 2.55380e-04 & 3.54681e+00 \\
90 &          5.98509e+03 & 9.89803e-04 & 9.52360e+08   &      1.06450e-03 & 6.30069e-04 & 2.14443e-04    &      2.02532e+03 & 6.82640e+02 & 6.71286e+02    &      9.92275e-01 & 5.34894e+03 &       5.43532e-04 & 2.80534e-04 & 3.62371e+00 \\
100 &           6.05968e+03 & 1.70592e-03 & 1.17648e+09   &      1.12368e-03 & 6.64594e-04 & 2.26760e-04    &      2.61392e+03 & 8.85020e+02 & 8.64331e+02    &      1.39551e+00 & 5.52812e+03 &       6.75030e-04 & 3.33160e-04 & 3.74625e+00 \\
125 &           5.68808e+03 & 6.19470e-03 & 1.79932e+09   &      1.36901e-03 & 7.47169e-04 & 3.08284e-04    &      4.37038e+03 & 1.50275e+03 & 1.43336e+03    &      3.03975e+00 & 5.85614e+03 &       1.05685e-03 & 4.77779e-04 & 3.89427e+00 \\
150 &           4.95001e+03 & 1.68525e-02 & 2.48830e+09   &      1.74338e-03 & 8.22466e-04 & 4.58198e-04    &      6.46259e+03 & 2.26630e+03 & 2.09701e+03    &      5.82976e+00 & 6.06477e+03 &       1.50504e-03 & 6.38359e-04 & 3.88637e+00 \\
175 &           4.19048e+03 & 3.54534e-02 & 3.21691e+09   &      2.16305e-03 & 8.89217e-04 & 6.35206e-04    &      8.81058e+03 & 3.15609e+03 & 2.82509e+03    &      9.95230e+00 & 6.20314e+03 &       2.00676e-03 & 8.09822e-04 & 3.80167e+00 \\
200 &           3.53715e+03 & 6.22028e-02 & 3.96223e+09   &      2.54485e-03 & 9.46684e-04 & 7.98006e-04    &      1.13505e+04 & 4.15136e+03 & 3.59604e+03    &      1.54740e+01 & 6.30063e+03 &       2.54990e-03 & 9.86974e-04 & 3.68920e+00 \\
225 &           3.00949e+03 & 9.60903e-02 & 4.70645e+09   &      2.85177e-03 & 9.94791e-04 & 9.28042e-04    &      1.40336e+04 & 5.23216e+03 & 4.39558e+03    &      2.23622e+01 & 6.37337e+03 &       3.12432e-03 & 1.16581e-03 & 3.57259e+00 \\
250 &           2.59110e+03 & 1.35378e-01 & 5.43687e+09   &      3.08241e-03 & 1.03413e-03 & 1.02427e-03    &      1.68237e+04 & 6.38073e+03 & 5.21448e+03    &      3.05105e+01 & 6.43012e+03 &       3.72211e-03 & 1.34366e-03 & 3.46122e+00 \\
275 &           2.25923e+03 & 1.78035e-01 & 6.14489e+09   &      3.25018e-03 & 1.06572e-03 & 1.09285e-03    &      1.96938e+04 & 7.58209e+03 & 6.04680e+03    &      3.97658e+01 & 6.47577e+03 &       4.33732e-03 & 1.51881e-03 & 3.35792e+00 \\
300 &           1.99378e+03 & 2.22075e-01 & 6.82487e+09   &      3.37078e-03 & 1.09079e-03 & 1.14103e-03    &      2.26238e+04 & 8.82402e+03 & 6.88860e+03    &      4.99551e+01 & 6.51324e+03 &       4.96554e-03 & 1.69010e-03 & 3.26300e+00 \\
350 &           1.60318e+03 & 3.07855e-01 & 8.08784e+09   &      3.52011e-03 & 1.12596e-03 & 1.19869e-03    &      2.86057e+04 & 1.13924e+04 & 8.59058e+03    &      7.24549e+01 & 6.57048e+03 &       6.24865e-03 & 2.01775e-03 & 3.09612e+00 \\
400 &           1.33509e+03 & 3.83750e-01 & 9.21271e+09   &      3.59939e-03 & 1.14751e-03 & 1.22790e-03    &      3.46848e+04 & 1.40297e+04 & 1.03064e+04    &      9.68155e+01 & 6.61114e+03 &       7.55350e-03 & 2.32135e-03 & 2.95575e+00 \\
500 &           9.98702e+02 & 4.95975e-01 & 1.10594e+10   &      3.66770e-03 & 1.16905e-03 & 1.25159e-03    &      4.69490e+04 & 1.93852e+04 & 1.37502e+04    &      1.47964e+02 & 6.66215e+03 &       1.01896e-02 & 2.84581e-03 & 2.73866e+00 \\
600 &           7.99431e+02 & 5.63451e-01 & 1.24444e+10   &      3.69035e-03 & 1.17775e-03 & 1.25863e-03    &      5.92022e+04 & 2.47475e+04 & 1.71850e+04    &      1.99654e+02 & 6.69043e+03 &       1.28283e-02 & 3.26054e-03 & 2.58552e+00 \\
800 &           5.75423e+02 & 6.24987e-01 & 1.42502e+10   &      3.70056e-03 & 1.18356e-03 & 1.26080e-03    &      8.34395e+04 & 3.53375e+04 & 2.39875e+04    &      3.00624e+02 & 6.71747e+03 &       1.80581e-02 & 3.82784e-03 & 2.39757e+00 \\
1000 &          4.52004e+02 & 6.47048e-01 & 1.52910e+10   &      3.70106e-03 & 1.18535e-03 & 1.26010e-03    &      1.07299e+05 & 4.57299e+04 & 3.07007e+04    &      3.97914e+02 & 6.72894e+03 &       2.32138e-02 & 4.16766e-03 & 2.29514e+00 \\
2000 &          2.21503e+02 & 6.74660e-01 & 1.69581e+10   &      3.70084e-03 & 1.18829e-03 & 1.25838e-03    &      2.23610e+05 & 9.61271e+04 & 6.35592e+04    &      8.57330e+02 & 6.74228e+03 &       4.83715e-02 & 4.73019e-03 & 2.13973e+00 \\
3000 &          1.47397e+02 & 6.92956e-01 & 1.73097e+10   &      3.70278e-03 & 1.18973e-03 & 1.25860e-03    &      3.38143e+05 & 1.45600e+05 & 9.59938e+04    &      1.30189e+03 & 6.74440e+03 &       7.31498e-02 & 4.85155e-03 & 2.10827e+00 \\
4000 &          1.10542e+02 & 7.11733e-01 & 1.74364e+10   &      3.70441e-03 & 1.19062e-03 & 1.25896e-03    &      4.52156e+05 & 1.94803e+05 & 1.28304e+05    &      1.74234e+03 & 6.74511e+03 &       9.78154e-02 & 4.89550e-03 & 2.09705e+00 \\
 \hline
  \end{tabular}
  }
\end{table*}

\begin{table*}[b!]
  \caption{Parameters entering the viscous heat equations for diamond. As discussed above, due to the symmetries of diamond's crystal,  $\kappa^{ij}_{\rm P}=\kappa_{\rm P}\delta^{ij}$, $\kappa^{ij}_{\rm C}=\kappa_{\rm C}\delta^{ij}$, $K^{ij}_{S}=K_{S}\delta^{ij}$, ${D}_U^{ij}={D}_U\delta^{ij}$, ${F}_U^{ij}={F}_U\delta^{ij}$, $A=A^i\;\forall \;i$, $W^j_{0i}=W^j_{i0}= W\delta^{ij}$, where $i,j=1,{\cdots},3$.}
  \label{tab:param_C}
  \centering
\resizebox{\textwidth}{!}{%
  \begin{tabular}{ccccccccccccc}
  \hline
  \hline
  {T [K]} & 
  {$\kappa_{\rm P}\;\big[\rm\frac{W}{m{\cdot}K}\big]$} & {$\kappa_{\rm C}\;\big[\rm\frac{W}{m{\cdot}K}\big]$} & ${K_S\;\big[\rm\frac{W}{m^2{\cdot}K}\big]}$ & 
  $\mu^{iiii}\;[\rm Pa{\cdot}s]$ & $\mu^{ijij}\;[\rm Pa{\cdot}s]$  &
    $M^{iiii}\;[\rm\frac{ Pa{\cdot}s}{m}]$ & $M^{ijij}\;[\rm\frac{ Pa{\cdot}s}{m}]$ &
    ${D}_U\;[\rm ns^{-1}]$ &  ${F}_U\;[\rm \frac{m}{s}]$ &
    A $\big[\rm\frac{ pg}{\mu m^3}\big]$ &  $C$ $\big[\rm\frac{ pg}{\mu m{\cdot}ns^2{\cdot}K}\big]$ & W $\big[\rm \frac{\mu m}{ns}\big]$\\
  \hline
50 &          3.43111e+04 & 2.07528e-05 & 2.73909e+07   &      2.95594e-03 & 2.39102e-03  &      5.06687e+00 & 1.68663e+00 &     8.37822e-02 & 1.20887e+04 &     6.98185e-07 & 6.53393e-06 & 6.98908e+00 \\
60 &          3.28294e+04 & 4.82483e-05 & 4.87721e+07    &     3.42258e-03 & 2.65063e-03   &     1.10924e+01 & 3.61332e+00 &    2.36470e-01 & 1.17603e+04 &      1.55771e-06 & 1.18825e-05 & 6.79196e+00 \\
70 &          3.05315e+04 & 9.23559e-05 & 7.95828e+07    &     3.77282e-03 & 2.80007e-03   &     2.16728e+01 & 7.01404e+00 &    5.51176e-01 & 1.13049e+04 &      3.15801e-06 & 1.99510e-05 & 6.51375e+00 \\
80 &          2.77912e+04 & 1.57229e-04 & 1.22126e+08    &     3.97491e-03 & 2.85772e-03   &     3.90900e+01 & 1.27110e+01 &    1.04179e+00 & 1.07756e+04 &      5.98689e-06 & 3.16787e-05 & 6.17705e+00 \\
90 &          2.48308e+04 & 2.52625e-04 & 1.78772e+08    &     4.03466e-03 & 2.83659e-03   &     6.62241e+01 & 2.18227e+01 &    1.66305e+00 & 1.02476e+04 &      1.07215e-05 & 4.80997e-05 & 5.82219e+00 \\
100 &           2.18775e+04 & 3.96722e-04 & 2.51659e+08    &     3.98459e-03 & 2.75499e-03   &     1.06364e+02 & 3.57240e+01 &    2.35506e+00 & 9.77500e+03 &      1.81905e-05 & 7.02007e-05 & 5.48421e+00 \\
125 &           1.54475e+04 & 1.17600e-03 & 5.13987e+08    &     3.60462e-03 & 2.41425e-03   &     2.85397e+02 & 1.01522e+02 &    4.21125e+00 & 8.92029e+03 &      5.47561e-05 & 1.55097e-04 & 4.80253e+00 \\
150 &           1.08162e+04 & 3.12615e-03 & 8.93341e+08    &     3.14294e-03 & 2.05193e-03   &     6.09964e+02 & 2.29408e+02 &    6.31694e+00 & 8.41644e+03 &      1.26987e-04 & 2.86510e-04 & 4.32960e+00 \\
175 &           7.74443e+03 & 7.08919e-03 & 1.37511e+09    &     2.76384e-03 & 1.77050e-03   &     1.10184e+03 & 4.34626e+02 &    8.87028e+00 & 8.09518e+03 &      2.43489e-04 & 4.63817e-04 & 3.97834e+00 \\
200 &           5.78523e+03 & 1.38027e-02 & 1.93367e+09    &     2.49842e-03 & 1.58204e-03   &     1.76124e+03 & 7.22624e+02 &    1.19199e+01 & 7.86336e+03 &      4.07776e-04 & 6.82343e-04 & 3.69265e+00 \\
225 &           4.53806e+03 & 2.36643e-02 & 2.54139e+09    &     2.32532e-03 & 1.46502e-03   &     2.57437e+03 & 1.09085e+03 &    1.53786e+01 & 7.67879e+03 &      6.19378e-04 & 9.35200e-04 & 3.44745e+00 \\
250 &           3.72328e+03 & 3.66215e-02 & 3.17342e+09    &     2.21560e-03 & 1.39570e-03   &     3.52071e+03 & 1.53178e+03 &    1.91087e+01 & 7.52304e+03 &      8.75213e-04 & 1.21427e-03 & 3.23270e+00 \\
275 &           3.16907e+03 & 5.21840e-02 & 3.80954e+09    &     2.14710e-03 & 1.35666e-03   &     4.57803e+03 & 2.03552e+03 &    2.29796e+01 & 7.38778e+03 &      1.17073e-03 & 1.51104e-03 & 3.04395e+00 \\
300 &           2.77499e+03 & 6.95343e-02 & 4.43449e+09    &     2.10528e-03 & 1.33661e-03   &     5.72518e+03 & 2.59170e+03 &    2.68886e+01 & 7.26885e+03 &      1.50073e-03 & 1.81740e-03 & 2.87843e+00 \\
350 &           2.25727e+03 & 1.05819e-01 & 5.61142e+09    &     2.06829e-03 & 1.32779e-03   &     8.21663e+03 & 3.82299e+03 &    3.45670e+01 & 7.07083e+03 &      2.24339e-03 & 2.43160e-03 & 2.60738e+00 \\
400 &           1.93101e+03 & 1.39434e-01 & 6.65793e+09     &    2.06410e-03 & 1.33876e-03   &     1.08779e+04 & 5.16074e+03 &    4.18582e+01 & 6.91554e+03 &      3.06547e-03 & 3.01541e-03 & 2.40095e+00 \\
500 &           1.53218e+03 & 1.92815e-01 & 8.33667e+09     &    2.09261e-03 & 1.37770e-03   &     1.64273e+04 & 7.98902e+03 &    5.51336e+01 & 6.69637e+03 &      4.83923e-03 & 4.01807e-03 & 2.12089e+00 \\
600 &           1.28760e+03 & 2.33780e-01 & 9.54694e+09     &    2.13344e-03 & 1.41760e-03   &     2.20429e+04 & 1.08752e+04 &    6.69946e+01 & 6.55639e+03 &      6.67950e-03 & 4.78618e-03 & 1.94940e+00 \\
800 &           9.89166e+02 & 2.97535e-01 & 1.10517e+10     &    2.20690e-03 & 1.48108e-03   &     3.30993e+04 & 1.65761e+04 &    8.81536e+01 & 6.39925e+03 &      1.03530e-02 & 5.78811e-03 & 1.76384e+00 \\
1000 &          8.07834e+02 & 3.45553e-01 & 1.18801e+10    &     2.26142e-03 & 1.52541e-03   &     4.38462e+04 & 2.21156e+04 &    1.07517e+02 & 6.31958e+03 &      1.39372e-02 & 6.35996e-03 & 1.67254e+00 \\
2000 &          4.26034e+02 & 4.74596e-01 & 1.31490e+10    &     2.38976e-03 & 1.62563e-03   &     9.50713e+04 & 4.84227e+04 &    1.96731e+02 & 6.20573e+03 &      3.09257e-02 & 7.26149e-03 & 1.54525e+00 \\
3000 &          2.90112e+02 & 5.43241e-01 & 1.34078e+10    &     2.43788e-03 & 1.66217e-03   &     1.44824e+05 & 7.38984e+04 &    2.83004e+02 & 6.18362e+03 &      4.73219e-02 & 7.44892e-03 & 1.52096e+00 \\
4000 &          2.20060e+02 & 5.95281e-01 & 1.35003e+10    &     2.46306e-03 & 1.68105e-03   &     1.94151e+05 & 9.91319e+04 &    3.68665e+02 & 6.17580e+03 &      6.35407e-02 & 7.51621e-03 & 1.51241e+00 \\
 \hline
  \end{tabular}
  }
\end{table*}

\begin{table*}[htbp!]
    \caption{Parameters entering the viscous heat equations for silicon. As discussed above, due to the symmetries of silicon's crystal, $\kappa^{ij}_{\rm P}=\kappa_{\rm P}\delta^{ij}$, $\kappa^{ij}_{\rm C}=\kappa_{\rm C}\delta^{ij}$, $K^{ij}_{S}=K_{S}\delta^{ij}$, ${D}_U^{ij}={D}_U\delta^{ij}$, ${F}_U^{ij}={F}_U\delta^{ij}$, $A=A^i\;\forall \;i$, $W^j_{0i}=W^j_{i0}= W\delta^{ij}$, where $i,j=1,{\cdots},3$.}
    \label{tab:param_Si}
  \centering
\resizebox{\textwidth}{!}{%
  \begin{tabular}{ccccccccccccc}
  \hline

  \hline
   {T [K]} & 
  {$\kappa_{\rm P}\;\big[\rm\frac{W}{m{\cdot}K}\big]$} & {$\kappa_{\rm C}\;\big[\rm\frac{W}{m{\cdot}K}\big]$} & ${K_S\;\big[\rm\frac{W}{m^2{\cdot}K}\big]}$ & 
  $\mu^{iiii}\;[\rm Pa{\cdot}s]$ & $\mu^{ijij}\;[\rm Pa{\cdot}s]$ &
    $M^{iiii}\;[\rm\frac{ Pa{\cdot}s}{m}]$ & $M^{ijij}\;[\rm\frac{ Pa{\cdot}s}{m}]$& 
    ${D}_U\;[\rm ns^{-1}]$ &  ${F}_U\;[\rm \frac{m}{s}]$ &
    A $\big[\rm\frac{ pg}{\mu m^3}\big]$ &  $C$ $\big[\rm\frac{ pg}{\mu m{\cdot}ns^2{\cdot}K}\big]$ & W $\big[\rm \frac{\mu m}{ns}\big]$\\
  \hline
50 &          2.23795e+03 & 4.59329e-02 & 2.20201e+08   &      1.68379e-03 & 8.58240e-04 &       3.33315e+02 & 2.76288e+02    &      5.11570e+00 & 2.70364e+03 &      3.27665e-04 & 2.11441e-04 & 9.02735e-01 \\
60 &          1.81398e+03 & 7.09541e-02 & 3.15143e+08   &      1.58868e-03 & 8.96618e-04 &       5.52469e+02 & 4.76459e+02    &      6.62366e+00 & 2.71505e+03 &      5.54460e-04 & 3.02489e-04 & 8.85284e-01 \\
70 &          1.47266e+03 & 9.57185e-02 & 4.08143e+08   &      1.49577e-03 & 9.08048e-04 &       8.04123e+02 & 7.07599e+02    &      8.78307e+00 & 2.72606e+03 &      8.14113e-04 & 3.91395e-04 & 8.73874e-01 \\
80 &          1.19453e+03 & 1.18329e-01 & 4.97327e+08   &      1.40019e-03 & 8.93142e-04 &       1.07864e+03 & 9.57983e+02    &      1.16582e+01 & 2.73491e+03 &      1.09495e-03 & 4.77916e-04 & 8.61085e-01 \\
90 &          9.74026e+02 & 1.37274e-01 & 5.82183e+08   &      1.30833e-03 & 8.61466e-04 &       1.36970e+03 & 1.22015e+03    &      1.52308e+01 & 2.74122e+03 &      1.38989e-03 & 5.62856e-04 & 8.44738e-01 \\
100 &           8.03719e+02 & 1.51720e-01 & 6.62519e+08   &      1.22579e-03 & 8.22564e-04 &       1.67298e+03 & 1.48947e+03    &      1.94247e+01 & 2.74509e+03 &      1.69464e-03 & 6.46690e-04 & 8.25091e-01 \\
125 &           5.32796e+02 & 1.69276e-01 & 8.42648e+08   &      1.06829e-03 & 7.27720e-04 &       2.46564e+03 & 2.17760e+03    &      3.19146e+01 & 2.74649e+03 &      2.48348e-03 & 8.49829e-04 & 7.69340e-01 \\
150 &           3.87535e+02 & 1.70507e-01 & 9.92662e+08   &      9.66838e-04 & 6.55421e-04 &       3.28405e+03 & 2.87136e+03    &      4.60065e+01 & 2.74093e+03 &      3.29333e-03 & 1.03633e-03 & 7.16183e-01 \\
175 &           3.02197e+02 & 1.66898e-01 & 1.11455e+09   &      9.00619e-04 & 6.04637e-04 &       4.11081e+03 & 3.56215e+03    &      6.06265e+01 & 2.73283e+03 &      4.11117e-03 & 1.19918e-03 & 6.71955e-01 \\
200 &           2.47534e+02 & 1.63344e-01 & 1.21236e+09   &      8.55850e-04 & 5.69080e-04 &       4.93707e+03 & 4.24698e+03    &      7.52349e+01 & 2.72447e+03 &      4.92977e-03 & 1.33675e-03 & 6.36972e-01 \\
225 &           2.09966e+02 & 1.60900e-01 & 1.29062e+09   &      8.24499e-04 & 5.43728e-04 &       5.75879e+03 & 4.92514e+03    &      8.96094e+01 & 2.71681e+03 &      5.74536e-03 & 1.45094e-03 & 6.09681e-01 \\
250 &           1.82675e+02 & 1.59448e-01 & 1.35343e+09   &      8.01853e-04 & 5.25234e-04 &       6.57443e+03 & 5.59687e+03    &      1.03682e+02 & 2.71013e+03 &      6.55621e-03 & 1.54510e-03 & 5.88350e-01 \\
275 &           1.61970e+02 & 1.58683e-01 & 1.40415e+09   &      7.85055e-04 & 5.11434e-04 &       7.38370e+03 & 6.26280e+03    &      1.17454e+02 & 2.70443e+03 &      7.36173e-03 & 1.62269e-03 & 5.71533e-01 \\
300 &           1.45710e+02 & 1.58355e-01 & 1.44545e+09   &      7.72310e-04 & 5.00922e-04 &       8.18690e+03 & 6.92362e+03    &      1.30950e+02 & 2.69961e+03 &      8.16196e-03 & 1.68684e-03 & 5.58128e-01 \\
350 &           1.21755e+02 & 1.58409e-01 & 1.50744e+09   &      7.54699e-04 & 4.86338e-04 &       9.77722e+03 & 8.23261e+03    &      1.57251e+02 & 2.69211e+03 &      9.74789e-03 & 1.78475e-03 & 5.38497e-01 \\
400 &           1.04880e+02 & 1.58914e-01 & 1.55067e+09   &      7.43511e-04 & 4.77032e-04 &       1.13499e+04 & 9.52838e+03    &      1.82836e+02 & 2.68671e+03 &      1.13174e-02 & 1.85414e-03 & 5.25164e-01 \\
500 &           8.25113e+01 & 1.60439e-01 & 1.60486e+09   &      7.30895e-04 & 4.66472e-04 &       1.44577e+04 & 1.20928e+04    &      2.32571e+02 & 2.67977e+03 &      1.44204e-02 & 1.94232e-03 & 5.08869e-01 \\
600 &           6.82316e+01 & 1.62305e-01 & 1.63594e+09   &      7.24532e-04 & 4.61092e-04 &       1.75321e+04 & 1.46339e+04    &      2.81113e+02 & 2.67571e+03 &      1.74906e-02 & 1.99348e-03 & 4.99732e-01 \\
800 &           5.08841e+01 & 1.66661e-01 & 1.66806e+09   &      7.18968e-04 & 4.56298e-04 &       2.36251e+04 & 1.96779e+04    &      3.76331e+02 & 2.67145e+03 &      2.35755e-02 & 2.04682e-03 & 4.90446e-01 \\
1000 &          4.06552e+01 & 1.71502e-01 & 1.68337e+09   &      7.16939e-04 & 4.54479e-04 &       2.96784e+04 & 2.46951e+04    &      4.70285e+02 & 2.66941e+03 &      2.96202e-02 & 2.07239e-03 & 4.86079e-01 \\
1200 &          3.38805e+01 & 1.76561e-01 & 1.69180e+09   &      7.16146e-04 & 4.53716e-04 &       3.57109e+04 & 2.96982e+04    &      5.63606e+02 & 2.66828e+03 &      3.56437e-02 & 2.08651e-03 & 4.83688e-01 \\
1400 &          2.90538e+01 & 1.81710e-01 & 1.69692e+09   &      7.15858e-04 & 4.53394e-04 &       4.17311e+04 & 3.46933e+04    &      6.56567e+02 & 2.66759e+03 &      4.16548e-02 & 2.09512e-03 & 4.82240e-01 \\
  \hline
  \end{tabular}
  }
\end{table*}
%


\section{Second sound} 
\label{sec:second_sound_appendix}
In this section we show that drifting second sound~\cite{PhysRev.148.778, hardy1970phonon, cepellotti2016second_sound, Chen_science_2019}, \textit{i.e.} thermal transport in terms of a temperature damped wave, is described by the viscous heat equations.
\added{We will first show the emergence of second sound following a bottom-up approach, \textit{i.e.} analyzing the conditions under which the two viscous heat equations decouple with the first equation~(\ref{eq:macro_diff_eq1}) reducing to a damped wave equations for the temperature field.
Afterwards, we will analyze second sound within the viscous heat equations following a top-down approach, \textit{i.e.} requiring the temperature field to have the mathematical form of a damped wave and analyzing if and under which conditions this can emerge from the viscous heat equations. }

\subsection{\added{Second sound from the viscous heat equations (bottom-up approach)}} 
\label{sub:red_second_sound_from_the_viscous_heat_equations}

For simplicity, we consider a system such that the tensors $W^i_{0j}$ and $\kappa^{ij}$ appearing in the viscous heat equations~(\ref{eq:macro_diff_eq1},\ref{eq:macro_diff_eq2}) are isotropic; the generalization to an anisotropic case is analogous to what is reported here.

Without loss of generality, we consider $\hat{x}$ as the direction of second sound propagation.
For simplicity we consider an isotropic system; the general derivation can be obtained straightforwardly generalizing the procedure reported here.
In an isotropic system, the drifting heat flux $\bm{Q}^D(\bm{r},t)$ is collinear with the drift velocity and the heat flux due to local temperature changes $\bm{Q}^\delta(\bm{r},t)$ is collinear with the temperature gradient. Thus, it follows that the only nonzero component of the drift velocity must be along the second sound propagation direction $u^{x}=u$ (for simplicity we omit all tensor indexes in the rest of this section, since the only the component having all the indexes equal to $x$ is needed for this discussion).
With these conditions, the viscous heat equations~(\ref{eq:macro_diff_eq1},\ref{eq:macro_diff_eq2})  become:
\begin{equation}
  C \frac{\partial T(x,t)}{\partial t} 
  + W \sqrt{\bar{T}A C} \frac{\partial u(x,t)}{\partial x} 
  - \kappa \frac{\partial^2 T(x,t)}{\partial x^2} = 0 \;,
  \label{eq:macro_diff_ss_1} 
\end{equation}  
\begin{equation}
\begin{split}
     A  \frac{\partial u(x,t)}{\partial t}
    +\sqrt{\frac{C A}{\bar{T}}} {W} \frac{\partial T(x,t)}{\partial x}&  -  \mu \frac{\partial^2 u(x,t)}{\partial x^2}\\ 
   & =  -A \;{D}_U u(x,t)\;.
\end{split}
        \label{eq:macro_diff_ss_2}
\end{equation}
In order to observe second sound, temperature changes need to propagate following a damped wave equation.
To this aim, we require that drift velocity and temperature are related as
\begin{equation}
  W \sqrt{\bar{T}AC} \frac{\partial u(x,t)}{\partial x}=C\tau_{ss}(1-f)\frac{\partial^2 T(x,t)}{\partial t^2}-Cf\frac{\partial T(x,t)}{\partial t}\;.
  \label{eq:second_sound_condition}
\end{equation}
where $\tau_{ss}$ is the second sound relaxation time and $0{<}|f|{<}1$ is a constant, both to be determined.
To better understand this requirement, we insert Eq.~(\ref{eq:second_sound_condition}) in Eq.~(\ref{eq:macro_diff_ss_1}), finding the desired temperature damped-wave equation:
  \begin{equation}
  \frac{\partial^2 T(x,t)}{\partial t^2} + \frac{1}{\tau_{ss}}\frac{\partial T(x,t)}{\partial t} 
  - \frac{\kappa}{C\tau_{ss}(1-f)} \frac{\partial^2 T(x,t)}{\partial x^2} = 0 \;.
  \label{eq:macro_diff_ss_temp}
\end{equation}

Next, we show that condition~(\ref{eq:second_sound_condition}) implies that also the drift velocity field follows a damped-wave equation.
To this aim, we take the derivative with respect to $x$ of Eq.~(\ref{eq:second_sound_condition}),  finding
\begin{equation}
  W \sqrt{\bar{T}AC} \frac{\partial^2 u(x,t)}{\partial^2 x}=\underbrace{\left(C\tau_{ss}(1-f)\frac{\partial^2 }{\partial t^2}-Cf\frac{\partial }{\partial t}\right)}_{\hat{O}}\frac{\partial T(x,t) }{\partial x}\;,
  \label{eq:second_sound_condition_2}
\end{equation}
where $\hat{O}$ is a differential operator.
Next, by applying the operator $\hat{O}$ to both sides of equation~(\ref{eq:macro_diff_ss_2}) and using  condition~(\ref{eq:second_sound_condition_2}), we obtain
\begin{equation}
\begin{split}
     \frac{\partial^3 u(x,t)}{\partial t^3}&- \frac{f}{1-f}\frac{1}{\tau_{ss}}\frac{\partial^2 u(x,t)}{\partial t^2}
    +\frac{{W}^2}{\tau_{ss}(1-f)} \frac{\partial^2 u(x,t)}{\partial x^2} \\
    & -  \frac{\mu}{A} \frac{\partial^4 u(x,t)}{\partial t^2\partial x^2}  
    +\frac{f}{1-f}\frac{1}{\tau_{ss}}\frac{\mu}{A}\frac{\partial^3 u(x,t)}{\partial t\partial x^2}\\ 
    & =  -{D}_U \left(\frac{\partial^2 u(x,t)}{\partial t^2}-\frac{f}{1-f}\frac{1}{\tau_{ss}} \frac{\partial u(x,t)}{\partial t}\right)\;.
\end{split} 
\raisetag{16mm}
        \label{eq:macro_diff_ss_drift}
\end{equation}
If we consider only the lowest-order derivatives in Eq.~(\ref{eq:macro_diff_ss_drift}), we obtain a simplified equation that holds in the close-to-equilibrium regime where variations in space and time are small. 
In particular, neglecting higher-than-second order derivatives gives:
\begin{equation}
\begin{split}
    \frac{\partial^2 u(x,t)}{\partial t^2}&
    +\underbrace{\frac{{D}_U f}{f{-}(1{-}f)\tau_{ss}{D}_U }}_{c_1}  \frac{\partial u(x,t)}{\partial t}\\
    &-\underbrace{\frac{W^2}{f{-}(1{-}f)\tau_{ss}{D}_U}}_{c_2} \frac{\partial^2 u(x,t)}{\partial x^2}  = 0\;.
\end{split} 
        \label{eq:macro_diff_ss_drift_simplified}
\end{equation}
Therefore, if
\begin{equation}
  f>\frac{\tau_{ss}{D}_U}{1+\tau_{ss}{D}_U} \;,
  \label{eq:condition_damped_wave}
\end{equation} 
then both constants $c_1$ and $c_2$ are positive and 
the evolution of the drift velocity is that of a damped wave equation like the one for temperature.

The coefficients $\tau_{ss}$ and $f$ are determined solving Eqs.~(\ref{eq:macro_diff_ss_temp},\ref{eq:macro_diff_ss_drift_simplified}) and imposing the second sound condition~(\ref{eq:second_sound_condition}).
The solutions are of the form
\begin{equation}
\label{sec_sou_eq_1}
  T(x,t)=\frac{1}{2\pi}\int C_T(k) e^{-\frac{t}{2\tau_{ss}} }e^{i(kx-\bar{\omega}(k)t)} dk
\end{equation}
and 
\begin{equation}
\label{sec_sou_eq_2}
  u(x,t)=\frac{1}{2\pi}\int C_u(k) e^{-\frac{t}{2\tau_{ss}}}e^{i(kx-\bar{\omega}(k)t)} dk
\end{equation}
with:
\begin{align}
  &\bar{\omega}(k)=\sqrt{v_{ss}^2k^2-\frac{1}{4\tau^2_{ss}}}\; ;\\
  &v_{g}(k) = \frac{\partial \bar{\omega}(k)}{\partial k}=\frac{k v_{ss}}{\sqrt{k^2+(2\tau_{ss}v_{ss})^{-2}}}\;\label{eq:prop_vel} ;\\
  &f={D}_U\tau_{ss}\label{eq:condition_f}\; ; \\
  &\tau_{ss}=\frac{C({W})^2 }{\kappa ({D}_U)^2+ {D}_UC({W})^2}\label{eq:condition_tau}\; ;\\
  &v_{ss}=\frac{\kappa{D}_U+C{W}^2}{C{W}}\; .\label{eq:vss}
\end{align}
The condition~(\ref{eq:condition_f}) is derived from the requirement $c_1=\frac{1}{\tau_{ss}}$, and is consistent with the damped wave requirement~(\ref{eq:condition_damped_wave});  condition~(\ref{eq:condition_tau}) is derived from the requirement that $c_2=\frac{\kappa}{C\tau_{ss}(1-f)}$ and the second sound velocity~(\ref{eq:vss}) has been obtained substituting Eq.~(\ref{eq:condition_f}) and Eq.~(\ref{eq:condition_tau}) into Eq.~(\ref{eq:macro_diff_ss_temp}).
We note that for ${D}_U\to 0$ (that is, negligible crystal momentum dissipation) $v_{g}(k)\to v_{ss}\to {W}$, \textit{i.e.} the second sound propagation velocity approaches the drifting second sound velocity defined by Hardy~\cite{hardy1970phonon}) and $\tau_{ss}\to {D_U}^{-1}$.
Finally, the second sound condition~(\ref{eq:second_sound_condition}) imposes the following relation between the coefficients:
\begin{equation}
  \begin{split}
    C_u(k)&{=}\frac{-i}{k}\bigg\{\frac{\tau_{ss}}{{W}}\sqrt{\frac{C}{\bar{T}A}}\bigg[(1-\tau_{ss}{ D}_U)\left(\frac{1}{2\tau_{ss}}{+}i\bar\omega(k)\right)^2\\
    &+{D}_U\left(\frac{1}{2\tau_{ss}}{+}i\bar\omega(k)\right) \bigg]\bigg\}C_T(k)
  \end{split}
  \label{eq:relation_coefficients}
\end{equation}
$C_T(k)$ must be set according to initial conditions and the form of $C_u(k)$ follows from equation~(\ref{eq:relation_coefficients}).
We note from Eq.~(\ref{eq:relation_coefficients}) that, when second sound occur, temperature and drift velocity are both damped waves with a phase shift of $\pi/2$.


\subsection{\added{Second sound from the viscous heat equations (top-down approach)}}
In the previous section, we obtained second sound properties by finding the conditions under which the viscous equation for temperature (Eq.~(\ref{eq:macro_diff_eq1})) becomes a damped wave equation.
However, we can also obtain a second sound equation taking inspiration from the approach outlined in Ref.~\cite{cepellotti2016second_sound},
\textit{i.e.} by looking for the conditions upon which the microscopic degrees of freedom of the transport equation evolve as a damped wave.
In particular, we want to find the conditions such that the solution of Eqs.~(\ref{eq:macro_diff_eq1},\ref{eq:macro_diff_eq2}) are
\begin{gather}
\label{sec_sou_guess_1}
  T(x,t) = \bar{T} + (\delta T)  e^{i(kx-\bar{\omega}(k) t)} e^{- t / \tau_{ss} } \;, \\
\label{sec_sou_guess_2}
  u(x,t) = u_0 e^{i(kx-\bar{\omega}(k) t)}  e^{- t /\tau_{ss} }  \;,
\end{gather}
where $\delta T$ and $u_0$ are in general complex numbers to allow for a phase difference between the two waves.
We note in particular that this guess for solution requires that both temperature and drift velocity oscillate/decay at the same frequency/rate, which is consistent with the conditions~(\ref{sec_sou_eq_1},\ref{sec_sou_eq_2}) obtained in the previous section.

Using this guess for the solution, the derivation of the dispersion relation and the decay time easily follows.
To this aim, we substitute Eqs.~(\ref{sec_sou_guess_1},\ref{sec_sou_guess_2}) in the viscous heat equations~(\ref{eq:macro_diff_eq1},\ref{eq:macro_diff_eq2}) and find:
\begin{gather}
 - i C \tilde{\omega}(k) \delta T + W \sqrt{\bar{T}AC} iku_0 + \kappa k^2 \delta T = 0  \;, \label{f_18} \\
 - i A \tilde{\omega}(k) u_0 + \sqrt{\frac{CA}{\bar{T}}} W i k \delta T + \mu k^2 u_0 = - A D_U u_0 \label{f_19} \;,
\end{gather}
where we introduced a complex frequency $\tilde{\omega}(k) = \bar{\omega}(k) - \frac{i}{\tau_{ss}}$ to simplify the calculation.
The real part of this complex frequency is the oscillation frequency of second sound, whereas the imaginary part describes its decay time.
Next, we rewrite Eq.~(\ref{f_19}) as:
\begin{equation}
u_0 = - \delta T \frac{ i k\sqrt{\frac{CA}{\bar{T}}} W} { \mu k^2 + A D_U - i A \tilde{\omega}(k) } \;,
\end{equation}
and substitute this expression into Eq.~(\ref{f_18}), finding
\begin{equation}
  - i C \tilde{\omega}(k) + \frac{  CA W^2 k^2} { \mu k^2  + A D_U - i A \tilde{\omega}(k) } + \kappa k^2 = 0 \;,
\end{equation}  
that gives:
\begin{equation}
(-iC \tilde{\omega}(k) + \kappa k^2 ) ( A D_U + \mu k^2  - i A \tilde{\omega}(k) ) + CA W^2 k^2 = 0 \;.
\end{equation}
This is a quadratic equation that determines the dispersion relations for $\tilde{\omega}_k$, given by:
\begin{align}
     \tilde{\omega}^2(k) 
  &+i \tilde{\omega}(k) \bigg[\bigg(\frac{\mu}{A} + \frac{\kappa}{C}\bigg) k^2 + D_U \bigg] +
  \nonumber \\
&- \bigg(W^2  + \frac{\kappa D_U }{C} + \frac{\kappa \mu k^2}{C A} \bigg)k^2 = 0 \;.
\end{align}
This equation can be solved to obtain the complex frequency $\tilde{\omega}(k)$ and thus the oscillation frequency and decay time of second sound as a function of the wavevector $k$.
Solving for this quadratic equation, we obtain:
\begin{align}
\tilde{\omega}(k)
=&
- \frac{i}{2} \bigg(\frac{\mu}{A} k^2 + D_U + \frac{\kappa}{C} k^2 \bigg)  \nonumber \\
&\pm \frac{1}{2} \bigg[ -\bigg(\frac{\mu}{A} k^2 + D_U + \frac{\kappa}{C} k^2 \bigg)^2 \nonumber \\
&+ 4 \bigg(W^2 k^2 + \frac{D_U \kappa k^2}{C} +  \frac{\kappa \mu k^4}{A C} \bigg) \bigg]^{\frac{1}{2}}
\;.
\end{align}
In order to compare this result with the expression for second sound derived in the previous section, it is worth recalling that the semiclassical description of thermal transport used throughout this work holds for long-wavelength perturbations.
Therefore, we simplify the previous expression retaining terms to smallest order in $k$, finding:
\begin{align}
\tilde{\omega}(k) \approx & - \frac{i D_U}{2}   \\
&\pm \sqrt{ - \bigg(\frac{D_U}{2}\bigg)^2
 +  k^2 \bigg(W^2 + \frac{D_U \kappa}{2C} - \frac{\mu D_U}{2 A} \bigg) } \nonumber 
\;.
\end{align}
We can readily see that in the limit of small wavevectors ($k\to0$) the non-trivial solution is $\tilde{\omega}_k \approx - i D_U$, that is, the second sound oscillation frequency goes to zero, and has a decay time set by the crystal momentum dissipation rate: $\tau_{ss} \approx \frac{1}{D_U}$.
To estimate the behavior of the oscillation frequency, is instead convenient to recall the hypothesis of small Umklapp rates.
In fact, if we set $D_U=0$, we find $\tilde{\omega}(k) \approx \pm W k$, that is, second sound disperses linearly with $k$, and has a velocity $v_g(k) = \frac{\partial \tilde{\omega}_k}{\partial k} \approx W$.
These limiting approximations ($v_{g}\approx W$ and $\tau_{ss}\approx\frac{1}{D_U}$) are consistent with the results found in the previous section.

\section{Analytical 1D example}
\label{sec:App_Sussman_Tellung}
The viscous heat equations~(\ref{eq:macro_diff_eq1},\ref{eq:macro_diff_eq2}) can be solved analytically in a handful of toy models.
Here, for simplicity, we neglect dissipation of momentum by Umklapp processes \added{($D_U^{ij}=0$)} and consider \added{steady-state} heat diffusion along the transversal direction of a thin film, so that the problem becomes effectively a 1D problem, with $x$ labeling the orthogonal direction position.
\added{In addition, since we are considering a 1D geometry, we label $A=A^x$, $W=W_{0x}^x$, $\mu=\mu^{xxxx}$ and assume that temperature and velocity fields depend only on the position $x$.
Under these considerations the viscous heat equations~(\ref{eq:macro_diff_eq1},\ref{eq:macro_diff_eq2}) reduce to:}
\begin{align}
 & \sqrt{\bar{T} A C} W \frac{\partial u(x)}{\partial x}
  - \kappa \frac{\partial^2 T(x)}{\partial x^2} = 0 \;, \\
    & \sqrt{\frac{C A}{\bar{T}}} W  \frac{\partial T(x)}{\partial x}
    - \mu \frac{\partial^2 u(x)}{\partial x^2} = 0 \;.
\end{align}
To solve the problem, we specify the following no-slip boundary conditions on a 1D geometry having length $2l$:
\begin{equation}
u(x=\pm l) = 0 \;,
\end{equation}
and
\begin{equation}
T(x=\pm l) = \bar{T} \pm \delta T \;,
\end{equation}
that is, we assume boundaries at thermal equilibrium.

We look for solutions of the form:
\begin{equation}
    u(x) = d \cosh(bl) + a \cosh{bx} \;,
\end{equation}
\begin{equation}
    T(x) = \bar{T} + c \sinh{bx} \;.
\end{equation}
After some algebra, one finds the solution
\begin{equation}
    u(x) = \delta T \sqrt{\frac{\kappa}{\mu\bar{T}}} 
    \bigg( \frac{\cosh(bx)}{\sinh(b l)} - \coth(bl) \bigg) \;,
    \end{equation}
\begin{equation}
    T(x) = \bar{T} + \delta T \frac{\sinh(bx)}{\sinh(b l)}  \;,
      \end{equation}
\begin{equation}  
    b = \sqrt{\frac{ A C W^2}{\mu k}} \;.
\end{equation}
This analytical solution shares several qualitative similarities with the numerical example discussed in the main text, and more clearly highlights how the factor $1/b$ represents a length scale at which surface scattering affects thermal transport, which is in turn dependent on both conductivity and viscosity.
Moreover, we note that the mathematical form of the solution has the same qualitative behavior of the problem studied by Sussmann and Thellung \cite{Sussmann_Thellung_1963}, which serves as a verification of the present model.
At variance with their work however, the prefactors introduced here allow us to go beyond the Debye approximation.

\section{Estimate of the characteristic drift velocity} 
\label{sec:App8_estimation_of_the_typical_velocity}

In this section, we estimate the characteristic value of the drift velocity ($u_0$) in the high ($u_H$) and low ($u_L$) temperature regimes. 
These characteristic values are determined substituting in the viscous heat equations~(\ref{eq:macro_diff_eq1},~\ref{eq:macro_diff_eq2}) the characteristic values of the temperature (and related derivatives) and solving them approximatively for the velocity. 
With this aim, we start estimating the characteristic temperature gradient in the setup of Fig.~\ref{fig:2heat_flux} when a temperature difference $\bar{T}\pm \delta T$ is imposed on the two opposite sides (at $x=0$ and $x=15\;\mu m $).

In this setup we clearly distinguish two regions. Choosing \replaced{$l=\frac{l_{\rm TOT}}{10}$ as a length unit, where $l_{\rm TOT}$ is the total length of the sample (e.g. $l_{\rm TOT}=15\mu m$ for the sample in Fig.~\ref{fig:2heat_flux})}{$l=1\mu m$}, the left-hand-side region has length $l_L=2 \;l$ and width $w_L=2 \;l$, while the right-hand-side region has length $l_R=8 \;l$ and width $w_R=6\; l$.
Energy conservation requires that the current in the left-hand side must be equal to the right-hand side. 
Therefore, the heat flux on the left $Q_L$ must be three times the heat flux on the right side $Q_L = 3 Q_R$.
Using Fourier's law $Q=-k\nabla T$, and supposing that the thermal conductivity is constant throughout the sample, it follows that the temperature gradients in the two regions are related as $\nabla^x T_L = 3 \nabla^x T_R$.
Requiring the total temperature drop to be equal to the temperature difference imposed by the boundary conditions, we can write
\begin{equation}
  - 2 \delta T 
  = \Delta T^x_L +\Delta T^x_R 
  = l_L \nabla^x T_L + l_R \nabla^x T_R 
  = 14 l \nabla^x T_R \;.
  \label{eq:relations_t_drop}
\end{equation}
Focusing from now on on the larger region on the right, it follows that the temperature drop taking place is approximately given by $\Delta T_R = - \frac{8}{7} \delta T$.

To determine a characteristic value of drift velocity $u_0$, we substitute $\nabla T^x_R\added{\simeq\frac{\Delta T_R}{l_R}}$ in Eq.~(\ref{eq:macro_diff_eq2}) and for simplicity consider \replaced{isotropic}{crystals of cubic} symmetry\added{, valid e.g. for silicon, diamond and for graphite in the in-plane directions.}
$u_0$ can be determined focusing on the steady state limit of Eq.~(\ref{eq:macro_diff_eq2}).
We simplify its estimate considering separately the limits of low and high temperatures.

In the high temperature limit, the term related to momentum dissipation ($\propto{D}_U^{ij}$) is much larger than the viscous term ($\propto \mu^{ijkl}$) (see Fig.~\ref{fig:trends}), therefore Eq.~(\ref{eq:macro_diff_eq2}) can be approximated as:
\begin{equation}
   \sqrt{\frac{C A^x}{\bar{T}}}  W_{x0}^x \nabla^x T^x
    \simeq   -  A^x {D}_U^{xx} u^x \;.
    \label{eq:simplifyed_eq_for_v}
\end{equation}
Using the estimated temperature gradient $\added{\nabla T^x_R\simeq\frac{\Delta T_R}{l_R}}$, the high-temperature characteristic value of drift velocity $u_H$ is found to be
\begin{equation}
  u_H=\sqrt{\frac{C}{\bar{T}A^x}}\frac{{W}_{x0}^x }{{D}_U^{xx}}\frac{8}{7}\frac{\delta T}{l_R} \;.
  \label{eq:u_H}
\end{equation}

At low temperatures, viscosity dominates over the momentum dissipation term (see Fig.~\ref{fig:trends}), so that Eq.~(\ref{eq:macro_diff_eq2}) is approximated as
\begin{equation}
    \sqrt{\frac{C A}{\bar{T}}} {{W}^x_{x0} \frac{\partial T(x,y)}{\partial x} }  \simeq \mu^{xxxx} \frac{\partial^2 u^x(x,y)}{\partial x^{2} } + \mu^{xyxy} \frac{\partial^2 u^x(x,y)}{\partial y^{2} } \;,
    \label{eq:estimate_2nd_a}
 \end{equation} 
where we considered only the two largest components of the viscosity tensor.
To estimate the average value of these second derivatives, we note that, as shown in Fig.~\ref{fig:2heat_flux}, $\bm{u}(x,y)$ has a bell-like profile in the sample interior, which vanishes at the boundaries.
We thus proceed with a few assumptions that allow us to make an estimate of $u_L$.
\added{First, we suppose that the average values of the second derivatives of the drift velocity in Eq.~\ref{eq:estimate_2nd_a} are of the same order of magnitude, that is $\left< \frac{\partial^2 u^x}{\partial x^{2} }\right> \sim \left<\frac{\partial^2 u^x}{\partial y^{2} }\right> $, which can be checked numerically.}
It follows that Eq.~(\ref{eq:estimate_2nd_a}) can be simplified as:
\begin{equation}
    \left< \frac{\partial^2 u^x}{\partial y^{2} }\right>\simeq\sqrt{\frac{C A}{\bar{T}}} \frac{{W}^x_{x0} \nabla^x T_R}{\mu^{xyxy}\left(1+\frac{\mu^{xxxx}}{\mu^{xyxy}}\right)} = -a \;.
    \label{eq:estimate_2nd}
\end{equation} 
Next, we note that the variation of $u$ is stronger along the $y$ coordinate.
To mimic the Poiseuille-like shape, we assume the velocity profile to be constant along the $x$ direction, and parabolic along the $y$ direction with vanishing velocity at the boundaries  ($y = 0\;\mu m$ and $y = w_R$), so that  $\bm{u}(x,y)\simeq (- a\cdot y (y-w_R),0,0)$.
\begin{figure}[t]
  \centering
  \includegraphics[width=\columnwidth]{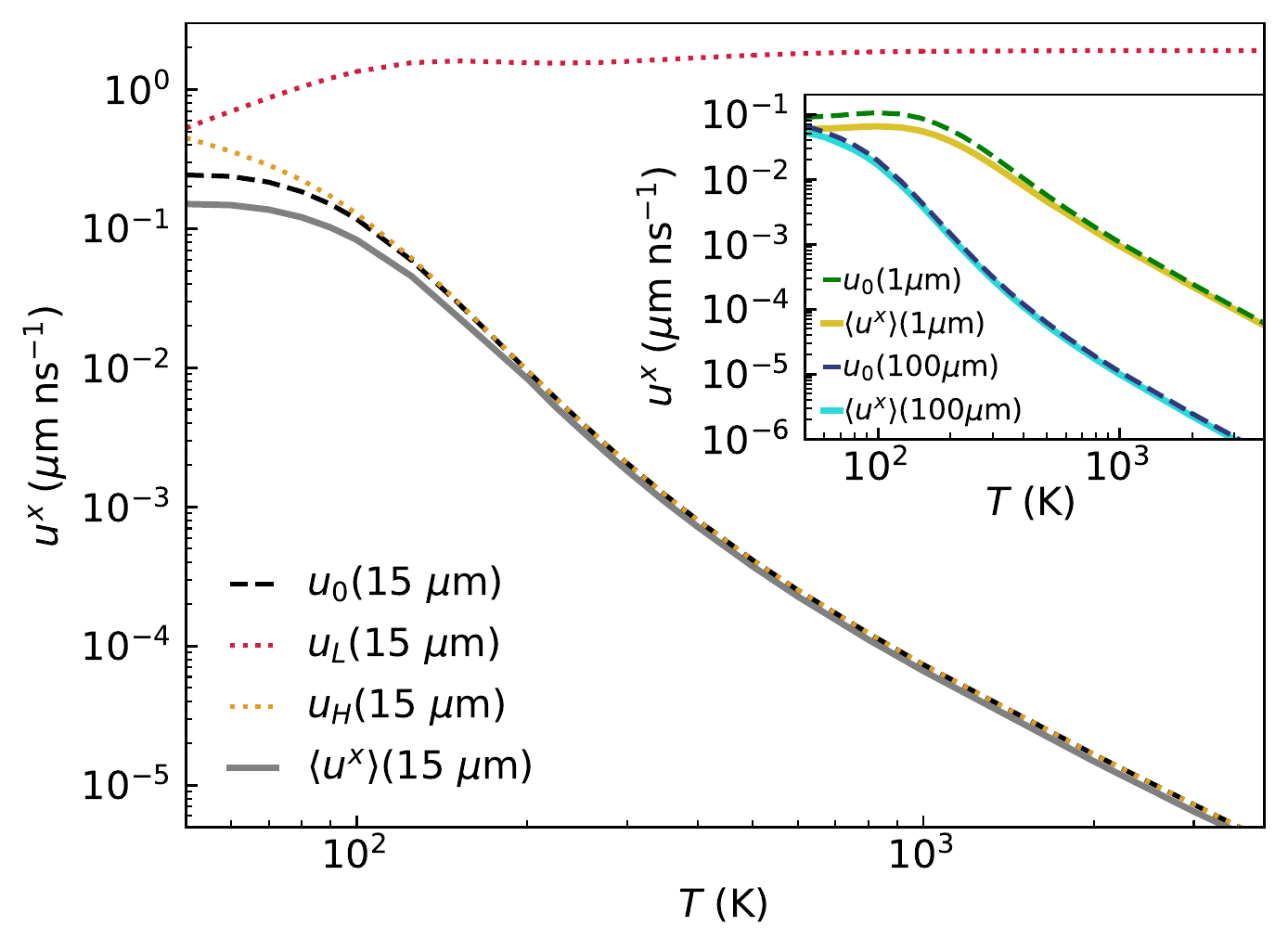}
  \caption{\added{Estimates of the characteristic value of the drift velocity $u_0$ (dashed lines) as combination of the asymptotics trends given by $u^H$ (Eq.~(\ref{eq:u_H}), dotted orange) and $u^L$ (Eq.~(\ref{eq:u_L}), dotted red) for graphite samples having total length $l_{\rm TOT}=l_L+l_R=15\;\mu$m (main plot), 1 $\mu$m and 100 $\mu$m (inset). $\left<u^x\right>$ (solid lines) is the average value of the drift velocity computed from the numerical solution of the viscous heat equations in the region $l_L<x<(l_L+l_R)$.}}
  \label{fig:estimate_u}
\end{figure}
With these approximations, we can estimate the average value of the parabolic velocity profile, \text{i.e.} the characteristic value of the drift velocity at low temperatures $u_L$, as:
\begin{align}
   u_L &= \left<u\right> = \frac{1}{w_R}\int_{0}^{w_R} u(y) dy = \frac{a\; w_R^{\replaced{2}{3}}}{6}  
   \label{eq:u_L}
 \end{align} 
 where we used $w_R=6\; l$ and $l_R=8\;l$.
We thus recover the expression for $u_L$ given in section~\ref{sec:deviations_from_fourier_s_law} in the main text.
The average value of $u$ is interpolated in between the high and low temperature limit using Matthiessen's rule, as: $u_0^{-1}=u_H^{-1}+u_L^{-1}$.
The estimates of $u_H$, $u_L$ and $u_0$ for \replaced{graphite}{diamond and silicon} are reported in Fig.~\ref{fig:estimate_u}.
We also compare this rough estimate with the average value of ${u}^x(x,y)$ computed \added{from the numerical solution of the viscous heat equations on the region $x>\frac{1}{5}l_{\rm TOT}$} of the geometry discussed in the main text \added{and denoted denoted with $\left<u^x\right>$.}
Despite the qualitative arguments used to derive $u_0$, the estimate is able to capture qualitative trends and approximately reproduce \added{average} results from the numerical solution of the viscous heat equations.

\section{Hydrodynamic behavior in macroscopic diamond samples} 
\label{sec:second_sound_in_very_large_diamond_samples}
\begin{figure}[t]
  \centering
  \includegraphics[width=\columnwidth]{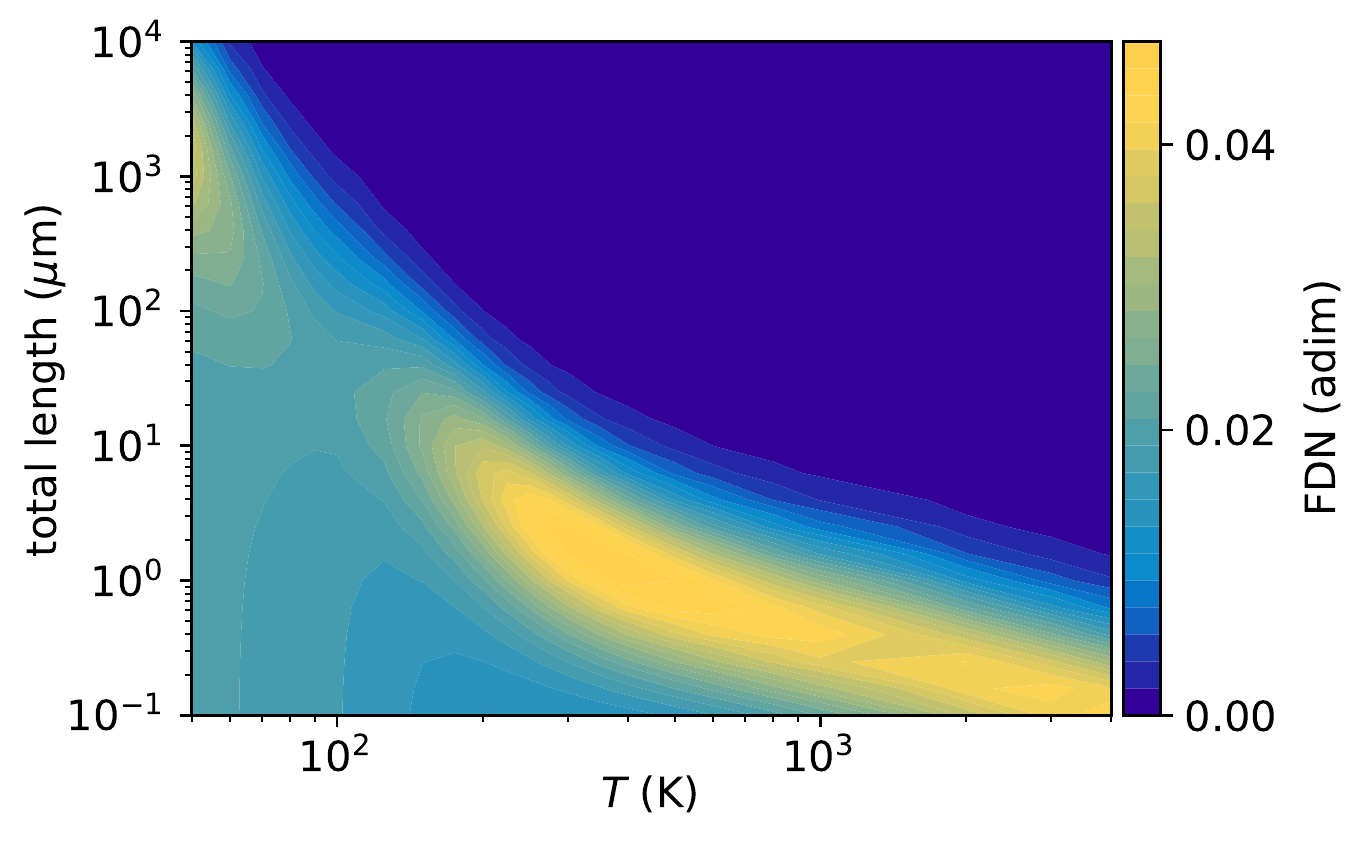}
  \caption{\added{Hydrodynamic behavior in crystalline diamond at natural isotopic abundance: increasing the dimension of the diamond sample can yield an increase of the hydrodynamic effects at low temperatures.}}
  \label{fig:diamond_large}
\end{figure}

\added{In this section we show how extending the sizes in Fig.~\ref{fig:deviation_Fourier}e up to 10 mm yields another peak for FDN for dimensions around 1 mm and temperatures around 50-60 K. This result has been confirmed by full solutions of the viscous heat equations similar to these reported in Fig.~\ref{fig:deviation_Fourier}b.}
\added{
The increase of FDN at low temperature and large size reported in Fig.~\ref{fig:diamond_large} is in qualitative agreement with the predictions for the second-sound window performed using the reduced isotropic crystal model or the Callaway model~\cite{second_sound_diamond}.
However, in contrast with this work, Ref.~\cite{second_sound_diamond} reports that an isotopic concentration much lower than the natural isotope abundance is necessary for the observation of second sound in diamond.}

\added{The analysis here and in Fig.~\ref{fig:deviation_Fourier} are limited to a minimum reference temperature of $\bar{T}=50$ K because, for lower temperatures, anharmonicity becomes decreasingly smaller while size effects become increasingly larger, resulting in convergence issues for the numerical calculations (which would also require a refined, computationally-expensive, treatment of surface effects~\cite{cepellotti2017boltzmann}).
Most importantly, the temperature/size range reported in Fig.~\ref{fig:deviation_Fourier} highlights the most accessible experimental conditions under which hydrodynamic behavior can appear in diamond, namely micrometer-sized diamond samples (thus much less expensive than millimeter-sized samples) and non-cryogenic temperatures.
}

\section{Computational details}
\added{
First-principles calculations are performed with the Quantum ESPRESSO distribution \cite{giannozzi2009quantum,giannozzi2017advanced}.
For all the materials analyzed, the LDA functional is used due to its capability to accurately describe the structural and vibrational \cite{RevModPhys.73.515} properties of graphite \cite{doi:10.1021/nl502059f}, diamond \cite{fugallo2013ab} and silicon \cite{cepellotti2016thermal}, and its compatibility with the D3Q code~\cite{paulatto2013anharmonic,paulatto2015first} for first-principles calculations of anharmonic (third-order) interatomic force constants.
Details on the computation of the second- and third-order interatomic force constants are reported in Ref.~\cite{cepellotti2016thermal} for silicon, in Ref.~\cite{fugallo2013ab} for diamond and in Ref.~\cite{doi:10.1021/nl502059f} for graphite.
The LBTE's scattering matrix $\Omega_{\state \statep}$ is computed as in Ref.~\cite{fugallo2013ab} and accounts for third-order anharmonicity \cite{paulatto2013anharmonic} and isotopic disorder \cite{Garg_PhysRevLett,tamura1983isotope} at natural abundance.
Thermal conductivity and viscosity calculations for silicon and diamond are performed using $27{\times}27{\times}27$ q-point grids and a Gaussian smearing of $2\;{\rm cm}^{-1}$ and $8\;{\rm cm}^{-1}$, respectively. Thermal conductivity and viscosity calculations for graphite are performed using a $49{\times}49{\times}3$ q-point grid and a Gaussian smearing of $8\;{\rm cm}^{-1}$.
The use of an odd, Gamma-centered q-points mesh is crucial to correctly account for the parity symmetries of the scattering operator.}


%

\end{document}